# Urban Food Self-Production in the Perspective of Social Learning Theory: Empowering Self-Sustainability

**Ewa Duda**
**Adamina Korwin-Szymanowska**

The Maria Grzegorzewska University Press

# Urban Food Self-Production in the Perspective of Social Learning Theory: Empowering Self-Sustainability

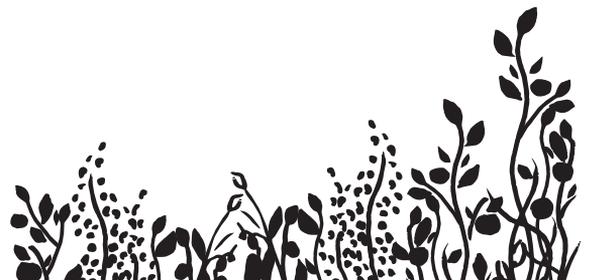

# Urban Food Self-Production in the Perspective of Social Learning Theory: Empowering Self-Sustainability

**Ewa Duda**
**Adamina Korwin-Szymanowska**

Scientific reviewers:
Full Professor Anna Odrowąż-Coates, Dr. hab.
Professor Aleksandra Tłuściak-Deliowska, Dr. hab.
Mari Hanssen Korsbrekke, Dr.

Funding information:
This publication was funded by the EEA / Norway Grants 2014–2021 and the state budget of Poland through the National Centre for Research and Development under grant agreement no. NOR/IdeaLab/SmartFood/0005/2020, awarded to Ewa Duda.

Editor:
Ewa Duda

Cover:
Andrzej Gmitrzuk

Composition:
Grafini DTP

Publisher:
Wydawnictwo Akademii Pedagogiki Specjalnej –
Maria Grzegorzewska University Press

# Table of contents

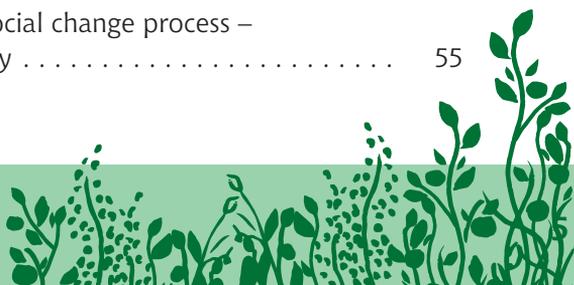

# Introduction

The introduction to our monograph presented below aims to encourage the reader to delve into its contents. We hope that it will serve as a friendly guide through the book and help understand the authors' intentions behind the work on the text.

Why have we written this book? The monograph is one of the results of an innovative project in which the authors had the pleasure of participating. This project was indeed unique. From the initiative of the National Centre for Research and Development, an unusual idea emerged to create a space for researchers to conduct studies that have little chance of being realised within regular scientific grant competitions. Thus, the IdeaLab concept was born. The first stage of the application process involved responding to an invitation to apply for participation in the IdeaLab workshops funded by the European Economic Area (EEA) and Norwegian Financial Mechanism 2014–2021, held in a town outside Warsaw on March 2–6, 2020. As a result of the competition, one of the monograph's authors participated in these workshops.

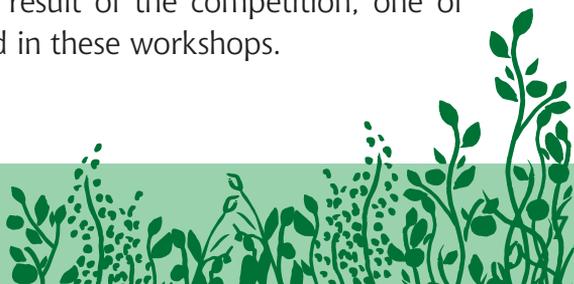

The workshops aimed to develop unique ideas for interdisciplinary research projects focused on services and solutions for the cities of the future. Through an unconventional organisational approach, the workshops created a space for researchers from Poland and Norway, representing various fields, to develop solutions addressing the needs and unforeseen challenges of future cities. Thus, from the meeting of the selected researchers and the atmosphere of collaboration and creativity, the idea for the SmartFood project: Engaging citizens in food diversity in cities was born. In June 2020, we submitted a full project proposal, which received a positive funding decision in July 2021.

On September 1, 2021, we began implementing the project within a consortium that included the Research and Innovation Centre Pro-Akademia, the Project Promoter responsible for urban food production, nutrition, energy systems, and project management; Cracow University of Technology, responsible for the supply of water to the novel food production system, rainwater management, resource efficiency improvements, and co-production of prototypes; Maria Grzegorzewska University, responsible for educational aspects of the project implementation, adaptation of the system to the needs of the population, impact assessment, and user engagement; the Norwegian Institute for Air Research (NILU), responsible for electronics and communication technology for various applications, including sensor development, data processing, and application and game development; BI Norwegian Business School, responsible for assessing the impact of smart city technology on citizen wellbeing, conducting field experiments, and designing rewarding incentives; and Western Norway Research Institute, responsible for co-producing a climate change adaptation strategy to ensure the project's environmental impact, as well as designing a social integration strategy.

The project turned out to be not only ambitious but also difficult to implement. It involved selecting a community of apartment block residents. This unique project on a national scale aimed to install twenty hydroponic cabinets in the corridors of the selected apartment block, where residents would grow edible plants for one year. To our knowledge, no such project had been undertaken in Poland before. While hydroponics is not an unknown technology, it has so far been used by companies rather than individual users. Several communities applied for the project, but after the project team inspection, it turned out that most of them could not participate due to the technical conditions of the building that did not meet the required project standards. Firstly, the size of the staircases was a problem, making it impossible to install the cabinets. Secondly, the project required the installation of a photovoltaic system on the building's roof and a rainwater collection system to supply the cabinets with energy, lighting, and water. It was not possible to install such systems everywhere. As a result, one community was selected, which we wanted to get to know in the first phase of the research. We conducted in-depth interviews with them to learn about their attitudes towards food cultivation and their expectations for the project. However, the residents' association withdrew from the project, fearing the loss of the building's warranty due to the required installations' interference during the project.

Thus, we began the search for another community, this time in Warsaw. The search was successful, and a community was selected, with whose residents we conducted a second series of interviews. In this way, we gathered research material for the presented monograph. The research was therefore part of the project and aimed to understand the people who joined this unique initiative, designed to provide innovative socio-technological solutions for sustainable food production and consumption toward a sustainable, smart city of the future. This goal is to be achieved by involving the local community in self-sufficient food production and changing household behaviours to (1) improve health, (2) reduce greenhouse gas emissions and energy waste, and (3) enhance social integration and (4) increase environmental awareness among residents.

The next two chapters are dedicated to presenting the theoretical frameworks adopted in our study. Due to the desire to provide the read-er with a comprehensive account of the theoretical frameworks that underpin our study and the analytical techniques employed in the data analysis, these two chapters have been separated from the subsequent chapters that describe the methodology adopted. The second chapter is devoted to the theory of social learning, as the experiment planned in the project is to be carried out within the social environment of a single apartment block. We are therefore interested in the learning process in a social dimension. The third chapter is dedicated to the theory of diffusion of innovations, considering the participants of the research sample as one of the initial links in the implementation of hydroponic systems for urban food cultivation.

In the fourth chapter, we present the methodological assumptions of our research. This chapter presents a more detailed justification for undertaking the research topic, outlines the research questions that determined the direction of the analysis, and provides an overview of the location context of the research. It also describes the characteristics of the research sample and the method of analysing the collected research material. Then, in the fifth and sixth chapter, we present the results of the qualitative analyses conducted. The seventh chapter is devoted to discussing the obtained results in relation to the adopted theoretical frameworks and previous research on the subject.

We hope that the monograph will be received kindly. We believe it will make a valuable contribution to the scientific discourse on education by raising awareness on promotion of urban food cultivation and that it will serve as a foundation and inspiration for further interesting research. We wish you pleasant reading.

*Ewa Duda and Adamina Korwin-Szymanowska*

# Growing food independence: Contemporary approaches to self-food production

## 1.1. The role of plants in human life

Plants are an integral part of life on Earth. Their development has been a complex and dynamic process (Pennington et al., 2002; Hoson, 2014). The transition of plants from water to land was a key event in history, leading to the development of Earth, the emergence of diverse ecosystems, and the evolution of various species (Kenrick et al., 1997), which consequently resulted in interdependence between diverse organisms (Fenster et al., 2004). The establishment of mutualistic relationships influenced the reproductive strategies of plants and their ecological success (Kiers et al., 2010), enabling plants to dominate the globe and become an essential component of life for many living organisms.

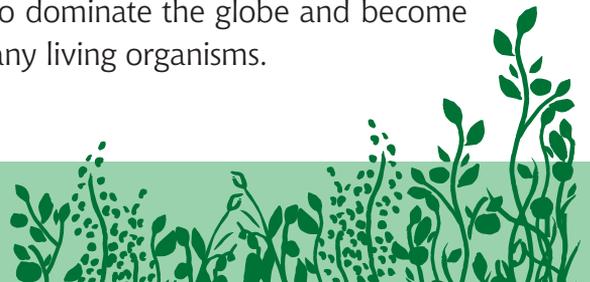

The history of plant cultivation by humans is tied to their survival in the world. The transition from a hunter-gatherer lifestyle to conscious cultivation of plants marked a crucial change in human survival strategies, as well as the creation of mutual relationships and dependencies between humans and plants. Throughout history, plants have been a primary source of food, providing essential nutrients and ensuring proper micro and macronutrients. Plants were also an integral part of traditional medicine and healing practices, deeply rooted in human history. The phytochemical compounds of plants have been used in treatments supporting human health. Herbal medicine has remained an important aspect of basic healthcare for people worldwide for centuries (Dogor et al., 2018). Furthermore, plants have found their place in cultural practices, including religious and shamanic rituals, allowing for mystical experiences and transcendental states.

The extensive use of plants in human life led to deliberate cultivation aimed primarily at ensuring human well-being and food security, interpreted by Sowa and Bajan (2019) as having reliable access to sufficient, affordable, and nutritious food. This prompted people to engage in self-sufficient food production, leading to the creation of specific cultivation spaces, eventually known as home gardens. Over the centuries, home gardening and domestic food production have played a crucial role in human life, fitting into the natural cycle of life, and offering numerous physical, psychological, and social benefits. From a biological perspective, home gardens contribute to human health by providing access to fresh, nutritious, and valuable products. The diversity of plant species grown in gardens supports the nutritional needs of individuals and their families, ensuring a continuous supply of food resources (Ivanova et al., 2021). The cultivation process involves physical activity essential for maintaining health and well-being (van den Berg et al., 2010). In addition to biological aspects, it is worth noting that home gardening has a significant impact on mental health and psychological well-being. The sensory experiences and therapeutic nature of gardening contribute to building a sense of connection with nature and fulfilling needs, providing a sense of purpose (Newton et al., 2021). Gardening is associated with stress reduction, improved mood, and increased mental resilience, allowing for better coping with everyday challenges (Zhang et al., 2021).

The practice of self-sufficient food production through home gardening also fosters a sense of empowerment and self-sufficiency, contributing to the overall well-being of individuals (Patalagsa et al., 2015). This empowerment is particularly significant as it enables individuals to actively engage in activities that support their physical and economic well-being. Moreover, the environmental and ecological dimensions of home gardening are integral parts of the natural life cycle, reflecting a harmonious relationship between humans and nature and emphasising the interconnections between human life and the environment. Simultaneously, home gardening aligns with ecosystem biodiversity and sustainable development, adhering to principles of environmental management and resource conservation (Calvet-Mir et al., 2012).

The concept of self-sufficient food production has evolved over the centuries, initially being closely linked to the issue of self-sufficiency (Bikernieks, 2022). Historical analyses conducted by Luan et al. (2013) indicate that self-sufficiency depends not only on production but also on consumption. Both aspects contribute to an integral process of survival by ensuring food security, as mentioned by Baer-Nawrocka and Sadowski (2019). According to Pradhan et al. (2014), this contributes to certain changes in consumer behaviours and producer practices that promote the consumption and production of local and regional food, thereby connecting self-sufficiency with locality. The process of self-sufficient food cultivation contributes to biodiversity by diversifying crop selection at the household level (Simelton, 2011; Kc et al., 2015). Considering contemporary trends, self-sufficient food cultivation and the need for self-sufficiency create a space for reflection on this issue.

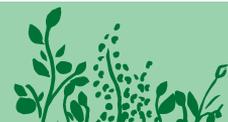

## 1.2. Contemporary challenges and trends in the context of urban food self-sufficiency

Self-sufficient food production refers to individuals, households, or communities who grow, produce, and procure their own food to meet their nutritional needs and achieve self-sufficiency or semi-sufficiency in food supply. This concept encompasses various activities such as home gardening, urban gardening, urban agriculture, and local food production, all aimed at increasing food security while promoting a sustainable lifestyle, caring for personal health, pleasure and reducing dependence on external food sources. Self-sufficient food production includes growing fruits, vegetables, grains, and other food products, which are typically pesticide-free. In a broader context, this ensures high food quality, stability of access, locality, and quick delivery times, which can result in reduced food waste. This concept is closely linked to regional food self-sufficiency and the development of local food economies, which play a significant role in promoting healthy and resilient food systems. Such a view of self-sufficient food production integrates various perspectives and contexts important for sustainable development and community well-being.

Currently, there are many overlapping trends and concepts within self-sufficient food production. Among them are urban farming and urban gardening, which involve growing and producing food in urban areas. According to Martellozzo et al. (2014), these practices are often undertaken by households or communities to increase food security and promote sustainable living. This can be achieved through complete self-sufficiency in food or partial self-production and consumption. Food prosumers produce food for their own needs, fitting into the contexts of sustainable consumption and self-sufficiency in food (Trębska et al., 2022), as well as consuming what is produced within the global agricultural economy.

The introduction of gardening into cities and towns supports local food production by reducing long supply chains and enhancing health benefits, flavour, and authenticity of products (Autio et al., 2013). It also meets various human needs – from physical health related to natural nutrition or allergy prevention to psychological and social needs associated with behaviours promoting sustainable development. Adopting such practices often stems from beliefs about the necessity of protecting the planet, driven by ecological awareness that dictates pro-environmental behaviours (Salciuviene et al., 2022).

This locality in food production is often viewed through the lens of food self-sufficiency, which was particularly emphasised during the Covid-19 pandemic and the war in Ukraine, where food security, consumption, production, and supply chains were disrupted. Phenomena occurring in the contemporary world influence processes related to self-sufficient food cultivation. It encompasses a whole spectrum of socio-cultural, political, ecological, economic, technological, and informational transformations, reflecting the multifaceted nature of food production and its connections with various fields. Changes in one area have direct or indirect consequences for the other spheres. The acceleration of changes in reality fits into the postmodern characteristic of the world, marked by speed, uncertainty, ambiguity, instability, and contradiction, where all forms of antinomies coexist, preventing the adoption of a single pattern of human functioning.

This is reflected in megatrends, which have been defined and described since 2018 by the forecasting institution called *Infuture Institute* in the form of a Trend Map, presenting their consequences for communities and economies in the current, short-term (up to 5 years), medium- (up to 15 years), and long-term (15+ years) perspectives. In 2023, Infuture Institute analysts identified five key megatrends in the areas of society, environment, technology, and legal changes, and the economy, which simultaneously intersect and exclude each other. These trends include *demographic shifts, mirror world, symbiocene, bioage, and multipolar world.*

According to the authors of the report, *demographic* changes have been contributing to the alteration of social structures worldwide for years. The authors argue that technological advancements, medical developments, increasing life expectancy, and low birth rates that do not ensure generational replacement, along with the liberalisation of social norms and customs, lead to diverse consequences. These consequences can be short-term, such as the need to address on-going issues like resolving intergenerational conflicts in the workplace, and long-term, such as creating economic visions for ageing societies. Demographic changes can influence self-sufficient food cultivation within the trend associated with the phenomenon of ageing societies (the "silver tsunami"), which involves an increase in the number of elderly people in European societies. This leads to changes in the demographic and social structure, necessitating reflection on how to meet the needs of both entire communities and individuals.

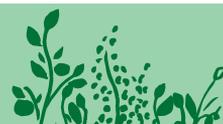

Self-sufficient food cultivation is also linked to the trend of achieving mental well-being, which involves actions aimed at mental health and combating loneliness, social isolation, and solitude. The concept of the *mirror world*, which aims to fully replicate physical reality in the digital realm, involves transferring all real-life activities into the digital space. Work, business, entertainment, education, development, and relationships should find their way in the digital world. The authors highlight that the Covid-19 pandemic has significantly accelerated this process, but it requires further technological development and financial investment over time.

The above-mentioned mirror world can currently impact self-sufficient food cultivation by providing access to knowledge on how to do it. The advent of the Internet has revolutionised the way individuals access information, learn new skills, and engage in community life. In the context of self-sufficient food production, the Internet has become a vast repository of knowledge, offering a wide range of resources, from instructional videos and online courses to forums and social media groups dedicated to sharing knowledge and experiences related to cultivation. The availability of information on the Internet has democratised knowledge about self-sufficient food production, allowing individuals to independently acquire knowledge, learn, and implement various methods regardless of their place of residence, geographical location, or access to traditional educational resources.

Online platforms provide a wealth of diverse resources related to organic farming, permaculture, food preservation, and urban gardening, enabling individuals to gain knowledge and skills that were previously limited to formal agricultural education or local farming communities. Social media platforms, in turn, play a significant role in disseminating information and influencing individuals' attitudes and behaviours regarding self-sufficient food production. Thanks to the information available there, people can gain knowledge and access to best practices or innovative cultivation methods, which can encourage them to explore this topic independently.

The Internet also facilitates the exchange of ideas and knowledge among various communities interested in self-sufficient food production. Online forums, social media groups, and virtual communities provide individuals with the opportunity to seek advice, share experiences, and learn from one another. This collaborative and interactive nature of the Internet fosters a culture of continuous learning and knowledge sharing, empowering individuals on their path to self-sufficiency.

In light of the above concept of the mirror world and the importance of the Internet, it can be inferred that lack of access to the digital world may deepen digital inequalities. However, do people truly feel the need to transfer their lives online? The period of the Covid-19 pandemic offers an example that both confirms and contradicts this thesis. On one hand, part of human life, including relationships and work, moved online; on the other, it intensified the need to return to roots and nature.

The Covid-19 pandemic led to increased interest in self-sufficient food cultivation and participation in gardening practices, as reflected in studies showing a rise in the number of home gardeners before and after Covid-19 (Park et al., 2021). This increased interest has been attributed to the pandemic's impact on food security, movement restric-tions, and disruptions in the supply chain to urban centres. Covid-19 significantly affected various aspects of social life, including food production and security. Disruptions in agricultural supply chains caused supply and demand shocks, negatively impacting food security (Al Nemer, 2023). During this time, trends in self-sufficient food production gained importance leading to increased demand for home-cooked food (Balagtas et al., 2023). This shift in consumer demand influenced how people acquired and consumed food, leading to renewed interest in gardening and home food production (Mullins et al., 2021) as a way to support community resilience (Khan et al., 2020).

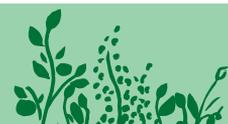

Post-Covid studies have shown that home gardening, aside from consumption, was also associated with attempts to stabilise mental health, physical, and psychological well-being, particularly among the elderly (Corley et al., 2021; Egerer et al., 2022). The therapeutic effects of gardening were highlighted as a means to alleviate stress and strengthen bonds with nature during the pandemic (Egerer et al., 2022; Zhang et al., 2021). The horticultural therapeutic effect of gardens has been demonstrated to have a scientific basis, with a particular focus on the non-pathogenic bacterium *Mycobacterium vaccae*, which is found in soil. This bacterium has been extensively studied for its potential impact on human health. Studies on mice and rats have shown that it creates an anti-inflammatory environment in the central nervous system, alleviating neuroinflammatory and behavioural effects of stress and increasing the resilience of mice to stress (Frank et al., 2018; Sanchez et al., 2022). Additionally, Mycobacterium vaccae has anti-inflammatory and immunoregulatory properties, making it a potentially useful remedy against the negative effects of stressors (Fonken et al., 2018; Foxx et al., 2021).

The pandemic emphasised the importance of home gardening in ensuring household food security and quality of life, particularly in urban areas (Dissanayake et al., 2020), where access to fresh food was limited due to transportation stoppages and movement restrictions. Lockdown gardening proved to be a crucial element in enhancing local food production while mitigating the negative effects of global food shocks and price fluctuations (Perera et al., 2021). Additionally, the pandemic prompted households to re-evaluate the functionality of their home food environment, often leading to changes in dietary practices, such as reducing food waste and cooking at home more frequently (Qian et al., 2020), aligning with household economics.

Another megatrend identified by the Infuture Institute, ushering humanity into a new era, is the Symbiocene. This proposed new geological era is characterised by harmonious coexistence and mutual benefits among all living beings, providing a potential solution to the climate crisis (Mead et al., 2023). It emphasises the interrelationships between individuals, people, and nature, striving to shift from the dominance of the Anthropocene to appreciating diversity, multilateralism, and cooperation. This concept is rooted in the idea of symbiosis, where living beings coexist, in contrast to the human-centred focus of the Anthropocene (Prescott et al., 2017). The term *Generation Symbiocene* has been coined to humanise the changes needed to transition from the present to the future, emphasising the need for an emotional revolution (Albrecht, 2019; Albrecht, 2020). Moreover, the Symbiocene is characterised by human intelligence and praxis that mimic the symbiotic and mutually reinforcing forms and reproductive processes of life found in living systems (Rahayu, 2023). Its construction is based on the latest scientific discoveries on symbiosis and its key role in sustaining life, highlighting the importance of ecological balance and planetary health (Albrecht, 2019).

The Symbiocene represents a shift toward a more sustainable and mutually beneficial relationship between humans and the environment, offering a new perspective on addressing contemporary challenges. The concept was introduced by Glenn Albrecht, who coined the term on his blog in 2011 (www.symbioscene.com/invitation-to-the-symbiocene). The departure from nature, excessive exploitation of natural resources, and environmental destruction have led to a redefinition of human-nature relations, moving away from anthropocentrism toward an ecocentric approach, which justifies the existence of the natural environment as an equal component of a holistic ecosystem. According to Albrecht, the Symbiocene will be characterised by symbiocentric human intelligence that replicates symbiotic and mutually reinforcing forms and processes of life present in all living systems in all aspects of social life. This perspective emphasises the interconnections between human well-being and ecosystem health, highlighting the importance of recognizing and preserving the services provided by natural systems for the benefit of all life forms (Hernández-Blanco et al., 2022).

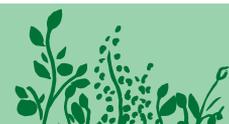

The Symbiocene's assumptions include the flourishing of positive Earth-related emotions, aiming to social and ecological homeostasis (Albrecht, 2019; Benatar et al., 2018). This transition requires a transformation of political power, political economy, and policy based on ethical commitments, emphasising the interconnections between human well-being and ecosystem health, as well as transforming education. According to the Infuture Institute, a symbiocentric approach entails legal changes, such as granting legal personality to animate and inanimate nature, economic changes aimed at transitioning to a circular economy or doughnut economics, and operational changes in management, service design, and products that will consider not only human interests. Hence, within this megatrend, the authors of the report identify the following trends: non-human rights, bio-architecture, resource crisis, e-resource recovery, disconnection effect, Internet of beings, resilience, social economy, circular economy, self-sufficiency, and mental well-being.

The trend strongly resonating with the Symbiocene is the Bioage, which enters the space of sustainable reality transformation. According to analysts at the Infuture Institute, the breakthrough lies in the development of technology, which at this stage integrates nature with synthetic materials created by humans or replaces and enhances it through direct intervention in nature. Regardless of the ethical aspects of such actions, this requires the development of a range of technological processes, such as genetic engineering or implantology, which not only allow for the replacement of nature but primarily for its genetic alteration. "This requires extensive implantation, the use of bio- and nanotechnologies, genetic engineering, tissue engineering, solutions from the field of synthetic biology, and so-called Human Enhancement Technologies (HET) or those eliminating the ageing process. The Bioage is an era in which humans, through technology, are able to improve and design what is biological – themselves and all other living organisms" (https://infuture.institute/mapa-trendow), hence, the development of biomaterials, bioenergy, bioarchitecture, and biocultural systems. These changes are linked to the need to protect natural resources, leading to the creation of legal regulations in which nature gains embodied and personified dimensions. This also enters the field of bioethics.

The last highlighted megatrend is the multi-polarisation of the world, which creates a perspective of interpreting the world through the prism of crises. Currently, we are dealing with geopolitical, climate, resource, economic, health, medical, and social crises, which underlie the disruption of societal functioning coherence, leading to the polarisation of attitudes, behaviours, or values in every area of our lives, manifesting as antinomies, making it impossible to establish a common vision for dealing with the problems of reality. This leads, among other things, to the effect of detachment, which the authors of the Infuture Institute report describe as the separation of areas that were traditionally inseparably linked (e.g., the severance of the human-nature connection and simultaneous deprivation of subjectivity from nature through its free modification within bioengineering), hyperlocality, i.e., meeting the needs of communities through locality and shortened supply chains, which in the perspective of 15–20 years is expected to lead to a greater development of self-sufficiency as a trend supporting individual autonomy in various areas of life.

The above-presented issues affect the narrative of the context of self-sufficient food production. The entire consideration of self-food cultivation encompasses a diverse range of issues, from sustainable development and food security to socio-cultural identity and eco­nomic paradigms. The contemporary landscape of self-sufficient food production is shaped by factors such as climate change, economic in­stability, resource constraints, globalisation, and population growth, which affect the ability of the global food system to meet human nutritional needs (Dorward et al., 2016). One of the key challenges in self-food cultivation is the concept of sustainable development, which refers to the ability to meet the needs of contemporary socie­ties without compromising the ability of future generations to meet their own needs. It involves responsible and sustainable use of re­sources, environmental management, social justice, and economic profitability to ensure long-term prosperity for future generations. Although this approach assumes an anthropocentric view of the nat­ural environment, it is worth emphasising that it takes into account efforts to maintain a balance between human needs and natural re­sources that will enable addressing the complex challenges of reality and the future while promoting resilience, adaptability, and ethical decision-making in various aspects of human life, including agricul­ture, industry, and education.

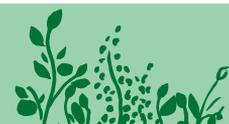

## 1.3. Development of urban agriculture in the perspective of global trends

Urban agriculture has a long history and significant place in human life. It emerged with the development of cities (de Bon et al., 2010), and its diverse forms evolving over time were usually responses to var­ious challenges, reflecting the adaptive capabilities of urban agricul­ture (Schoen et al., 2021) considering the needs of individuals, regional possibilities, climate, available technologies, and cultural preferences (Lovell, 2010). The term "urban agriculture" can be defined as a de­liberate effort undertaken by individuals or communities to increase self-sufficiency and prosperity through the cultivation of plants and animals in urban or suburban areas (Hardman et al., 2022; Komalawati et al., 2022). It also includes diverse and sustainable small-scale agriculture within city limits, emphasising the integration of agriculture with urban economic and social systems (Dobbins et al., 2021). Urban agriculture is perceived not only as a sustainable practice bringing social, economic, and environmental benefits but also as contributing to food security, community prosperity, and urban resilience to various fluctuations (Othman et al., 2018), especially in difficult times such as the Covid-19 pandemic (Komalawati et al., 2022) or wartime.

There are many diverse forms of urban agriculture, some of which will be discussed in the following paragraphs.

## Vertical farming

Vertical farming utilises vertical space for plant cultivation in multi-story buildings, often using hydroponic or aeroponic systems, allowing for spatial production maximisation (Zaręba et al., 2021; Avgoustaki et al., 2020; Beacham et al., 2019). Due to vertical space utilisation, vertical farming has become an intriguing alternative to traditional agriculture, offering a range of potential benefits for sustainable production. According to Jürkenbeck et al. (2019), vertical farming can provide higher yields per square metre than conventional agriculture, increasing land use efficiency for crop production and offering the opportunity to grow food in areas with unfavourable climatic conditions. Figure 1 illustrates an example of this utilisation of building space, situated in Nicosia, which employs a vertical cropping system. Moreover, as observed by Martin et al. (2019), vertical farming reduces the environmental footprint associated with plant production or transportation, allowing for the strengthening of local, resilient food production, emphasising its potential for sustainable production.

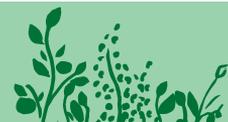

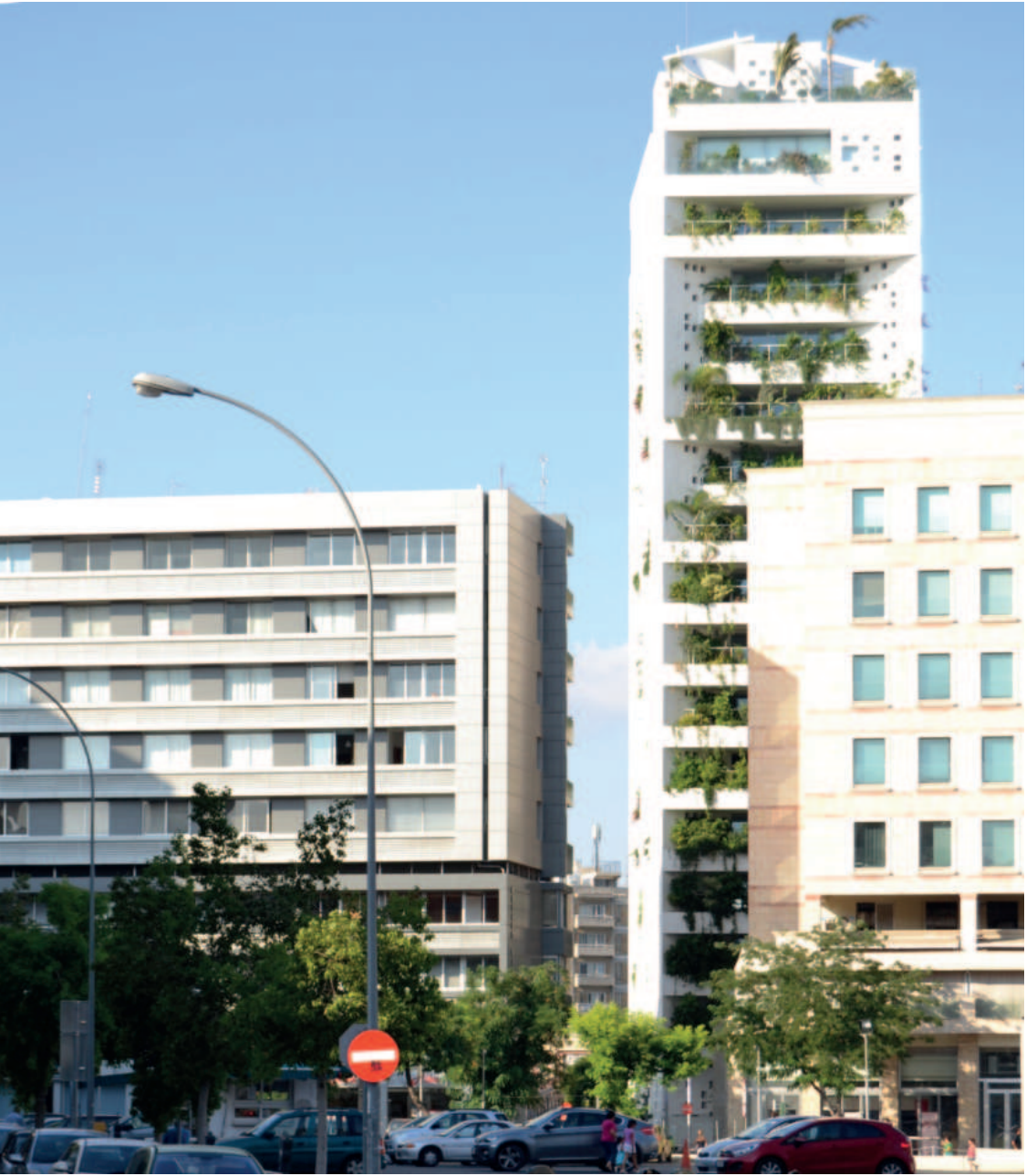

**FIGURE 1.** Nicosia, Cyprus. Example of vertical farming
Photo by Ewa Duda

## Rooftop agriculture

Another interesting form of urban agriculture is rooftop cultivation. Rooftop agriculture, also known as green or living roofs, is an innovative and sustainable approach to shaping the urban landscape, involving the cultivation of vegetation on building rooftops. They offer a range of environmental, social, and economic benefits, contributing to sustainable development and urban resilience. Green roofs can be divided into two main types: extensive and intensive.

Extensive green roofs are characterised by a thin soil layer and low-maintenance vegetation such as sedums, mosses, and grasses. They are lightweight, making them suitable for retrofitting existing buildings. Extensive green roofs are known for their ability to retain rainwater. Some research indicates that they can retain a significant percentage of incoming rainfall, contributing to reducing runoff and improving water management in urban areas (VanWoert et al., 2005). They also offer thermal insulation, reducing the urban heat island effect and reducing energy consumption in buildings (Kim et al., 2004). Extensive green roofs are often used to increase biodiversity as they provide habitat for wildlife.

On the other hand, intensive green roofs are characterised by a thicker soil layer and greater plant species diversity, including shrubs, trees, and even agricultural crops. They require more maintenance and structural support, making them suitable for larger buildings and new construction projects. Intensive green roofs offer greater opportunities for urban agriculture, providing space for vegetable gardens, orchards, and recreational areas. Studies have shown that intensive green roofs have the potential to sequester more carbon dioxide and mitigate climate change compared to extensive green roofs, due to their deeper soils and diverse vegetation (Ismail et al., 2019). They also offer aesthetic and recreational benefits, serving as outdoor green spaces for residents and contributing to psychological well-being (Rahman, 2023).

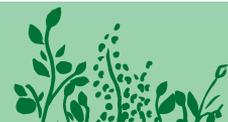

Regardless of the adopted solution, it is worth emphasising the benefits of shading and thermal and acoustic insulation associated with roof greening (Lee et al., 2013). Plants on rooftops improve the microclimate, reduce carbon dioxide, and release oxygen, while also purifying the air from dust, pollen, and pollutants. They also help retain more water, which evaporates into the atmosphere, increasing air humidity and preventing it from entering the sewage system. It is also worth emphasising the aesthetic qualities of such buildings. Rooftop gardens can also create space for wild pollinators, which is a valuable practice for their protection and promoting urban biodiversity while maintaining the diversity of planted plants. An example of a roof garden is shown in Figure 2.

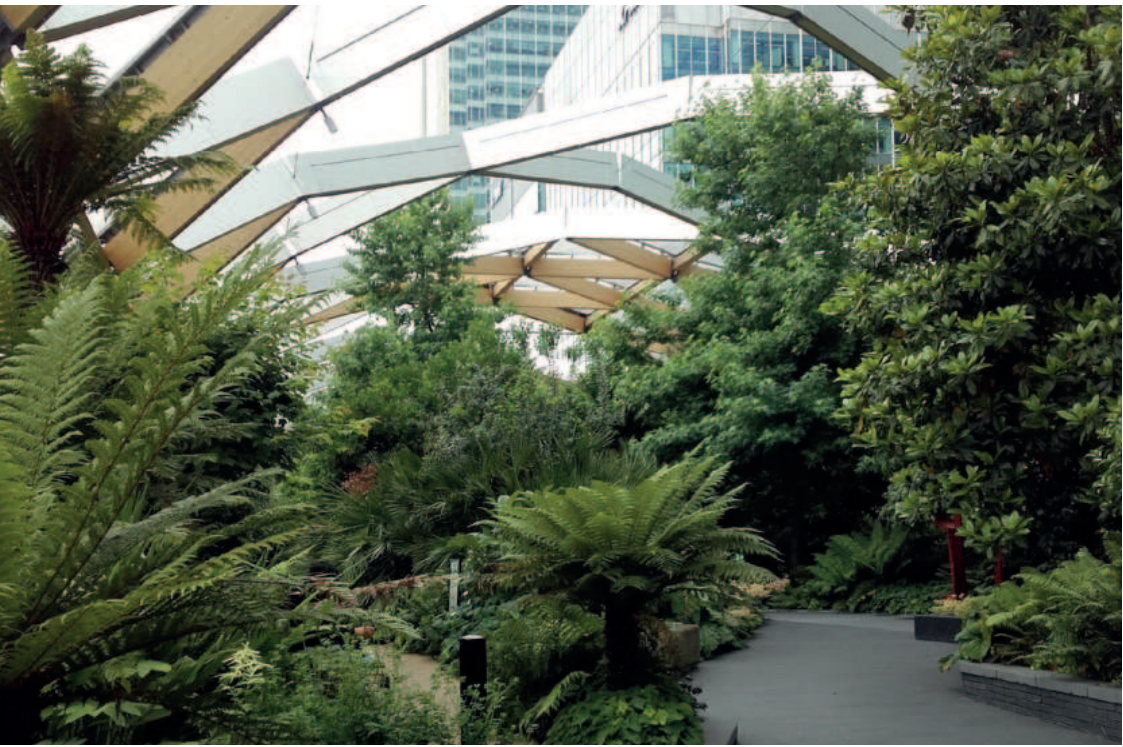

**FIGURE 2.** London, United Kingdom. Crossrail Place Roof Garden
Photo by Ewa Duda

## Greenhouse farming

The evolution of greenhouse cultivation from simple, covered rows of field crops to highly sophisticated controlled environment agriculture (CEA) facilities has created an image of urban plant factories (Shamshiri et al., 2018). Greenhouse farming addresses the problem of limited space by utilising technological innovations related to soilless production, such as hydroponics, aeroponics, and aquaponics, which increase the potential for sustainable food production in urban environments (Al-Kodmany, 2018; Sanyé-Mengual et al., 2019). Cur-rently, the idea of greenhouses in its basic understanding as a place that creates a very good microclimate for the cultivation and growth of plants through optimal temperature, proper humidity levels, and protection against variable weather conditions, as well as partially against pests, is focused on increasing control over these parameters, introducing the concept of smart greenhouse agriculture, which involves digitising agriculture using modern information and communication technologies for the automation of agricultural processes (Orakwue et al., 2022). However, it is worth emphasising that greenhouses in their basic understanding are still used in household farms. An example of this is shown in Figure 3.

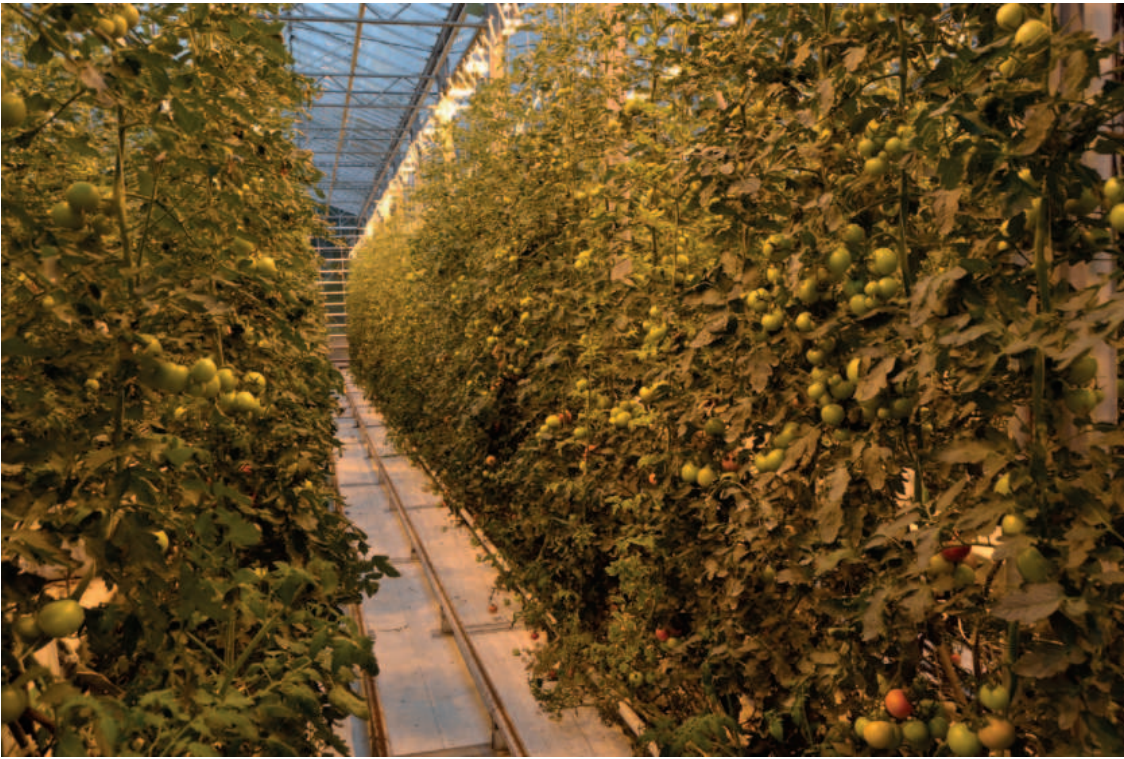

**FIGURE 3.** Friðheimar, Iceland. Tomatoes grown all year round, despite long dark winters, with artificial lighting in greenhouses powered by geothermal energy
Photo by Ewa Duda

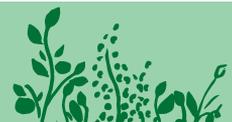

## Container farming

Container farming is an innovative and sustainable approach to urban agriculture, involving plant cultivation in shipping containers (Xi et al., 2021). This method primarily provides a controlled environment for plant production, enabling year-round cultivation regardless of external weather conditions by precisely managing factors such as temperature, humidity, and light, optimising plant growth and yields. This is particularly beneficial in urban areas where arable land may be limited. A huge advantage of containers is their portability, allowing for great flexibility in their placement, making them suitable for urban environments with limited space. Container farms can be established on rooftops, vacant lots, or other urban spaces, contributing to the efficient use of urban land for agricultural purposes. Container crops are usually integrated with technologies such as hydroponic or aeroponic systems, enabling efficient water use and reducing the overall environmental footprint of plant production. Such plants do not need to be imported from distant parts of the globe which solves the problem of climate change, and they do not need to be treated with antifungal or preservative agents, which can prevent al-lergies. An example of a building where green containers for planting greenery have been designed, next to the traditional balcony, is shown in Figure 4. Furthermore, the roof area within the same building is utilised for greenhouse farming.

Another illustration of container farming is presented in Figure 5, which depicts the utilisation of above-ground containers within a railway station space.

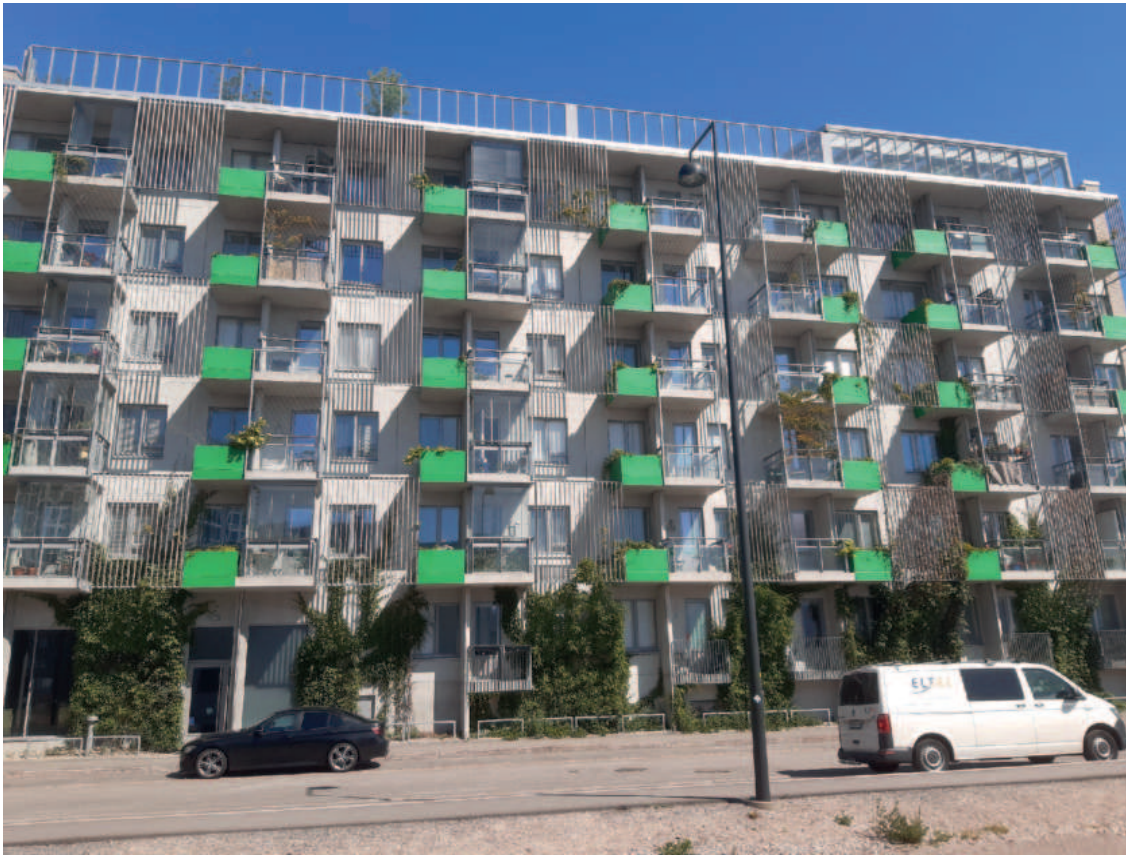

**FIGURE 4.** Helsinki, Finland. Built-in planting containers in the body of the building

Photo by Ewa Duda

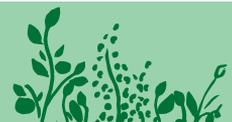

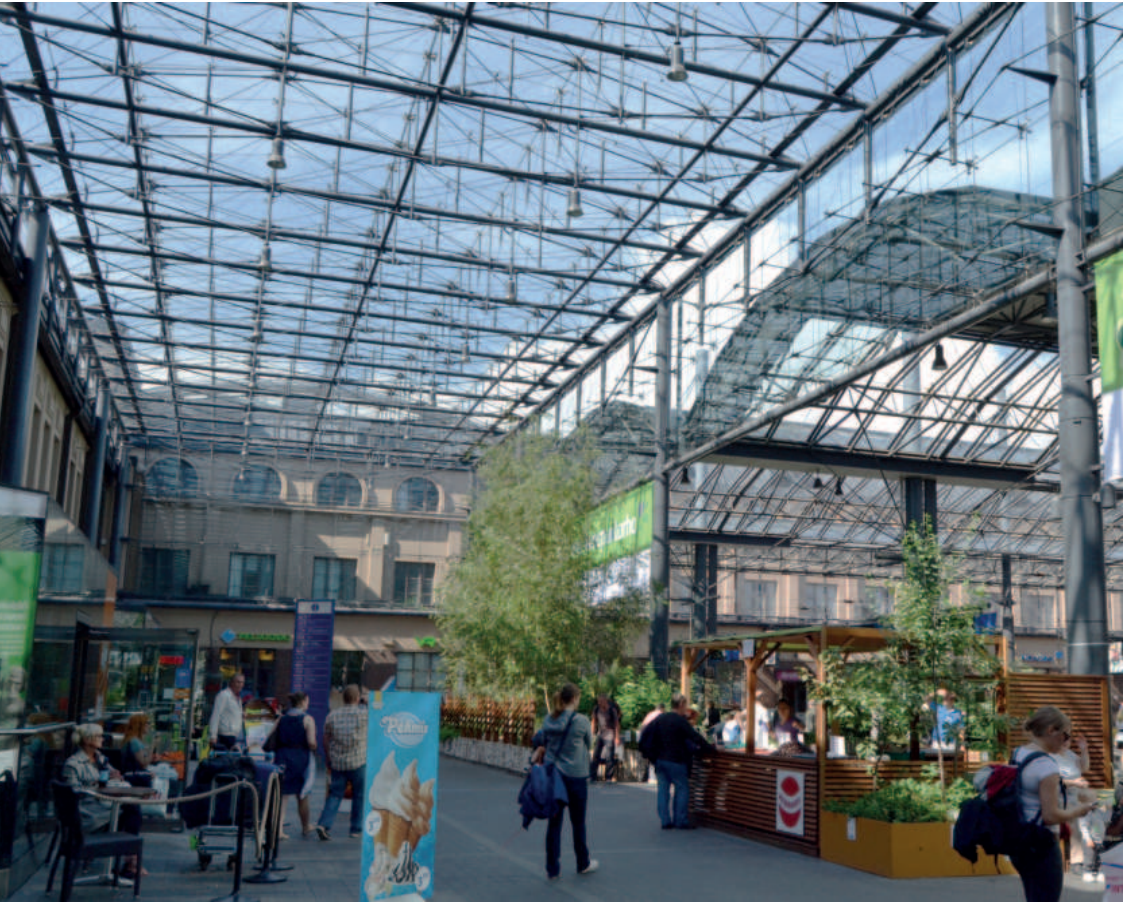

**FIGURE 5.** Jyväskylä, Finland. Ground farming containers at the railway station

Photo by Ewa Duda

## Community gardens

Community gardens have recently become an immensely popular form of gardening. They are the subject of extensive research in various disciplines, but their interest strongly fits into the social sciences (Guitart et al., 2012). Despite their community nature, it is worth emphasising their individual dimension, which allows us to see them as a place of self-realisation – meeting individual needs that lead to well-being, psychological resilience, stress reduction, and enhanced self-esteem (Okvat et al., 2011; McVey et al., 2018; Koay et al., 2020). Community gardens provide a feeling of pride, joy from achievements, or satisfaction with life (Koay et al., 2020). They are particularly important for people with mental illnesses (Wood et al., 2022), improving not only physical but also mental health. They also provide opportunities for sharing knowledge, developing skills, and sharing talents (Jackson, 2017; van Holstein, 2017).

Community gardens play a crucial role in promoting social integration, supporting local cohesion; they help to build networks and a social capital. They are created and maintained by their members-participants, leading to the development of social relationships (Okvat et al., 2011), which build community prosperity and integrate people with each other (Egli et al., 2016). These gardens create spaces that support multidimensional locality in social, economic, and ecological aspects (Cornfield et al., 2023). Not only do they create a space for sharing values and support (Tracey et al., 2020), but also educational places where knowledge, information, skills, and experience can be shared. It is where educational processes take place, regardless of age, in which social capital is produced, available, and used by the social network of gardeners (Glover, 2004). They are bastions of democratic citizenship and political practice, contributing to strengthening community positions and engagement (Ghose et al., 2014).

Furthermore, community gardens also serve environmental functions. As Falkowska (2021, p. 31) states, "Community gardens did not arise naturally, yet they significantly influence the development of the ecosystem structure and contribute to the formation of a local biotope. Meanwhile, fauna acts as consumers. An example can be flowers in flower beds and insects that pollinate plants, enabling the production of honey or other gifts of nature". It can be stated that community gardens have a protective function, strengthening species biodiversity.

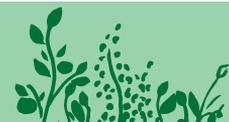

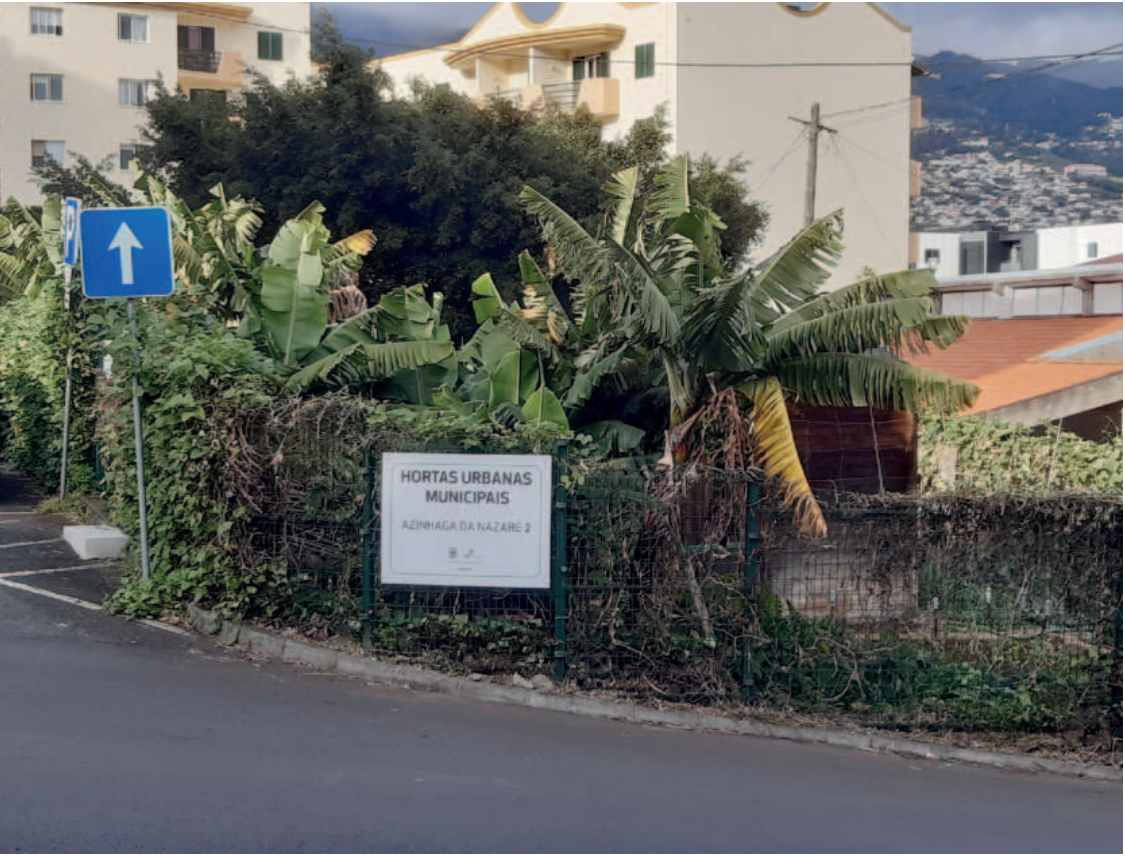

**FIGURE 6.** Funchal, Madeira. Urban municipal gardens with banana trees
Photo by Ewa Duda

An example of an urban municipal garden with banana trees growing is shown in Figure 6.

## Aquaponics

Aquaponics is a sustainable farming technique that integrates aquaculture (fish farming) and hydroponics (soilless plant cultivation) in a closed-loop system (Goddek et al., 2016). In aquaponics, fish waste serves as an organic source of nutrients for plants, while plants filter and purify the water, which is then returned to the fish tanks. This symbiotic relationship creates a self-sufficient ecosystem that requires minimal water and does not require chemical fertilisers. Aquaponics is known for its water efficiency and minimal environmental impact. It is often advertised as a method that mimics natural systems and is considered water-saving.

## Urban orchards

Creating green urban spaces, such as urban orchards, is the basis for integrating agriculture with sustainable urban development. They introduce fruit trees and shrubs into city boundaries, becoming centres of biological diversity (Davivongs et al., 2023). They not only green cities but, above all, improve the quality of life for residents and contribute to food security (Betz et al., 2017). They also play a key role in promoting food sovereignty, human health, and climate resilience, adapting to contemporary nature-based solutions (Lovell et al., 2021). Urban orchards thus support local community resilience (Sarker et al., 2019); they are a kind of response to the fluctuations and turmoil of our contemporary world.

Urban orchards are also often remnants of former agricultural areas, transformed into urban areas with the progressive process of urban sprawl. The maintenance of perennial fruit trees is a priority for local authorities and urban activists, as evidenced by initiatives in Warsaw's Żoliborz district (see Figure 7). These initiatives provide residents of apartment buildings with access to both greenery and fresh fruit, which is particularly valuable given that fruit trees are often not covered by protected felling regulations. From an ecological perspective, urban orchards become spaces for the life of other living organisms. They are valuable ecosystems in urban environmental protection, although they are often exposed to anthropogenic stress (Vahidi et al., 2018), hence supporting green urban areas with appropriate urban policies.

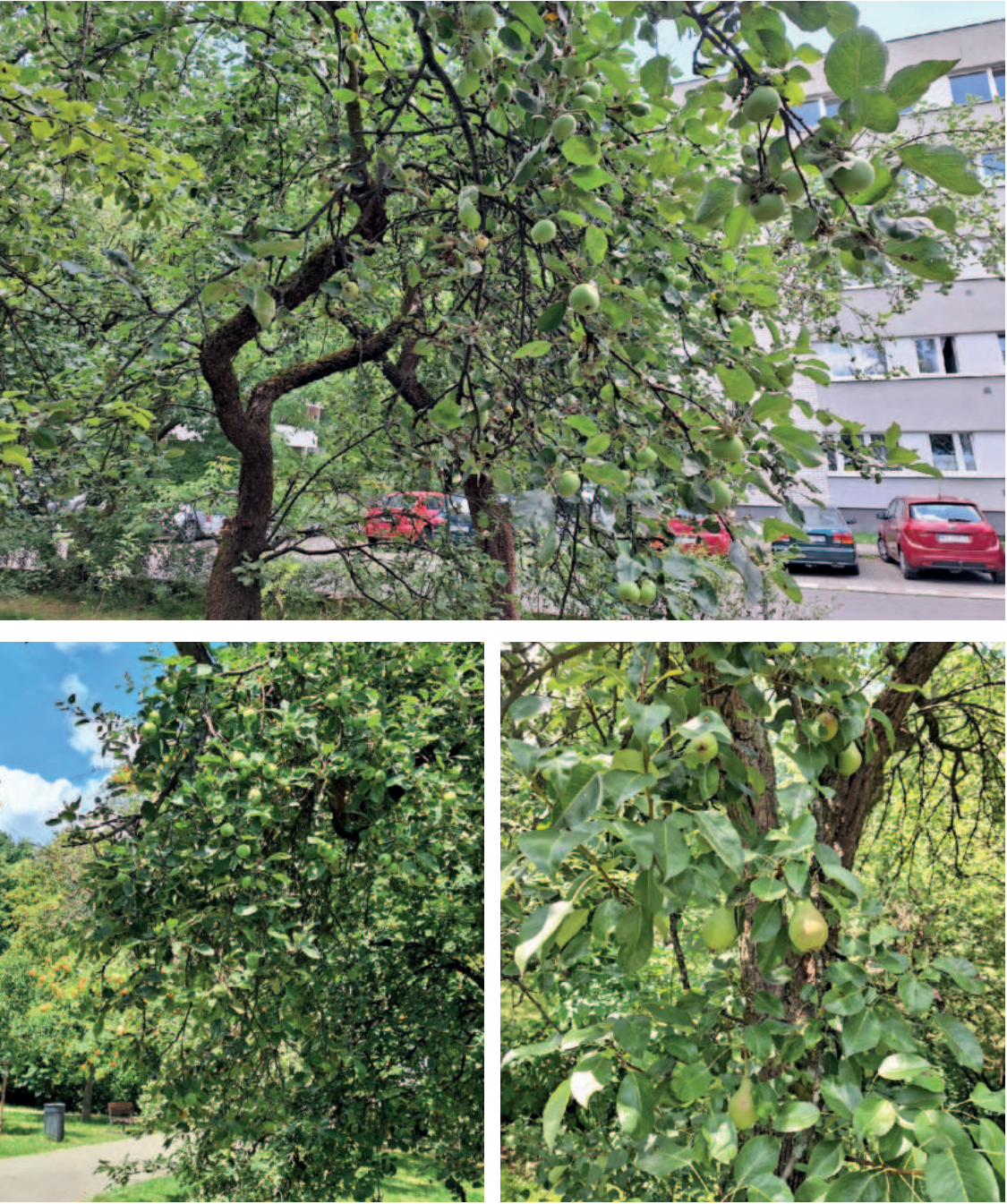

**FIGURE 7.** Warsaw, Poland. Fruit trees in Sady Żoliborskie Park
Photo by Ewa Duda

The above examples illustrate diverse and innovative approaches to urban agriculture, emphasising its potential for food security, sustainable environmental development, and community resilience in urban environments. This trend is gaining momentum worldwide, with an increasing number of city dwellers engaging in self-cultivation, motivated by a range of factors shaped at both individual and social or natural levels. The emergence of new types of urban agriculture and the application of technological solutions aimed at improving the cultivation process reflect the diversification and evolution of urban agriculture practices. In response to these trends, there is growing interest in integrating urban agriculture with initiatives for sustainable urban development. This includes identifying barriers to the expansion of small-scale agriculture in urban areas and developing policy interventions to leverage the role of urban agriculture in promoting sustainable urban development, poverty alleviation in cities, and increasing community prosperity.

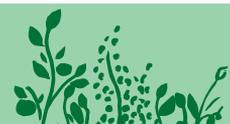

# Self-food production in the perspective of social learning theory

## 2.1. Introduction to social learning processes

Armitage et al. (2008) point to three fundamental, complementary learning theories: transformative learning, experiential learning, and social learning. According to Mezirow (1991), in the transformative learning process, the key role is played by the mechanism of effective change based on reflection and the individual's critical engagement. The second theory proposed by Kolb (1984), namely experiential learning, assumes that learning is a process that allows for the construction of knowledge based on experience and action. The third theory, Albert Bandura's social learning theory, is strongly grounded in psychological, pedagogical, and social sciences. According to its basic assumptions, the role of observation, imitation, and social interaction is emphasised as mechanisms that allow for the acquisition and modelling of new behaviours.

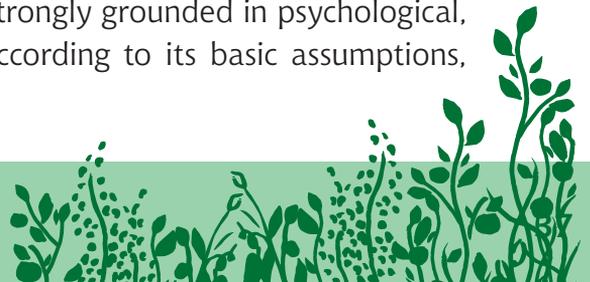

According to Bandura's social learning theory, people learn through observing the behaviours of others, meaning that the person serving as a model usually possesses specific characteristics, competencies, and skills that are crucial factors for effective learning from the observer's perspective. Observation can occur directly or through the media. As a result, the observer of specific behaviour can both imitate it and adapt it to their needs. A key factor influencing learning processes is the individual's motivation to imitate the behaviours of others. If the observed behaviour is rewarded and desired, there is a greater likelihood of its imitation. This reinforcement of behaviour can contribute to its repetition. Furthermore, the significance of modelling and observation increases not only in positive situations but also in situations where the individual lacks their own experiences, and thus has not developed their own reactions and behaviours to specific problems, difficulties, or issues. Analysing this perspective, it can be concluded that it is the social context that can support learning.

Bandura's social learning theory, which later became the basis for the theory of social cognition, laid the foundation for understanding how individuals learn through observation, modelling, and imitation of behaviours (Lyons et al., 2012). This theory emphasises the role of social observation and imitation in the learning process, highlighting the importance of social interactions in shaping behaviours. Over time, the theory of social learning has been enriched with observations from various fields such as psychology, sociology, and pedagogy, leading to a transition from individual cognitive constructivism to social constructivism, emphasising knowledge construction in social interactions (Gunn, 2017), indicating that learning is a process based on cooperation and dialogue, in which individuals collaborate with others to construct their knowledge and understanding of certain phenomena. The reorientation toward social aspects of learning processes empha-sised the importance of cooperation in learning processes and the role of social participation in knowledge construction, contributing to transformations in contemporary educational environments, creating space for the realisation of common aspirations, active engagement, mutual support, and experience that takes place in authentic social contexts (Gunn, 2017). According to Reed et al. (2010), social learning can be defined as a change that extends beyond the individual and is situated in a broader social context, reaching its full realisation in social interactions between individuals within specific social networks.

Social learning plays a fundamental role in learning processes, emphasising the importance not only of social interactions and engagement but also of observation as factors that influence the acquisition of knowledge, skills, and behavioural change. As Lyons and Berge (2012) argue, individual learning depends on many factors, both individual, related to personality and individual differences, and social, which are deeply connected to social experiences occurring in specific contexts. Therefore, the question arises: how do people learn about food cultivation from the perspective of social learning theory?

## 2.2. Implementation of social learning theory in the self-food production

The independent cultivation of food is a process influenced by a range of factors, both individual and societal. Situating this process within contemporary contexts and megatrends makes these issues significant components of the daily functioning of communities and individuals, hence the presence of numerous initiatives at the local, national, and European levels aimed at regulating and supporting these areas. One such initiative is the European "Farm to Fork" Strategy, which addresses the growing urban population's demand for sustainable, healthy, and local food to enhance the resilience of the food system in the European Union. International organisations such as Food and Agriculture

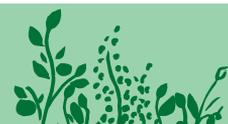

Organization (FAO) and Organisation for Economic Co-operation and Development (OECD) also advocate for strengthening food system resilience to changes and reducing their carbon and ecological footprint. As Caron et al. (2018) argue, this is particularly important as instability and numerous crises necessitate a response that combines aspects of food security, nutrition and human health, ecosystem vitality, climate change, and social justice.

Given the on-going urbanisation of rural regions across Europe, the transformation of agricultural land into urban and suburban areas becomes crucial for creating a system that supports food production and biodiversity conservation in urban fabric, which is systematically expanding its boundaries (Erisman et al., 2016). This implies, on the one hand, limiting space for cultivation and, on the other hand, increasing the actual presence of cultivation in urban space, which should also be part of urban planning (Cabannes et al., 2018; Sarker et al., 2019). Thus, increasing emphasis should be placed not only on creating space for independent food production but primarily on shaping awareness related to engagement in sustainable food production and consumption, which in turn aligns with broadly understood learning processes.

New behavioural patterns can be learned based on one's own experience or through the observation of others. As Bandura asserts, "From the perspective of social learning theory, it is not true that people are driven by internal forces or pushed by environmental stimuli. In fact, human psychological functioning involves continuous, reciprocal interactions between personal and environmental determinants" (2007, p. 29). According to the author, human nature is characterised by enormous possibilities for generating potential new behaviours under the influence of direct and vicarious experience. Although Bandura argues that interaction can be understood in various ways, it is generally accepted that behaviour results from interactions between the individual and the environment, meaning that people's learning ability can occur through observation, enabling the acquisition of large, integrated behaviour patterns without the need to shape them gradually through trial and error (Bandura, 2007, p. 27). Such streamlining of behaviour acquisition processes is crucial for human development and survival, as it allows for the reduction of error replication and its negative consequences.

There are several sources shaping and learning new behaviours. One of them is learning through the consequences of one's own reactions, which provides the following reactions:

a) informational – showing what needs to be done to achieve the desired outcome in the future,
b) motivational – providing the opportunity to formulate predictions about the consequences of behaviour, which may contribute to specific actions being taken,
c) reinforcing – in which reinforcement typically serves to provide information, motivate behaviour, and regulate it (Bandura, 2007).

The second source of behaviour is learning through modelling or observing others. The individual transforms the modelled activity into mental images and verbal symbols, which are then encoded in memory. This information later serves as cues for new behaviour. According to Bandura, modelling supports learning processes primarily through its informational function (Bandura, 2007, p. 38), facilitating the formation of symbolic representations of modelled activities, thus becoming the basis for appropriate behaviour. However, it is important to emphasise that attention processes play a significant role in this process, enabling the observation of modelled behaviour. Analysing the issue of attention in observational learning processes allows us to conclude that a significant determinant of attention is the attractiveness of the model presenting specific behaviours. As Bandura argues, the greater the interpersonal attractiveness of the model, the more frequent the replication of behaviours. However, it should be noted that mere observation of modelled behaviour will not result in new behaviour if it is not memorised by the individual, hence the importance attached to the processes of storing memorised model behaviours.

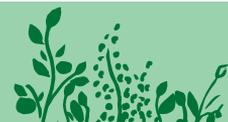

Bandura emphasises that "Through symbols, momentary modelling experiences can be stored in long-term memory. It is this advanced symbolic capacity that allows people to learn many other behaviours by observation" (Bandura, 2007, p. 40). Storing modelled behaviours is based on two representation systems – verbal and imagery. While verbal representations rely on verbal coding of modelled behaviours, imagery representations rely on visual imagery of modelled behaviours, which are created based on sensory stimulation activating impressions. Learning through modelling combines these two forms of representation, with visual imagination playing a particularly important role, especially in the early stages of human development. Repeating behaviours and practising them, both in the individual's imagination and in real actions, also plays a crucial role in memorising and storing them, which is justified by the neurobiological basis of learning processes. The third element of behaviour modelling is the transformation of symbolic imagery and verbal representations into appropriate action, i.e., behavioural replication consistent with modelled patterns, which Bandura divides into "cognitive organisation of reactions, their initiation, monitoring, and improvement based on informational feedback" (Bandura, 2007, p. 42). However, social learning theory distinguishes between acquiring behaviours and performing them, as people do not always exhibit behaviours they have learned. This is particularly important in terms of the effects of these behaviours. If they are rewarding, there is a greater chance of individuals carrying out modelled behaviours.

Considering social learning theory, it is worth reflecting on how people learn food cultivation. In light of the literature, various strategies enabling the acquisition of skills for independent food cultivation can be identified.

## Strategy 1: Learning through observing others

Observation plays a crucial role in independent food production. Observing family members, neighbours, or the local community engaged in activities related to plant cultivation allows individuals not only to acquire knowledge or practical skills but also to understand cultural norms and practices associated with food production (Rogoff, 2014). This creates excellent conditions for learning new behaviours related to food production.

Learning through observation plays a particular role in the early stages of human development – from birth to 6–7 years of age. During this period, children construct their knowledge about the world, including food cultivation. This lays the groundwork for understanding the processes involved in this area. The mere observation of garden plots, allotments, or windowsill gardening provides an opportunity to familiarise oneself with the actions required for independent food cultivation, while also allowing one to experience and observe how it is done. Children's participation in learning and discovering the regularities associated with independent plant cultivation allows for the construction of symbolic mental and verbal representations and creates an opportunity for potential action, i.e., behavioural replication consistent with modelled patterns, provided a conducive environment is created for carrying out these activities. Therefore, much depends on the environment in which the child is raised and the experiences and awareness of their parents. Kharuhayothin and Kerrane (2018) emphasise the role of parents in shaping dietary behaviours and practices related to food production. Parents' dietary patterns, based on their own experiences, may be transmitted to their children.

Learning through observation is also crucial in adult education. It involves adults acquiring knowledge, skills, and building understanding of reality through observation and reflection on it. This method emphasises learning through experience, where adults actively engage in observing processes, behaviours, or phenomena to gain insight and deepen their understanding (Wilson, 1993). This can promote critical thinking, problem-solving skills, and the practical application of knowledge. By observing specific situations, adults can integrate theoretical concepts with observations and practical results, enhancing their ability to transfer acquired skills and incorporating them into their own context and life narrative (Wilson, 1993).

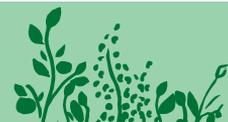

This approach also promotes a human-centred environment, where adults have the opportunity to learn at their own pace, reflect on observations, and apply new knowledge in a meaningful way, encouraging independence and enabling them to take responsibility for their educational journey (Alhosban et al., 2018).

## Strategy 2: Learning based on collaboration

Analysing collaborative learning emphasises the importance of social interactions, shared experiences, and group dynamics in learning processes. Social learning theory suggests that individuals learn by observing others, modelling behaviours, and participating in joint activities, which facilitate knowledge sharing and contribute to understanding processes (Coffield, 1999). By engaging in collaboration-based activities, both children and adults can benefit from different perspectives, feedback, and collaborative problem-solving, enriching their educational experience.

Collaborative learning environments promote active engagement, the development of communication skills, and teamwork, which are fundamental competencies important in personal and professional life. This is important because it creates a supportive and interactive environment where individuals can learn from each other, exchange knowledge, and collectively construct meaning through dialogue and reflection (Taylor & Cranton, 2013), supporting overall learning experience and nurturing a sense of belonging and community among people. This sense of community and belonging to a group of individ-uals with similar interests or experiences can be a particular support for those undertaking independent food cultivation. This support includes conversation, information exchange, problem-solving, mutual learning of practices, techniques, and cultural traditions related to cultivation, preparation, and consumption of food, creating an inclusive and participatory learning environment that satisfies the diverse needs of its participants. Through these shared experiences, individuals develop a sense of community belonging and a shared identity based on their shared engagement in food-related activities (Michalski et al., 2020). Joint activities related to independent food production, such as community gardening, cooking, or meal-sharing, provide community members with opportunities to meet, share knowledge, and build social bonds.

## Strategy 3: Immersion in social support networks

One way of social learning is to engage in social networks related to agriculture or food cultivation. These networks, as learning environments, can facilitate the improvement of food production skills by providing individuals with access to resources, knowledge, tips, and support from those involved in food cultivation.

Social support networks play a crucial role in food production in urban areas, enabling collaboration, knowledge sharing, and community engagement. They serve as platforms for individuals to learn from each other, exchange ideas, and collaborate on sustainable urban farming practices (Orsini et al., 2013).

Urban agriculture initiatives often rely on social support networks to provide various forms of assistance to individuals engaged in independent food production. These networks may offer training workshops on urban farming techniques, cooking classes, or sustainable production practices. Their essence often lies in empowering community members with the knowledge and resources needed for successful food cultivation (McClintock, 2013). Social support networks in urban agriculture can facilitate resource sharing, such as land, seeds, and tools, promoting community involvement in independent food production. By creating a supportive environment for knowledge exchange and skill development, these networks contribute to the resilience and sustainable development of urban food systems (Guitart et al., 2012), connecting individuals with gardening interests. Through collective efforts and shared experiences, community members can build social bonds.

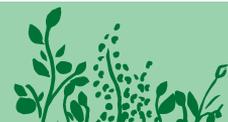

## Strategy 4: Personal experience, reflection, and feedback

Personal experience, self-reflection, and feedback play a crucial role in shaping individuals' behaviours and motivation to engage in activities. Bandura and Schunk (1981) emphasise the importance of personal experiences in shaping individual motivation, which regulates their behaviour. The experience gained from growing one's own food and observing others' actions can significantly influence interest in engaging in gardening activities, such as planting, nurturing crops, and harvesting. This process involves drawing conclusions to assess progress and outcomes of actions taken, indicating areas that are functioning correctly and those that need improvement. This builds beliefs in an individual's self-efficacy in action (Bongers, 2022). In addition to self-assessment, individuals also seek feedback from others perceived as experts in the field. The feedback on the quality of actions taken and their effects can be obtained, which also influences perceptions and shaping of one's own efficacy, engagement, learning, and improvement in food cultivation practices.

By reflecting on their own experiences, seeking feedback, and modifying behaviours related to growing their own food, individuals can enhance their skills, motivation, and interest in independent food cultivation.

## 2.3. Motivational factors in self-sustained food production

According to Łukaszewski, "the term motivation, as used in psychology, describes all mechanisms responsible for initiating, directing, sustaining, and terminating behaviour. It applies to both simple and complex behaviours, as well as internal, external, affective, and cognitive mechanisms" (2000, p. 427). This means that motivation refers to processes that drive individuals to initiate, sustain, and direct their behaviour towards achieving specific goals. It encompasses desires and needs that underlie human actions and decision-making processes. Motivation can be influenced by a range of factors, both internal and external, shaping individuals' attitudes, beliefs, and behaviours. Madsen (1980) identified four groups of motivational theories, indicating four fundamental sources of motivation. The homeostatic model assumes that motivation arises from the disruption of the body's equilibrium, which triggers cognitive and energetic processes aimed at restoring this balance. The stimulus model suggests that motivation originates from a stimulus that is cognitively processed, aiming for the appropriate response to alleviate these stimuli. The significance lies in the stimulus triggering appropriate energetic processes. The cognitive model posits that information processing is the source of motivation for a given behaviour. These models do not focus on the stimulus, as in the stimulus model, but the reaction's effect is confronted with cognitive structures. The final model – humanistic – assumes that the individual's interior plays a fundamental role in motivation. Based on this, reflection can be made on motivational factors playing a significant role in actions towards independent food cultivation.

Among the various motivating factors, the following can be distinguished and will be discussed in the paragraphs that follow.

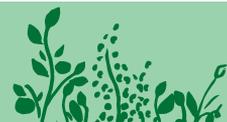

## Health motivations

Research by Zahaf and Ferjani (2016) has shown that concerns about one's health, high food quality, and taste are motivating factors for individuals to engage in independent food production. Consumer interest in organic food products is increasing, justified by a concern for one's health, as well as concerns about the quality of consumed products, which are related to issues of genetically modified organism (GMO) cultivation, pesticide and antibiotic use, and excessive industrialization of agriculture (Hamzaoui-Essoussi et al., 2012). Organic products are perceived as healthier, better in quality, significantly affecting individuals' well-being and satisfaction, thus potentially encouraging greater engagement in independent food cultivation. Additionally, Ashtab et al. (2021) emphasise that factors such as freshness, high food quality, and freedom from pesticides increase food security by providing individuals with access to non-industrial food networks, which primarily aim to produce safe, high-quality food by mimicking natural farming practices (Chareonwongsak, 2022). This is particularly important in times of various crises.

## Environmental motivations

One of the factors playing a fundamental role in independent food cultivation is environmental issues. Individual ecological awareness and awareness of sustainable development seem crucial here, reflecting individuals' understanding and concern for the natural environment and the belief in the need to protect it for current and future generations. Ecological awareness is commonly defined as the recognition by an individual of environmental problems and their willingness to contribute to their solution (Hollmann et al., 2012, cited in Paço et al., 2012). It is associated with awareness of various environmental challenges and the recognition of the need to address them (Hollmann et al., 2012). This awareness extends to the perception of the importance of environ-mental protection and taking action to mitigate the negative effects of human actions.

On the other hand, sustainable development awareness involves a broader perspective, emphasising the maintenance of ecological balance by integrating social, political, economic, and environmental issues (Rosário, 2021). It recognizes the interconnections between social, economic, and environmental aspects and the need to adopt practices promoting long-term, harmonious development. This concept often goes beyond individual actions and includes organisational and social efforts aimed at achieving a balance between economic development, social justice, and environmental protection (Rosário, 2021). Both ecological awareness and awareness of sustainable development play a crucial role in shaping more responsible behaviours, policies, and daily practices. Individuals and institutions aware of environmental and sustainable development issues are more likely to engage in pro-environmental behaviours, support sustainable initiatives, and contribute to the protection of natural resources and ecosystems (Kusturica et al., 2016). In this context, individuals' concerns about the state of the natural environment can motivate them to engage in independent food cultivation.

Blay-Palmer et al. (2019) emphasise that local food production can provide fresh, healthy food while reducing the environmental impact of transportation, thus supporting the implementation of sustainable development concepts. Additionally, it can mitigate negative ecological effects caused by human activity (LeVasseur et al., 2021). Such an approach not only helps to solve environmental problems but also increases local and social resilience in the face of challenges such as climate change (LeVasseur et al., 2021).

## Hedonistic motivations

The pleasure derived from both the cultivation process and the consumption of self-grown food can be a key motivator for individuals to undertake their own food cultivation. On the one hand, the process of growing plants is an extremely enjoyable experience for many people. The opportunity to observe plant growth, as well as the physical contact of individuals with the soil, which contains mood-enhancing Mycobacterium vaccae bacteria, can provide joy from gardening activities. Additionally, the pleasure associated with consuming self-grown, high-quality food can instil pride in individuals, fulfil the need for self-realisation, and please the palate.

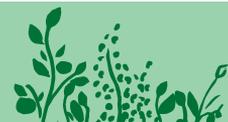

Independent food production can have a significant impact on human development by supporting a deeper connection with nature, promoting self-sufficiency, and improving well-being. Engaging in independent food production, such as growing fruits and vegetables or raising animals, allows individuals to develop a sense of autonomy and independence (Chareonwongsak, 2022). This process of independent food production not only contributes to food security but also nurtures a sense of fulfilment and empowerment, which are essential aspects of self-development (Chareonwongsak, 2022).

Furthermore, the act of growing plants can promote mindfulness and compassion for oneself (Neff et al., 2017). By actively engaging in the food production process, individuals can cultivate a deeper respect for the environment, the food they consume, and the effort required to produce it. This mindfulness can lead to a greater sense of gratitude and connection with the natural world, promoting personal development and well-being (Neff et al., 2017). As a result, independent food production can contribute to a healthier lifestyle, both physically and mentally. Consuming food grown with mindfulness and care can have a positive impact on physical health by providing essential nutrients, thereby reducing chemical intake (Ferruzzi et al., 2010). Consequently, this leads to a better understanding of ecological systems and sustainable practices. By engaging in food production, individuals become more aware of the impact of their dietary choices on the environment and may be more inclined to adopt sustainable practices, which, in turn, can lead to personal development in terms of ecological awareness and a sense of responsibility towards the planet.

## Personality motivations

Personality traits play a significant role in shaping individuals' attitudes and behaviours, even in the context of engaging in independent food cultivation. Research by Wiggins and Pincus (1994) has shown that personality structure can influence various aspects of behaviour. Individuals with specific personality traits may be more inclined towards independent food production due to factors such as motivation for self-improvement or behaviours associated with elevated self-esteem (Barry et al., 2011). The complex relationship between personality traits, eating behaviours, and well-being is also significant. According to Lisá (2020), self-efficacy, which refers to belief in success in specific tasks, plays a significant role in motivating individuals to engage in independent food production. When people feel confident in their skills and knowledge related to food production, they are more likely to take proactive actions towards food cultivation. The concept of self-regulation and the desire for personal development and improvement also play a role which can be perceived as a way to improve overall health and well-being, consistent with the motif of self-improvement identified in motivational theories.

## Safety and self-sufficiency motivations

One of the key motivating factors for individuals to engage in independent food production is the fulfilment of the need for safety and self-sufficiency, which refers to the ability to satisfy one's own needs without relying on external sources. By engaging in independent food production, individuals reduce their dependence on external food sources, which in today's chaotic times do not allow for stability. This autonomy in food production allows for greater control over the quality and diversity of consumed food, contributing to a varied and nutritious diet (Iloh et al., 2020), and reducing dependence on imported food items, which is essential for increasing food security at both individual and national levels. By focusing on local production of diverse food, individuals and communities can reduce the risk of food insecurity and build resilience to external food-related challenges (Chareonwongsak, 2022).

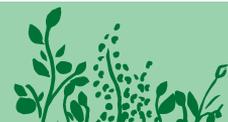

## Financial motivations

Financial motivation is a significant factor influencing individuals' decisions and behaviours related to engaging in actions associated with producing their own food. It can serve as a driving force for initiating or expanding the process. Potential savings and economic benefits associated with food cultivation can motivate individuals to invest time, resources, and effort in crop cultivation (Oishi, 2021). Financial considerations, such as reducing expenses on food items or generating additional income through the sale of surplus products, can encourage individuals to actively participate in independent food production (Oishi, 2021). Moreover, financial motivation can influence the sustainability and scale of undertaken initiatives. Grants or subsidies for innovative agricultural and gardening activities can help individuals overcome initial investment barriers and operational costs associated with food production. By providing financial support, governments or organisations can encourage more people to engage in independent food production, thereby promoting food security and self-sufficiency at both individual and societal levels.

## Cultural and traditional motivations

Cultural motivations and traditions associated with producing one's own food are embedded in the experiences of communities through shaping dietary choices and cultivation practices. These motivations not only influence what is cultivated and consumed but also contribute to a sense of belonging, shared values, and continuity within food production systems and family traditions. It is worth adding that the cur-rent trend of returning to nature has allowed for the rediscovery of traditional knowledge regarding the use of wild edible plants in our everyday lives, as well as emphasised the importance of traditional food cultivation practices (Parrotta et al., 2015).

## Axiological motivations

Values and beliefs held by individuals can significantly influence their decision to engage in independent plant cultivation for their own needs. Axiological motivations can encompass a range of components. Ethical considerations related to food production, such as respect for plants, animal welfare, fair labour practices, sustainable production, or locality in the context of shortened supply chains, shape the thinking and actions of individuals. From a psychological standpoint, values do not always correspond with behaviours, yet they constitute an important factor enabling decision-making. The alignment between values and actions can lead to a sense of fulfilment in contributing to a more sustainable food consumption. Environmental issues within axiology, addressing problems such as reducing carbon footprint, minimising food waste, or promoting biological diversity, can influence motivation related to food cultivation.

Another factor influencing the decision to engage in independent food cultivation is health-related beliefs. Axiological motivations associated with personal health and well-being can prompt individuals to cultivate food to ensure access to fresh, nutritious products and reduce dependence on processed and unhealthy food (Isaksson, 2014), as well as have therapeutic effects. Cultural values and traditions also play a significant role in shaping axiological motivations for independent food production. In many cultures, food is deeply intertwined with identity, heritage, and community values (Piñeiro et al., 2020). Engaging in food production can be a way for individuals to connect with their cultural roots, maintain traditional culinary practices, and transmit knowledge to future generations. Axiological motivations rooted in cultural values can encourage individuals to cultivate native plants, uphold culinary traditions, and strengthen social bonds through shared food experiences.

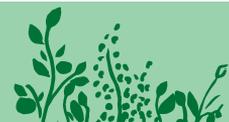

## Therapeutic motivations

Since ancient times, humans have used plants for various purposes, including therapeutic ones. Depending on the needs, independent food production can encompass physical, psychological, or social therapy. For example, food cultivation without the use of chemicals contributes to improving food quality, which in turn provides the body with better-quality nutrients, thereby affecting physical health. With diverse health issues, individuals can cultivate plants with biochemical properties that help combat illness, fitting into the age-old tradition of herbalism. Cultivating one's own food can also have psychotherapeutic effects. The widely spread field of horticultural therapy allows for a multifaceted approach to therapy, utilising gardening activities to improve mental and physical health. Horticultural therapy can be applied in various contexts, including the treatment of individuals with health problems, as well as for general body support. The activities undertaken in gardening, such as planting, tending to the garden, or simply spending time outdoors, contribute to stress reduction, improved concentration, well-being, increased self-esteem, and the development of social skills. Horticultural therapy is applied in various settings, such as healthcare facilities, rehabilitation centres, integration of war-experienced refugees, and even within local community actions. In this dimension, it also fits into social therapy, as being in green spaces integrates the group and influences relationships within the community.

# Diffusion of innovation as a strategy for the urban food self-production

## 3.1. Urban food self-production as a social change process – reorientation towards sustainability

Food self-production in urban areas is increasingly recognized as a social change process oriented towards sustainability. It encompasses a multifaceted approach that integrates elements of urban agriculture, community engagement, and sustainable food systems. This process involves individuals and communities taking an active role in producing their own food within urban environments, thereby promoting food security, environmental sustainability, and social well-being by various forms of urban agriculture, ranging from self-production allotments to high-tech companies, which offer environmental, social, and economic benefits (Sanyé-Mengual et al., 2019). It has evolved into an ideological movement promoting environmentally and socially sustainable choices, community networks, reconnections with nature, and social change (Mok et al., 2013).

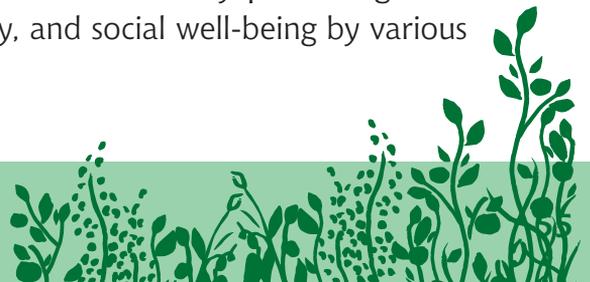

Urban food self-production can be viewed as a transformative process that is driven by various factors and initiatives, which lead to social change. By engaging in urban agriculture practices, such as rooftop farming, community gardens, or vertical farming, individuals and communities can enhance their food self-sufficiency, reduce their reliance on external food sources, and promote local food production (Orsini et al., 2013; Sanyé-Mengual et al., 2016). It fosters community resilience, social cohesion, and empowerment by creating opportunities for shared decision-making, knowledge exchange, and collective action. Through initiatives that promote food sovereignty and autonomy, urban residents can reclaim agency over their food systems, strengthen community ties, and address issues of food justice and equity (Giraud, 2021).

The transformative potential of urban food self-production lies in its ability to challenge conventional food systems, promote sustainable practices, and foster a sense of connection to the environment and local food culture. By integrating urban agriculture into urban planning and development, cities create more resilient and sustainable food systems that benefit both residents and the environment (Lovell, 2010). Although urban food self-production contributes to the creation of shared value, where economic, social, and environmental benefits are generated for individuals, communities, and society as a whole, it is perceived as a process which can be understood through various phases that involve a series of interconnected steps, actions, and time.

The process of change often begins with raising awareness and educating individuals and communities about the benefits and importance of urban food self-production. Educational initiatives, awareness cam-paigns, and knowledge-sharing activities play a crucial role in informing and engaging stakeholders in the transition towards sustainable urban agriculture (Al Mamun et al., 2023; Pulighe et al., 2020). Next, establishing supportive policies, regulations, and governance structures is essential for facilitating the adoption and implementation of urban food self-production initiatives. Clear guidelines, land-use regulations, and institutional frameworks can create an enabling environment for sustainable urban agriculture practices (Lavallée-Picard, 2018). If the regulations are clear, there is a greater probability to engage communities, foster social cohesion, and promote participatory approaches which are key aspects of the social change process. Community involvement in decision-making, planning, and implementation of urban agriculture projects enhances ownership, empowerment, and sustainability (Lavallée-Picard, 2018).

Another component tackles the problem of technological advancements, innovative farming practices, and smart agriculture solutions which can drive efficiency, productivity, and sustainability in urban food self-production. Leveraging technology for urban farming can enhance yields, resource efficiency, and environmental sustainability (Lavallée-Picard, 2018). Environmental factors such as land availability, soil quality, water management, and climate resilience are crucial for the success of urban food self-production. Sustainable practices, conservation efforts, and climate-smart agriculture play a vital role in mitigating environmental impacts and promoting resilience (Pulighe et al., 2019). Promoting a cultural shift towards sustainable food practices, healthy eating habits, and environmental stewardship is integral to the change process. Cultivating a culture of sustainability, food security, and community well-being fosters positive behavioural change and societal transformation.

However, what seems to be a crucial point of the implementation of urban food self-production initiatives, is a comprehensive framework for understanding how new ideas, practices, or technologies spread within societies and contribute to transformative processes. By promoting innovative agro-ecological practices, sustainable farming approaches, and social movements, urban food self-production can drive positive change and promote a more equitable and sustainable food system (Kiminami et al., 2022) which can be viewed as a public good. Firstly, urban agriculture has been recognized for its potential to reduce food miles, mitigate carbon emissions, and improve community relations, thereby contributing to environmental sustainability and public health (Bellemare et al., 2020). By promoting local food production and reducing the need for long-distance transportation of food, urban farming can help mitigate the environmental impact of conventional agriculture and contribute to a more sustainable food system. Moreover, it offers a range of socio-cultural benefits to communities, such as enhancing social cohesion, providing educational opportunities, and reducing social alienation associated with urban poverty (van Averbeke, 2018). These social benefits not only improve the quality of life for residents but also contribute to building stronger and more resilient communities. Therefore, urban farming can serve as a platform for community engagement and empowerment, fostering a sense of ownership and pride among residents (Lee et al., 2023).

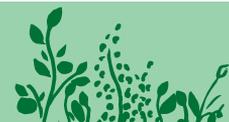

From an economic perspective, urban agriculture can contribute to food security, create employment opportunities, and generate income for households (van Averbeke, 2018). By using land in urban areas for agricultural purposes, urban farming can help address food insecurity issues and provide a source of nutritious and affordable food for local residents. The economic benefits derived from urban farming extend beyond material gain, encompassing social and psychological well-being (van Averbeke, 2018). In terms of environmental sustainability, urban agriculture plays a crucial role in promoting sustainable water management, biodiversity conservation, and ecosystem health (Dhakal et al., 2015). By integrating green spaces and agricultural practices in urban areas, urban farming can enhance urban biodiversity, provide habitat for pollinators like bees, and contribute to the overall ecological balance of cities. It can help reduce the urban heat island effect, im-prove air quality, and promote sustainable water use practices (Dhakal et al., 2015). Another critical issue is connected with the urban food self-production which has potential to contribute to the resilience of urban food systems by diversifying food sources, reducing dependency on external food supplies, and enhancing food sovereignty. In times of crisis or disruptions to global food supply chains, urban farming can play a critical role in ensuring food security at the local level and reducing vulnerabilities associated with centralised food distribution systems.

## 3.2. Diffusion of innovation in food self-production – towards social change

One of the key factors supporting the process of social change is the diffusion of innovation. Social learning theory explains how new ideas, practices, or technologies spread within social systems through shared experiences, interactions, and knowledge exchange. Social learning theory states that individuals learn by observing others, engaging in collective action, and reflecting on their experiences, which can influence the adoption and diffusion of innovations (Greenhalgh et al., 2004). As Bandura (2007) argues, "modeling also plays a fundamental role in spreading new social ideas and practices within a society or from one society to another. Effective innovation dissemination typically follows the pattern: new behavior is introduced by notable, exemplary case, then, it is rapidly assimilated, and later, either stabilizes or fades away, depending on its functional value" (p. 62).

The author identified two processes related to the social diffusion of innovation. The first is the acquisition of innovative behaviours, while the second is the adoption of innovative behaviours in practice. Symbolic modelling, especially in the early stages of innovation dissemination, is a primary way of acquiring innovative behaviours, which occurs through informing people about new practices and their likely benefits, without pointing out drawbacks or potential risks. This dissemination often happens through mass media channels such as the internet, newspapers, radio, or television. According to Robertson (1971), new behaviours are most commonly adopted by individuals who have been exposed to media sources. However, their implementation is influenced by many factors such as personal characteristics, social conditions, or economic circumstances. Symbolically introduced innovations are then spread during personal contacts with individuals who have previously adopted innovative behaviour. Direct modelling spreads through existing interpersonal communication networks. However, it is worth noting that if a new behaviour seems exceptionally appealing, then strangers may learn it through public dissemination. However, observers of new behaviour may be reluctant if they do not perceive the benefits it may bring.

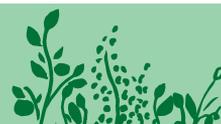

As Bandura (2007) states, "as acceptance spreads, the novelty gains further social endorsement. Models not only provide examples and legitimize innovations but also serve as advocates, encouraging others to adopt them. Acquiring innovations is necessary but not sufficient for their application in practice" (p. 63). Encouraging stimuli are needed to activate individuals to engage in new behaviours. The adoption of these new behaviours is heavily dependent on reinforcements, which are interpreted in terms of benefits. Since individuals cannot experience the benefits before trying the new behaviour, the dissemination of innovative behaviour is based on communicating expected reinforcements and substitute reinforcements, as well as conformity to commonly accepted values. It is essential to emphasise that innovations spread according to different patterns and at different rates because their adoption requires adaptation to various requirements. These requirements serve as factors influencing the diffusion of innovation. People will not adopt innovative behaviours, despite favourable attitudes, if they lack the skills, knowledge, or money needed to implement the behaviour. According to Bandura (2007), "the main determinants of adopting new behaviors are closely related to them – encouraging stimuli, expected satisfactions, observed benefits, experiencing their functional value, the risk associated with their adoption, self-eval-uation of such behaviors, and various social barriers and economic constraints. The composition of determining factors will change depending on the type of products (…). Behaviors involving adopting novelties are better analyzed in terms of the conditions influencing them than in terms of the types of people who exhibit such behaviors" (p. 65).

A different approach to the issue of novelty diffusion is presented in the Everett M. Rogers' theory of diffusion of innovations (1983). He is the creator of a ground-breaking concept that explains how new ideas, practices, or technologies spread within a social system over time, leading to social change. Rogers (1983) defines it as a process in which innovation is communicated over time and through specific channels among members of the social system. It is a particular type of communication related to the dissemination of new information, ideas, or technologies (Rogers, 1983, p. 5), in which individuals share information with others to achieve mutual understanding of a phenomenon. It is this novelty in the content of the message that gives diffusion a particular character because an inherent element of novelty is uncertainty resulting from a lack of information, predictability, or knowledge of the real consequences of its implementation. To reduce this uncertainty, explanatory information about the novelty is needed, which can lead to its adoption, contributing to social change, or its rejection.

The communication process depends on many factors, but one of the principles of human communication worth mentioning is similarity. Rogers argues that homophily, indicating the degree to which individuals interacting with each other are similar in terms of beliefs, education, or social status, contributes to more effective communication than heterophily, indicating differences between communicating individuals (Rogers, 1983, pp. 18–19). Moreover, the channels of communication through which messages reach from one person to another are essential. As Rogers (1983) claims, mass media channels are more effective in creating knowledge about innovations, while interpersonal channels are more effective in shaping and changing attitudes towards new ideas.

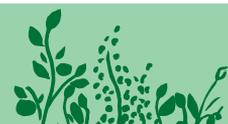

Rogers (1983) identified four basic components of innovation diffusion: (1) the innovation, understood by the individual as a novelty, (2) its communication through various channels of communication, (3) time, and (4) members of the social system among whom new ideas or technologies are disseminated. This perspective on understanding diffusion is reflected in the development of urban agriculture along with applied technological innovations related to soilless production, such as hydroponics, aeroponics, or aquaponics, which increase the potential for sustainable food production in urban environments (Al-Kodmany, 2018; Sanyé-Mengual et al., 2019).

According to Rogers (1983), innovation has five inherent attributes that allow to explain the varying degrees of adoption of innovation by community members. Among these attributes, the author lists: (1) relative advantage, (2) compatibility, (3) complexity, (4) trialability, and (5) observability. The relative advantage of technological innovation is the degree to which innovation is perceived by the individual as better than the idea it replaces. The key point is whether the individual sees the innovation as beneficial. The greater the benefit perceived by the individual, the faster its adoption will be. Compatibility of innovation means the degree to which the innovation is consistent with the values, experiences, and needs of potential users. An idea that is not compatible with the norms of the social system will not be quickly adopted and considered as a compatible innovation. It often happens that the adoption of innovation requires the prior acceptance of a new value system, which often takes time.

The third attribute of innovation is complexity, which indicates to what extent the innovation can be understood and used. This means that more complex innovations often require more time for implementation, as they usually require the development of new skills or understanding of their operation. Another attribute of innovation is its trialability. New ideas that can be tried out will generally be adopted more quickly than innovations that cannot be easily tested. Conversely, the observability of innovation indicates the degree to which the re-sults of innovation are visible to others. Therefore, the easier it is to observe the results of a given innovation, the greater the chance of its adoption by others.

Rogers' theory (1983) posits that the adoption of innovation follows a predictable pattern, although influenced by various factors that may lead to different outcomes: adoption or rejection of the innovation. The author distinguishes five stages of the innovation decision-making process, in which the individual progresses from awareness of the innovation to forming attitudes toward the innovation, making the decision to adopt or reject it, and implementing the new idea and confirming the decision made. The first stage concerns knowledge, which arises when an individual (or another decision-making unit) comes into contact with the innovation and gains some understanding of its functioning. The individual then forms a favourable or unfavourable attitude toward the innovation. Next, the person engages in actions that lead to the choice of adopting or rejecting the innovation. Implementation occurs when the individual puts the innovation into use.

Confirmation of the appropriateness of the innovation occurs when the individual receives reinforcement for the correctness of the decision made. If such reinforcement is lacking, the individual may reject the innovation (Rogers, 1983, pp. 20–22). The entire process is embedded in time as a factor that plays a significant role not only in the decision-making stage from knowledge to adoption or rejection of the innovation but also in the speed at which a given individual implements the innovation compared to other members of the community. This led Rogers (1983) to create a sort of typology of community members in terms of innovativeness, understood by him as the speed at which an individual implements innovation in relation to other community members, which led him to distinguish five social groups: (1) innovators, (2) early adopters, (3) early majority, (4) late majority, (5) laggards. Rogers assumes that instead of describing the individual as "less innovative than the average member of the social system," it is more convenient and effective to refer to individuals as later adopters (Rog-ers, 1983, p. 22). The basis of this typology is the assumption that the variable, which is the degree of innovativeness, will have a normal distribution (see: Figure 8), and each adoption of innovation in society is in a sense equivalent to an individual learning attempt, which places the theory of innovation diffusion in an educational perspective.

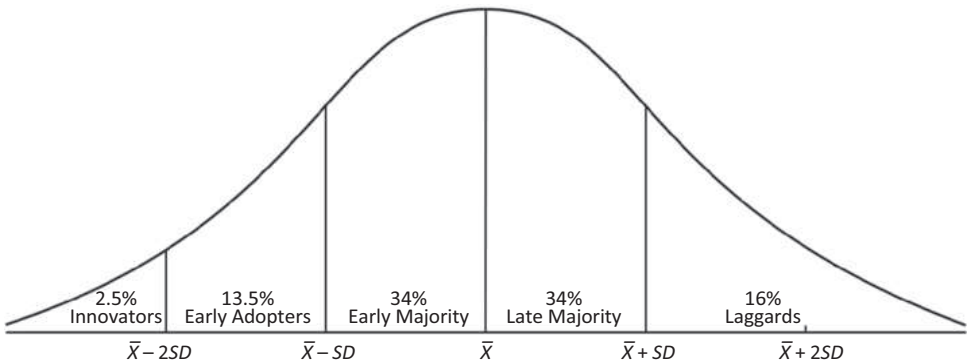

**FIGURE 8.** Adopter categorization on the basis of innovation
Source: Rogers, E. M. (1983). Diffusion of Innovation (p. 271)

Roger's categorisation of individuals is associated with ideal types, which creates the possibility of comparing members of a given community in terms of their degree of innovativeness.

## Innovators portrait

In Everett M. Rogers' theory of innovation diffusion, the category of "innovators" plays a vital role in the process of adopting new ideas, practices, or technologies within a social system. Innovators, as individuals who are among the first to adopt innovations, are active seekers of information about new ideas. They willingly test and experiment with new concepts because they are highly entrepreneurial and bold in their actions. Innovators are willing to embrace uncertainty and assume the potential risks associated with early adoption of new concepts (Rusek et al., 2017), although this often requires them to have significant financial resources that may be lost due to unsuccessful innovations. On one hand, innovators must be prepared for failures because it is an inevitable part of the process, and on the other hand, they must be able to cope with a higher level of uncertainty resulting from the implementation of innovations (Rogers, 1983, p. 22).

Understanding the characteristics and behaviours of innovators is essential for effectively introducing and promoting new ideas or technologies, as they can create momentum for the adoption of innovations within the social system, ultimately leading to broader acceptance among the early and late majority of society, and potentially even laggards (Rusek et al., 2017). Although innovators are not always respected in their communities, they establish supra-local relationships by forming cliques independent of geographical space.

## Early adopters

Early adopters represent a crucial group in the adoption process of new ideas, practices, or technologies within a social system. They are individuals who are quick to embrace innovations after the innovators have introduced them. They are considered to be opinion leaders within their social networks and are influential in spreading awareness and acceptance of an innovation to a wider audience (Berwick, 2003).

Characteristics of early adopters include being open to change, having a higher social status, and being more integrated into the social system compared to innovators. Early adopters are willing to take risks in adopting new innovations but are more discreet in their decisions compared to innovators. They serve as a bridge between the innovators and the early majority, helping to legitimise an innovation and encourage its adoption by a larger segment of the population (Berwick, 2003).

In the diffusion process, early adopters shape the trajectory of innovation adoption. Their willingness to try new ideas and technologies helps to create momentum for the diffusion process. Organisations of-ten target early adopters to gain feedback, generate positive word-of-mouth, and establish credibility for the innovation (Berwick, 2003).

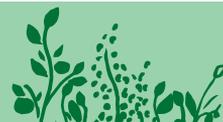

Understanding the characteristics and behaviours of early adopters is essential for effective introduction and promotion of innovations within the society. By engaging with early adopters and leveraging their influence, innovators can accelerate the diffusion of innovations and facilitate their acceptance by a broader audience (Berwick, 2003).

## Early majority

In Everett M. Rogers' diffusion of innovation theory, the "early majority" represents a significant group in the adoption process of new ideas, practices, or technologies. The early majority are individuals who follow the lead of the early adopters once the value and benefits of the innovation have been established. They are more deliberate in their decision-making process compared to the early adopters and are characterised by a cautious approach to adopting innovations (Lund et al., 2020).

The early majority represent a substantial portion of the population. Their acceptance of the innovation marks a critical point in the diffusion curve, signalling the transition from early adoption to mainstream acceptance. The early majority are influenced by the experiences and feedback of the early adopters, and their adoption behaviour is essential for the widespread dissemination of innovations within a social system (Lund et al., 2020).

Rogers' diffusion theory categorises adopters into diverse groups based on their timing of adoption, with the early majority falling between the early adopters and the late majority. The early majority are pivotal in the diffusion process as they help to legitimise the innovation and make it more acceptable to the broader population. Their adoption behaviour sets the stage for the eventual acceptance of the innovation by the late majority and potentially even the laggards (Lund et al., 2020).

By targeting the early majority, innovators can leverage their influence to achieve broader adoption and diffusion of innovations within a social system. The early majority's acceptance of the innovation is a critical step towards achieving widespread adoption and societal change (Lund et al., 2020).

## Late majority

The "late majority" are individuals who adopt innovations after the early majority has embraced them. They are characterised by a cautious and sceptical approach to change, often waiting until an innovation has become well-established and widely accepted before adopting it themselves.

The late majority represent a substantial part of the population. Their adoption behaviour marks the point at which an innovation reaches a critical mass within a social system. The late majority are influenced by the experiences and feedback of the early majority, and their acceptance of the innovation is essential for achieving widespread adoption and normalisation of the innovation.

The late majority are pivotal in the diffusion process as they help to bridge the gap between early adopters and the more resistant laggards. Their adoption behaviour is crucial for achieving full acceptance and integration of innovations within a social system.

By addressing the concerns and barriers that the late majority may have towards adopting innovations, innovators can facilitate the diffusion process and encourage broader acceptance of innovations within a social system. The late majority's eventual adoption of the innovation is a critical step towards achieving widespread societal change and impact.

## Laggards

The "laggards" represent a group that is typically the last to adopt new ideas, practices, or technologies in the social system. Laggards are char-acterised by their resistance to change, scepticism towards innovations, and a preference for traditional and established methods. They are often hesitant to adopt new technologies and tend to rely on tried and tested approaches, even when faced with evidence of the benefits of innovation (Matzler et al., 2014). Despite their reluctance to embrace change, laggards play a role in the introduction of innovation by providing a different perspective. Their cautious approach can serve as a counterbalance to the enthusiasm of early adopters and the early majority. Laggards' adoption behaviour, albeit delayed, can offer valuable insights into the potential challenges and barriers that innovations may face in gaining widespread acceptance within a social system (Matzler et al., 2014).

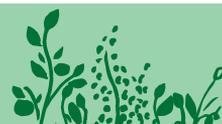

Rogers' diffusion theory put laggards at the end of the adoption curve. While laggards may be slow to adopt innovations, their eventual acceptance can contribute to the full saturation of an innovation within a social system. By observing the experiences of early adopters and the early majority, laggards may eventually be persuaded to adopt innovations, especially if they see tangible benefits and positive outcomes (Matzler et al., 2014).

The laggards play a significant role in accepting innovations as innovators can tailor their strategies to overcome their resistance, concerns, and reservations. Engaging with laggards and addressing their specific needs can help facilitate the diffusion of innovations and ensure broader acceptance in the social system (Matzler et al., 2014).

Both Bandura's Social Learning Theory and Rogers' Innovation Diffusion Theory explain how individuals change their behaviour in response to communication with others (Rogers, 1983, p. 305). Both theories emphasise that the exchange of information is a key factor contributing to behaviour change, which occurs within certain social networks. This fundamental assumption is also supported by Hamblin et al. (1979), who argue that "diffusion models portray society as a huge learning system where individuals are continually behaving and making decisions through time but not independently of one another. Everyone makes his own decisions, not just on the basis of his own in-dividual experiences, but to a large extent on the basis of the observed or talked about experiences of others" (Hamblin et al, 1979, by: Rogers, 1983, p. 305). However, there are a few differences between these two theories. In comparison to diffusion of innovation theory, social learning frameworks advocate for a more precise examination within diffusion studies, aiming to discern exactly what knowledge individuals acquire through their connections with adopters of innovations. This nuanced understanding may encompass various factors such as the time, financial investment, effort, expertise, and comprehension of technical terminology required for an individual to embrace an innovation. It would explore whether the innovation addresses the perceived issues or needs of the individual, its comparative advantages over existing practices, and the level of satisfaction experienced by adopter-peers.

The diffusion of innovation tends to assess the impact of modelling in a more generalised manner, often categorising individuals as either embracing or rejecting an innovation without delving into finer details. Moreover, a diffusion of innovation theory focuses on time as a variable which influences the change of human behaviour whereas social learning theory provides greater attention to behaviour change as a process. Next, both theories indicate that the individual does not always follow a model. The diffusion of innovation theory describes the phenomena as a re-invention, defined by Rogers as a degree to which an innovation is modified or changed by a user during the process of its implementation (Rogers, 1983, p. 175). Therefore, the resulting behaviour change may be a modification of that being modelled. Last, both social learning and diffusion of innovation theory focus on interpersonal information exchange as a starting point of behaviour change, taking into account other people as sources of change.

Both theories discussed above fit into the process of social change in terms of urban food self-production as together they create a framework for understanding how new ideas, practices, or technologies spread within societies and contribute to transformative processes. In the context of social change, the diffusion of innovation theory can shed light on how innovative practices or interventions aimed at addressing societal challenges are adopted and integrated into communities. Innovations that drive social change, such as sustainable urban agriculture, renewable energy technologies, or community-based healthcare initiatives, often follow a diffusion process involving various stages of adoption by various segments of society (Thurber et al., 2009). The theory of diffusion of innovation helps to explain how social change initiatives spread through society, starting from innovators and early adopters who embrace new ideas, to the early and late majority who gradually adopt these innovations, and finally to the laggards who may be more resistant to change.

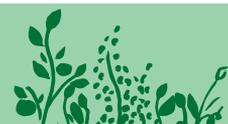

Understanding the characteristics and behaviours of each adopter category is essential for designing effective strategies to promote social change and achieve widespread adoption of innovative practices (Thurber et al., 2009). Moreover, the diffusion of innovation theory highlights the importance of communication channels, social networks, and contextual factors in facilitating the adoption of innovations for social change. By leveraging existing social structures, engaging opinion leaders, and tailoring communication strategies to different adopter groups, advocates of social change can accelerate the diffusion process and maximise the impact of their initiatives (Thurber et al., 2009).

## 3.3. Empowering food self-sufficiency in educational practice – implications for pedagogy

In the era of global crises and global trends, among which self-sufficiency appears to be crucial, independent food cultivation is essential in shaping human resources. This has significant implications for pedagogy and learning processes not only at the individual but also at the societal level, creating space for education and raising awareness of food systems functioning.

Urban food production systems have become important educational sites, providing opportunities for informal learning, and sharing experiences (Davila et al., 2015). This aligns with the concept of lifelong learning, which applies to any activity undertaken at any stage of life, allowing for acquiring new skills and knowledge in the realm of food sovereignty. These skills are most often acquired outside traditional learning environments (Jones, 2018), which is associated with the location and form of urban agriculture. Hydroponics, aquaponics, or container gardening rarely enter formal educational institutions, preschools, or schools. However, it is worth emphasising the significant resurgence of school and preschool gardens, whose functions are being rediscovered in contemporary Poland, especially in the context of creating spaces for child development.

The idea of creating school and preschool gardens is not new. As Ziemkowska (2023) argues, the educational value of the garden was already emphasised by Jan Amos Komenský, Johann Heinrich Pestalozzi, John Dewey, and Ovide Decroly. Komenský asserted that wisdom should be drawn not from books but "from heaven, earth, oaks, and beeches; in other words, to recognize and investigate things themselves, and not just others' observations and testimonies about things" (Komenský, 1956, p. 161), indicating the immense significance of the experience category and experiencing the surrounding reality in the educational process. The positive impact of nature contact on humans was incorporated into the pedagogical principles of Maria Montessori, Waldorf schools, and currently widespread forest kindergartens (Ziemkowska, 2023, p. 20). Historically, school gardens were perceived as spaces where observation, cognition, or knowledge acquisition skills were shaped, but above all, they fostered competencies enabling self-determination and care for one's livelihood.

Although over time, the formalisation of teaching and learning processes led to a limitation in using the potential of green spaces in education, it is worth emphasising that school and preschool gardens create space for interdisciplinary learning on physical, mental, emotional, and social levels; they allow for sensory integration, scientific research, and are part of ecological education. School gardens offer a unique opportunity for practical learning that can change students' social and environmental awareness (Davila et al., 2015). These initiatives can help children connect with nature, develop a sense of environmental responsibility, and support critical thinking about food systems (Davila et al., 2015). Children's participation in gardens positively influences their health, eating habits, and social skills (Turner et al., 2016). As Lee et al. (2018) argue, the use of gardens in preschools and schools improves research skills, develops mathematical abilities, creativity, and teaches interpersonal relationships.

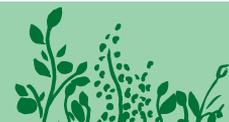

Gardening in the school environment promotes diverse and high-quality food production, contributing to global food security (Orsini et al., 2013; Sant'Anna de Medeiros et al., 2020; Eigenbrod et al., 2014). By engaging in urban food production, children can learn about and experience the cultivation process, develop a deeper understanding of food systems, food awareness, and nutrition, which form the basis of quality of life.

Incorporating gardening into the curriculum supports interdisciplinary learning, develops research skills, and promotes teamwork and social skills among students (Gardner et al., 2023; DeMarco et al., 1999). Such gardens are therefore valuable tools for promoting healthy eating habits, sustainable environmental development, and community engagement (Scherr et al., 2013; Bolshakova et al., 2018). The benefits of gardening fit into everyday educational practices. They can contribute to community development, social integration, and sustainable urban development (Casazza et al., 2016). By involving students in urban food production, schools can create experiential learning opportunities, support environmental education, and foster social responsibility (Álvarez-Herrero et al., 2021).

However, there are challenges associated with implementing and running school or preschool garden programs. The first issue concerns limited resources related to the education of teaching staff. Teachers themselves are not always adequately prepared to manage gardens in terms of knowledge and skills that could support children in maximising their potential (Hazzard et al., 2011; Nalumu et al., 2021). Secondly, preschools and schools are not financially prepared for the burdens associated with the basic equipment of such gardens. Maintaining preschool or school gardens requires care and gardening knowledge. According to teachers, the rush associated with implementing the curriculum often hinders field trips. Thirdly, it is not always possible to captivate children with nature-related topics. Building a relationship with nature requires time, space, daily exposure to nature, which over time leads to establishing a child-nature relation, thus understanding, and increasing ecological awareness. It is an important task of education in the face of current global crises.

Gardening, besides creating space for child development, is also an excellent form of therapy for adults and the elderly. Horticultural therapy, the use of gardening and plant-related activities for therapeutic purposes, supports older adults in building their physical, mental, and emotional well-being, becoming a valuable tool in promoting health and quality of life. Horticultural therapy can also support the resolution of mental health problems. One of the key benefits of horticultural therapy for older adults is its potential to improve cognitive function and emotional well-being. Engaging in gardening can stimulate memory, problem-solving skills, and creativity, which are important aspects of cognitive health in older adults (Detweiler et al., 2012). The sense of fulfilment resulting from plant care can increase self-worth and overall psychological well-being (Dayaningsih et al., 2021), while achieving tangible results in the form of grown plants. This element is the basis for creating space for self-sufficiency, both in terms of nutrition and economics, allowing for independent food production while overcoming various material, physical, and cognitive barriers to independent cultivation. Other people, who became a source of knowledge and skills in independent food cultivation, turn out to be extraordinary support in this process.

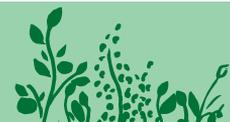

# Research assumptions

## 4.1. Rationale for addressing the research topic

Urban food cultivation is becoming an increasingly significant topic in the context of climate change, urban population growth, and food security. Conducting research on this subject is becoming an essential element of urban development, deepening knowledge regarding the benefits, challenges, and potential for the development of urban agriculture as an alternative form of food production (Gómez-Villarino et al., 2021). Urban agriculture plays a crucial role in promoting food security in cities by diversifying food sources, promoting healthier diets, and enhancing community resilience. The authors of the presented monograph aim to contribute to the scholarly discourse on urban food cultivation as a means to strengthen sustainable urban self-sufficiency, considering arguments for undertaking innovative initiatives to promote urban food cultivation and highlighting the significance of this issue for contemporary society.

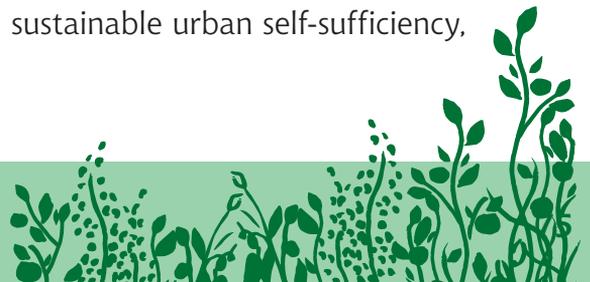

The monograph represents a significant excerpt of research conducted within the framework of the SmartFood project, focusing on food production and consumption in urban areas (Duda et al., 2023). The project's aim was to develop an innovative solution for co-creating tasty and nutritious food based on vegetables. Urban food cultivation requires access to adequate water and energy resources. For many urban agriculture projects, limitations in water and energy can pose significant barriers. There is a need to develop efficient water and energy management systems (Fox-Kämper et al., 2023). In response to this challenge, the food cultivated by the project's participants did not require the use of soil, potable water, or land, shifting food production to the corridors of residential buildings using hydroponic installations powered by renewable energy generated by rooftop photovoltaic panels and supplied with water from a rainwater harvesting system. This food production model considers four key dimensions of food production: availability, accessibility, stability, and utilisation (Ali et al., 2022). It proposes a response to contemporary urban challenges related to food security, such as limited access to fresh agricultural products, reliance on long supply chains, and susceptibility to price fluctuations. The proposed urban cultivation within the research project aims to contribute to the discourse on addressing these problems by increasing local food availability and accessibility.

The growth of urban populations is one of the key demographic trends in the contemporary world. The process of urbanisation, including the migration of people from rural to urban areas, has a profound impact on food production. Urban food cultivation emerges as a significant strategy to address the challenges of providing sufficient and healthy food for the growing urban population. The dynamic increase in urban population leads to a substantial increase in food demand. Traditional methods of agricultural production become insufficient, ex-posing cities to the risk of food shortages. Technological advancements, such as the development of hydroponic systems, open up new possibilities for food cultivation in urban spaces, leading to increased agricultural yields (Oliveira et al., 2020). Therefore, addressing the topic of alternative urban cultivation is crucial considering the consequences of demographic changes in cities and the associated need to seek innovative solutions that resonate and spread within urban communities.

In the face of climate change, traditional agricultural methods are becoming increasingly vulnerable to adverse weather phenomena such as droughts or extreme precipitation. Urban food cultivation can provide a local food source that is more resilient to extreme weather conditions. Research demonstrates that urban farming can achieve relatively high yields, better quality, and efficiency (Mishra et al., 2022). Urban agriculture, encompassing various forms of farming within cities as discussed more extensively in the first chapter of the presented monograph, emerges as a promising strategy to enhance food security by promoting local food production, distribution, and sustainable consumption patterns. This monograph delves into the relationship between food security and urban farming, analysing challenges, benefits, and scientific insights from the perspective of social learning among residents participating in a unique social experiment.

Urban farming can reduce reliance on distant food sources, contributing to a more resilient food supply system and reducing carbon dioxide emissions associated with machinery and vehicles used in food production and distribution (Kafle et al., 2023). It can support sustainable development and help protect the environment. Local food production requires less transportation and storage, resulting in lower greenhouse gas emissions. Concurrent educational efforts promoting consumption patterns based on urban farming and local products can significantly reduce carbon dioxide emissions generated by the food sector.

Another issue underscoring the importance of the monograph's theme is securing suitable land for urban farming, which remains a pri-mary challenge in densely populated cities, requiring creative urban planning solutions such as cultivation on unused land or alternative locations, including corridors between residential buildings. One of the major challenges for urban food cultivation is the lack of access to suitable land for agricultural production. Land in cities is often expensive and inaccessible to urban farmers, especially in city centres. The size and cost of land can discourage investment in urban farming; therefore, promoting hydroponic farming in residential buildings may prove to be a valuable alternative to existing traditional solutions. Urban farmers face limitations in accessing water, nutrients, and suitable growing spaces, necessitating innovative approaches to resource management. Ensuring food security in urban conditions requires attention to potential contaminants and harmful substances, necessitating effective monitoring and certification mechanisms (Suchá et al., 2022).

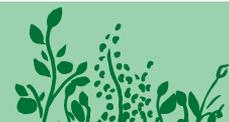

Another issue is the investigation of local community engagement in urban farming initiatives and the integration of food production into urban planning, which can promote equal access to healthy food. Freshly harvested products from urban gardens can be more nutritious and promote healthier diets among urban residents. Consuming local agricultural products can reduce the consumption of processed foods, which in turn can contribute to reducing diet-related diseases such as obesity, diabetes, or heart diseases. Urban farming can also stimulate community engagement and empowerment by involving residents in food production and distribution, thereby strengthening social cohesion. Hydroponic urban agriculture can promote community engagement and help to build interpersonal bonds. Residents engaged in food cultivation in their neighbourhoods can collectively contribute significantly to the local agricultural market. This not only promotes healthy physical activity but also enhances community and social integration (Engel-Di Mauro et al., 2021).

Unfortunately, in addition to potential benefits, there are also threats to the local community. Conflicts over urban space can lead to difficulties in developing urban food self-sufficiency projects, which is also a concern addressed by the authors of the monograph. Managing urban agricultural projects can be challenging due to the need to coordinate different interests and actors. Effective management of teams, finances, and educational activities may be crucial for the success of urban food cultivation (Lee et al., 2023). Therefore, joint efforts of scientists, policymakers, urban planners, and community stakeholders are essential for scaling up urban farming initiatives and maximising their impact on food security. By adopting an interdisciplinary approach and leveraging inherent technological advancements, we have the potential to develop alternative urban food systems that prioritise equal access to nutritious food and contribute to the well-being of urban populations, eliminating potential conflicts of interest among city residents.

The presented monograph focuses on the crucial role of education and social awareness in promoting urban food cultivation and increasing its acceptance and popularity among urban residents. Many people may not be aware of the benefits of urban food cultivation or may fear negative environmental impacts. Education and communication are key to changing attitudes and building support for urban agricultural projects. Urban food cultivation represents a promising solution to the challenges of global food demand, climate change, and urban population growth, but it also faces a series of challenges and barriers. Addressing these problems requires collaboration among various social, economic, and scientific sectors. Developing innovative technologies, changing urban policies, and promoting social education are necessary to promote sustainable agricultural practices in urban environments. Overcoming these challenges can bring numerous benefits to urban communities, including healthier diets, sustainable development, and a more resilient food system. Despite existing challenges, the benefits associated with urban agriculture are significant and require further scientific research and investment in the development of this form of food production. This monograph can therefore be a significant contribution to highlighting the impor-tance of researching urban food cultivation and pointing out its potential as a key element supporting sustainable development of urban communities and their self-sufficiency.

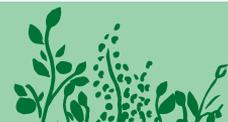

## 4.2. Research questions

The potential of local food production is attracting a growing amount of interest from the research community. There are increasing attempts to reduce the length of the supply chain (Cappellesso et al., 2019) and to raise the awareness of the population with regard to sustainable food consumption (Vermeir et al., 2020). Both producers and environmental activists, as well as educators with the backing of policymakers, strive to establish schemes that facilitate networks aimed at meeting the local food product demands of consumers.

Hydroponic farming, as a solution for mentioned problems, is gaining more and more adherents as an innovative approach to food production. This method involves growing plants without soil, instead using mineral nutrient solutions in a water solvent. It offers numerous benefits such as increased crop yield, reduced water usage, and minimised pesticide use (Khan et al., 2021). The integration of hydroponic food growing solutions into urban infrastructure is a focal point of interest for researchers, especially in its capacity to enhance food security in urban areas (Gentry, 2019), and in the context of small-scale farmers (Allaby et al., 2021).

However, the dedication of those responsible for implementing novel and efficient environmentally friendly practices, including hydroponic farming, is crucial. The matter of motivating individuals towards implementing innovation is thus an essential research query that has gained attention from subsequent scholars (Luehr et al., 2020). However, our literature review indicates that there is still a research gap in this field and suggests that more needs to be done to explore this. Therefore, the first main objective of our analysis is to gain a deep-er understanding of the reasons that motivate urban dwellers to implement sustainable food production solutions. We were guided by the following research questions:

RQ1: What factors motivate residents to engage in hydroponic food self-production?

RQ2: What is the role of learning in the decision-making process regarding participation in an experiment?

Motivation to act arises from diverse factors, yet it is also contingent on the task at hand. Some of the research focuses on motivational factors which specifically provide the outlook for urban gardening. Commonly mentioned motivations may be classified into the following categories: personal, social, environmental, and output. The first classification reflects individual motivations toward urban gardening, such as (1) well-being both individual – related to interaction with nature provided by gardening (Pourias et al., 2016) and related to the act of gardening itself – and communal (Drake et al., 2015); (2) recreation, relaxation, and leisure (Spilková, 2017), (3) health issues (Pourias et al., 2016), these research findings also presented in Duda (2024), (4) emancipation from urban life (Pourias et al., 2016), (5) education (Andersson et al., 2007; Bartner et al., 2010; Lewis et al. 2018), and (6) family history, childhood and passion for gardening (Kingsley et al., 2019).

As childhood experiences are important in developing a passion for gardening, our study also continued this theme. There is also some evidence that people with previous experience in farming or gardening are more eager to engage in urban gardening (Spilková, 2017). It was therefore our intention to gain a deeper understanding of the childhood and adulthood experiences of growing food held by those enrolling in a project aimed at urban food self-production. In this context, our research, which built upon previous findings (Duda & Korwin-Szymanowska, 2023), was guided by the following research questions:

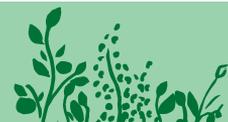

RQ3: What are the farming backgrounds of residents who wish to engage in hydroponic food subsistence production?

RQ4: What are their educational backgrounds in this field?

The social context seems to be even more interesting as it tackles the problem of meeting neighbourhood expectations and norms (Nassauer et al., 2009; Goddard et al., 2013), building community (Pourias et al., 2016), socialisation (Veen & Eiter, Eiter, 2018). The third group of motivational factors includes environmental concerns, interests in sustainability and impact on cities. The final category of motivation aspects is that of outcomes. These outcomes can be tangible, such as the production of food, improved food quality, food security, and self-production (Garcia et al., 2018), or intangible, such as the satisfaction derived from producing one's own food.

## 4.3. Characteristics of study location

The study was conducted within a project that aimed to establish alternatives for implementing innovative approaches to tackle the issue of climate change. As part of the Urban Living Lab (ULL, Duda et al., 2023) initiative, a group of inhabitants were chosen to grow food using hydroponic cabinets located inside the corridors of their apartment complex. The experiment aimed to ascertain whether self-production of food in urban areas could modify the eating habits of local residents, curtail food wastage and reduce carbon emissions. Given the considerable expense associated with the necessary installations, it was determined that the most cost-effective approach would be to undertake the project in an individual location, specifically within a single block of flats with a single community of residents. Given that the project leader resides in Łódź, the project was planned to be implemented there for logistical reasons.

As previously stated, the initial research sample was derived from the city of Łódź, a city with county rights, belonging to Łódź Voivode-ship, located in central Poland. The town was founded before 1332 and received city rights in 1423. The area of Łódź covers 293.3 km[1], with a population of 655 279 people, of which 54.4% are women. After Warsaw and Krakow, Łódź is the third city in Poland in terms of population. According to official data from the Statistics Poland, the population of Łódź decreased by 16.5% between 2002 and 2023. The average age of residents is 44.9 years, which is higher than the average age of residents in Poland as a whole, which is 42.1 years. In 2022, 22.5% of deaths in Łódź were due to cardiovascular disease, 22.4% of deaths were due to cancer, and 9.1% of deaths were due to respiratory disease. There are 15.65 deaths per one thousand population, which is significantly higher than the average for Łódź region (13.98 deaths) and the national average (11.86 deaths)[2].

The city is administratively divided into thirty-six subsidiary units (Figure 9). The individual residential areas unofficially operate within the five districts, a division abolished in the 1990s. They are: Bałuty--Centrum, Doły, Bałuty Zachodnie, Julianów-Marysin-Rogi, Łagiewniki, Radogoszcz, Wzniesień Łódzkich, Teofilów-Wielkopolska belonging to former district Bałuty; Chojny, Chojny-Dąbrowa, Górniak, Nad Nerem, Piastów-Kurak, Rokicie, Ruda, Wiskitno belonging to former district Górna, Józef Montwiłł-Mirecki's residential area, Karolew-Retkinia Wschód, Koziny, Lublinek-Pienista, Retkinia Zachód-Smulsko, Stare Polesie, Zdrowie-Mania, Złotno belonging to former district Polesie; Andrzejów, Dolina Łódki, Mileszki, Nowosolna, No 33, Olechów-Janów, Stary Widzew, Stoki-Sikawa-Podgórze,

Widzew-Wschód, Zarzew belonging to former district Widzew and Katedralna, Śródmieście-Wschód belonging to former district Śródmieście.

---

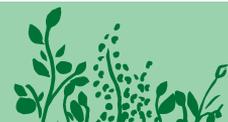

**FIGURE 9.** Administrative units of the Łódż city
Source: City of Lódź Office, https://bip.uml.lodz.pl/samorzad/rady-osiedli/, 15/03/2024

## Urban greenery in the city of Łódź

The largest complex of green areas in Łódź is Łagiewnicki Forest, covering 9.23% of the city area. It is also the largest urban forest in Poland and one of the largest in Europe[3]. Due to its character, single-fami-ly housing prevails in the Łagiewniki residential area, where most of the area is covered by the forest. Publicly accessible green areas account for 2.4% of the city's area, with the largest part located in the Polesie former district. In the Polesie area there are such facilities as: the Łódź Botanical Garden with an area of 67 ha, Józef Piłsudski Zdrowie Park, the largest city park in Łódź and one of the largest in Europe (with a total area of approx. 187 ha), J. Poniatowski City Park – the most valuable greenery in the city of Łódź, Klepacz Park located on the premises of the Łódź University of Technology, Lublinek Forest, with an area of 90 ha, located in the Lublinek residential area.

---

[3] Source: Urząd Miasta Łodzi, Departament Strategii i Rozwoju, Biuro Strategii Miasta (2020). Raport o stanie Miasta. https://bip.uml.lodz.pl/miasto/informacja-o-stanie-miasta/raport-o-stanie-miasta/, 15/03/2024.

There are 99 Urban Gardens in Łódź, with a total of 15'963 individual family allotments. The smallest Urban Garden (Waryński's Family Allotment Gardens, located in a residential area Widzew-Wschód) includes thirty-four individual allotments and the largest (Family Allotment Gardens "Księży Młyn", located in a residential area Stary Widzew) 902 allotments[4]. City residents are also encouraged to grow plants, including vegetables, herbs, flowers, within publicly accessible Urban Farms located, for example, in Greyer Gardens.

For the last 20 years municipalities of the city of Łódź have been introducing new ideas concerning the local greenery. As mentioned by Zdyb (2017, p. 74) it is related to the process of improving the quality of life of city residents through the process of revitalization in several spheres – social, spatial, and economic, which are aimed at general social revitalisation of degraded areas. One of the examples of such an action was the project "Green Polesie". The "Green Polesie" project was one of the initiatives in 2013 implemented in the area of Old Polesie in Łódź – a housing estate situated in the central part of the city in the Polesie district (Zdyb 2017, p. 79). The whole project was aimed at creating a greener and friendlier neighbourhood by implementing a participatory model of revitalisation which was based on social needs. The participation of residents helped to diagnose the needs and to recognise people's expectations in order to program the accurate and adequate solutions which would meet their demands. It also supports residents' involvement in the entire process of change. As a result, new pocket parks and some woonerfs were created (see Figure 10). As part of the works, the geometry of the streets was changed, and the space was supplemented with greenery and elements of small architecture. The diversity of forms of organisation of public spaces has created favourable conditions for building relations between people and spending leisure time in created spaces.

---

4 Source: http://lodz.pzd.pl/, 15/03/2024.

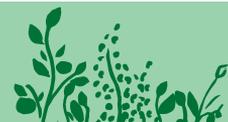

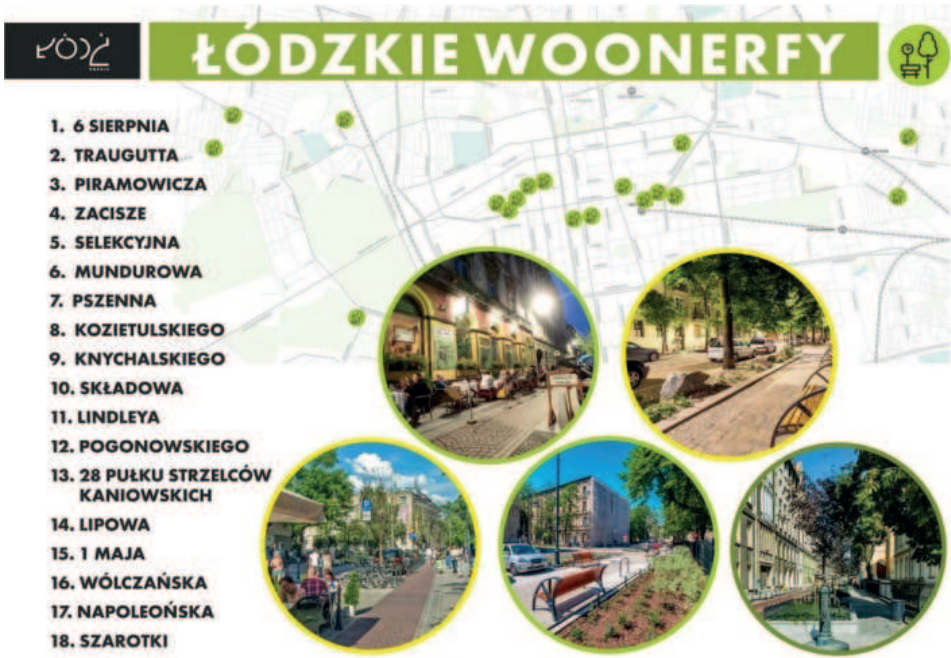

**FIGURE 10.** The map of woonerfs in Łódź

Source: City of Łódź Office, https://uml.lodz.pl/aktualnosci-lodzpl/artykul-lodzpl/woonerfy-po-lodzku-spacer-po-miescie-zielonych-ulic-zobacz-zdjecia-id42987/2021/8/26/, 15/03/2024

At the planning stage of the project activities, we assumed that the initiated social experiment would be conducted with one housing community. In order to ascertain the suitability of this assumption, we conducted a series of interviews with the participants who were qualified to take part in the study. However, after almost six months of working with a housing community from Łódź, the management of the community change and the new board member had objections to organising the project in their building. The concerns were related to a potential risk in the form of a possible loss of warranty due to the necessity to carry out construction works during the installation of the hydroponic cabins in conjunction with the photovoltaic and water installation. Consequently, the community, despite the discontent of the residents who had volunteered to participate in the experiment, decided to withdraw from the project. It was therefore necessary to alter the location of the study. A residential community in Warsaw was selected as the second location, following the fulfilment of the technical conditions for the installation used in the project. Warsaw is a city with county rights, situated within the Masovian Voivodeship in central Poland (Figure 11). It is the capital of the country.

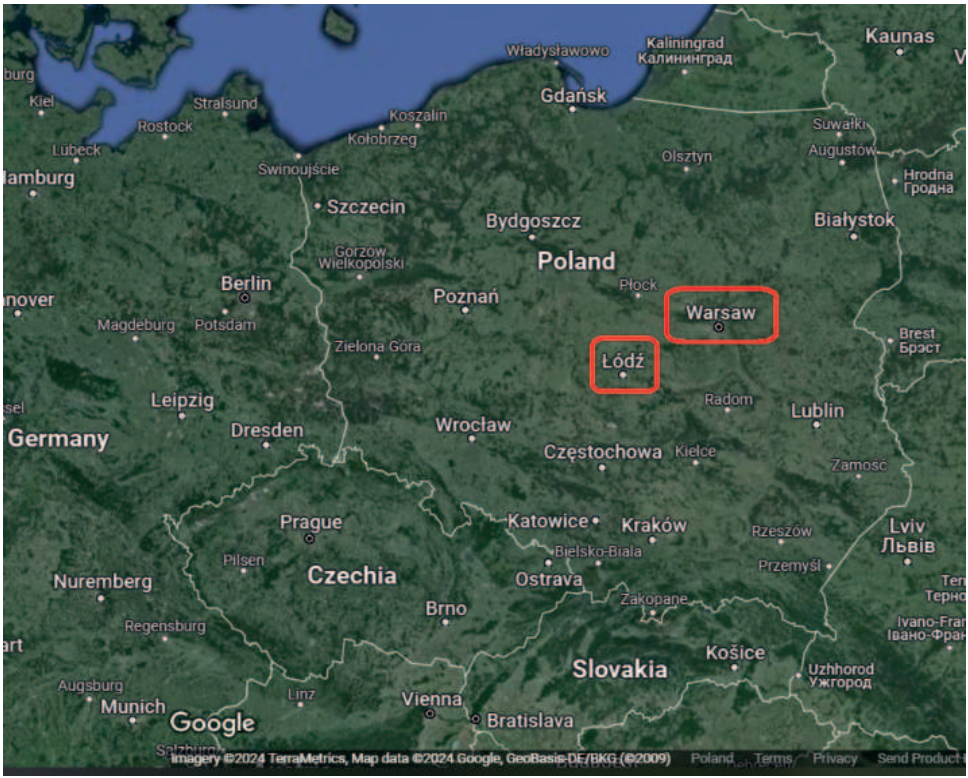

**FIGURE 11.** The location of studies
Authors' own elaboration using Google Maps

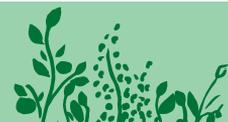

The town was founded before 1300 and received city rights in 1323. The area of Warsaw covers 517,2 km², with a population of 1 861 644 people, of which 53.8% are women. According to official data from the Statistics Poland, the population of Warsaw increased by 10.3% between 2002 and 2023. The average age of residents is 41.4 years, which is lower than the average age of residents in Poland as a whole, which is 42.1 years. In 2022, 28.3% of deaths in Warsaw were due to cardiovascular disease, 24.9% of deaths were due to cancer, and 9.0% of deaths were due to respiratory disease. There are 11.01 deaths per one thousand population, which is lower than the average for Warsaw region (11.36 deaths) and the national average (11.86 deaths)[5].

The city is administratively divided into eighteen subsidiary units (Figure 12): Bemowo, Białołęka, Bielany, Mokotów, Ochota, Praga-Południe, Praga-Północ, Rembertów, Śródmieście, Targówek, Ursus, Ursynów, Wawer, Wesoła, Wilanów, Włochy, Wola, Żoliborz.

## Urban greenery in the city of Warsaw

Urban greenery in the city of Warsaw plays a crucial role in enhancing the urban environment and quality of life for its residents. The presence of green spaces within cities has been linked to diverse benefits, including mitigating the urban heat island effect. Research has shown that increasing urban vegetation cover can effectively help reduce the surface urban heat island intensity during the day, particularly in the growing season (Soltani & Sharifi, 2017). However, studies have also highlighted challenges in maintaining biodiversity in urban green areas, with some

---

[5]  Source: https://www.polskawliczbach.pl/Warszawa, 15/03/2024.

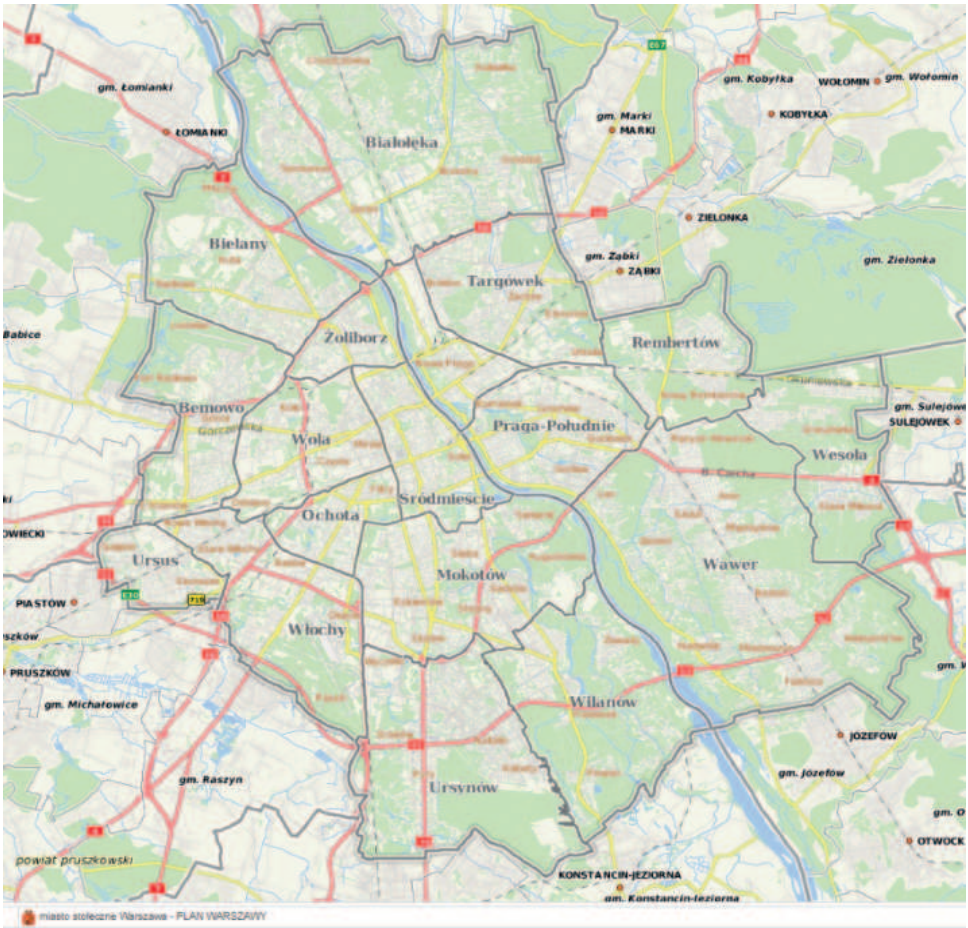

**FIGURE 12.** Administrative units of the Warsaw city

Source: https://mapa.um.warszawa.pl/mapaApp1/mapa?service=fast_mapa, 15/03/2024

green spaces in Warsaw being severely degraded or managed in a way that only supports a limited number of species (Ślipiński et al., 2012). Efforts to enhance urban greenery in Warsaw are evident in various initiatives and development paths aimed at greening the city centre and increasing the amount of green areas.

According to the report of the Warsaw city hall (*What kind of environment are we living in*) the capital city has extensive areas of natural or semi-natural value. These areas (Figure 13) include forests, meadows, wetlands, river valleys and agricultural areas with fertile soils. Together they form a vast ring of greenery around the city, serving a protective function against the spreading urban agglomeration.

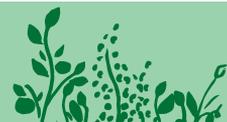

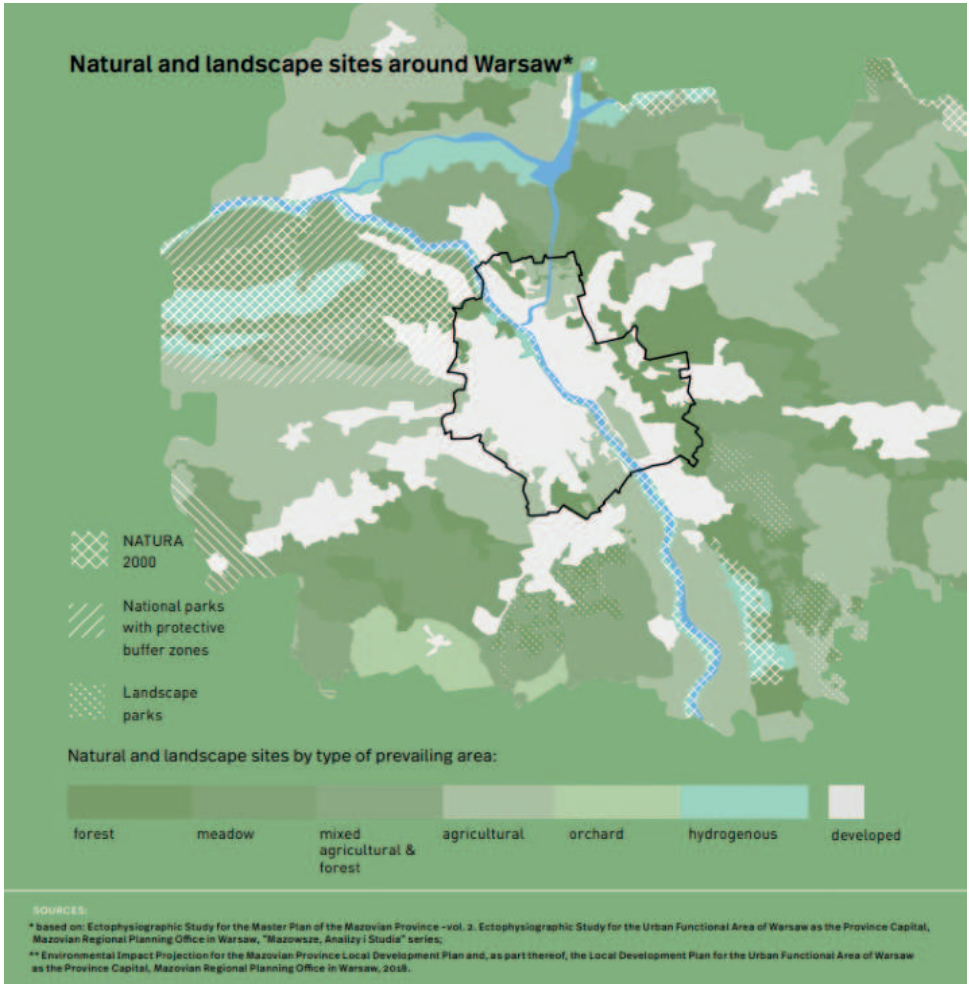

**FIGURE 13.** Natural and landscape sites around Warsaw

Source: Municipal Office for Spatial Planning and Development Strategy (n.d.). *Report Warsaw planning reports II. The environment we live in*, (p. 5). https://architektura. um.warszawa.pl/-/raporty-z-planowania-studium, 16.03.2024

They also provide places of recreation and leisure for residents. About 43% of these areas are environmentally valuable spaces covered by various forms of nature protection, such as a national park, seventy-four nature reserves, three landscape parks and areas belonging to the Nature 2000 network. Areas of high environmental value are also located within the administrative boundaries of the capital itself. Almost one third of the city area (27%) is covered by legal forms of nature protection. On the territory of Warsaw there are a fragment of the buffer zone of the Kampinos National Park, 6 Natura 2000 areas, twelve nature reserves, one landscape park, one area of protected landscape, six ecological utilities, five nature and landscape complexes. There are also other valuable areas in Warsaw that have not yet been legally protected, however, they are protected through city planning protection and maintaining their natural character that promotes biodiversity.

Warsaw is home to many wild animals. Forests, riverside areas, meadows and agricultural areas, parks with historic trees, green areas in housing estates, ancient necropolises, allotments, small streams, and ponds play a key role in providing shelter for animals. Areas with numerous endangered, protected, and rare plant species are also important vegetation habitats in Warsaw. Understanding the need of flora and fauna in the city is the basis of effective management and conservation of biodiversity. The Report indicates that the blue-green infrastructure of Warsaw defined as a network of natural and semi-natural areas and facilities in the city space, covered with vegetation or water, is needed for the good and healthy functioning of the city and its inhabitants: people and animals (Report, p. 9). The blue-green area serves as a space for recreation, improves the microclimate of the area, provides rainwater retention, absorbs pollution, and creates space for animal living. The above mentioned infrastructure has different land use: developed areas with a significant share of greenery (19%), forests (15%), agricultural and post-agricultural areas (13%), natural and semi-natural greenery, including biological restoration of water (10%), areas of allotment gardens (3%), surface waters (3%), areas of public greenery (2%), sports and recreation areas (1%).

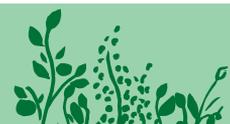

In the report, there are three categories of Warsaw green lands (See: Figure 14): core areas, supporting areas and supplementary are-

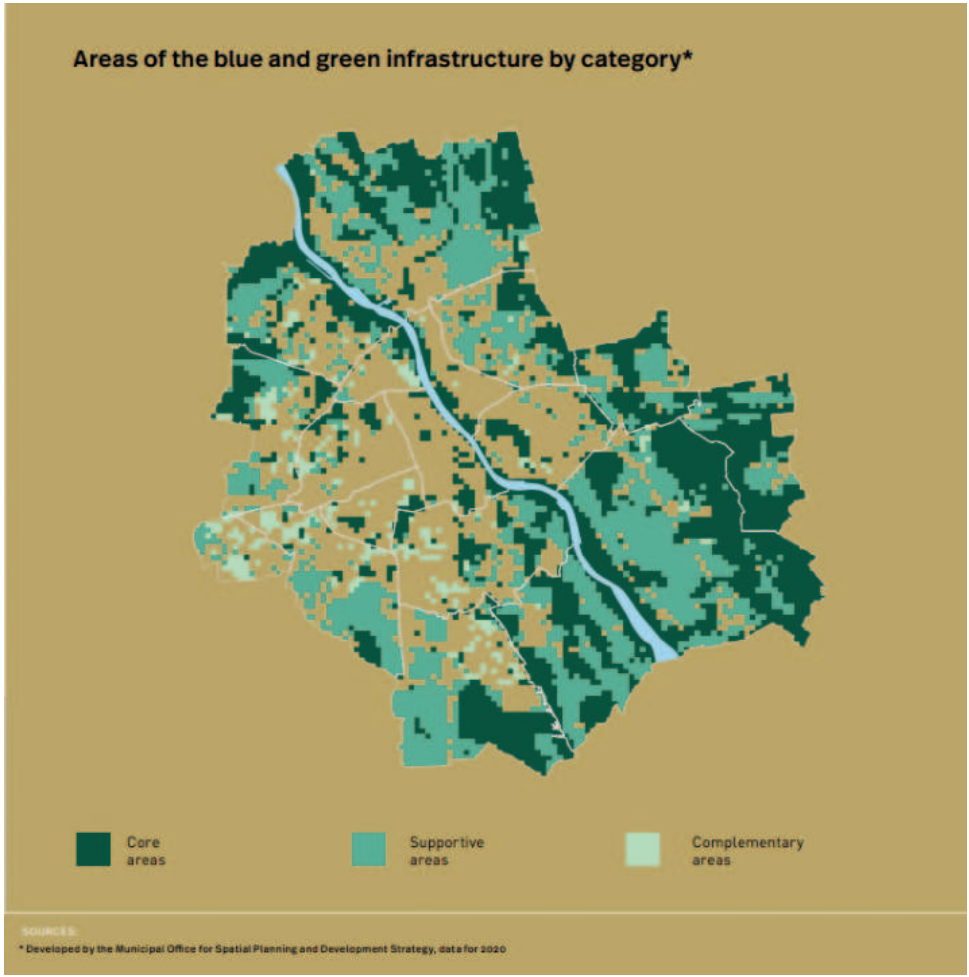

**FIGURE 14.** Blue-green infrastructure areas of Warsaw

Source: Municipal Office for Spatial Planning and Development Strategy (n.d.). *Report Warsaw planning reports II. The environment we live in*, (p. 12). https://architektura. um.warszawa.pl/-/raporty-z-planowania-studium, 16.03.2024

as. The first type is the most important for maintaining biodiversity or providing recreational functions. These include protected areas, forests, waters and their surroundings, wetlands, and areas of special risk of flooding, undeveloped fragments of the Skarpa Warsaw escarpment, as well as parks, greens, allotment gardens and other valuable areas, which are mainstays of biodiversity (Report, p. 11). Supporting areas have an immense potential to join basic blue-green infrastructure. They also contribute to the airflow within the urban environment and aid in managing rainwater runoff. These areas predominantly consist of farmland, former agricultural land, less dense woodland, cemeteries, and urbanised zones, all rich in vegetation.

Most interesting, however, is the analysis of Warsaw residents' access to green areas that provide recreation and physical activity. The public parks and greens, as well as forests and areas along the Vistula were taken into account as spaces available for recreation (through footpaths, bike paths, scooter paths or other facilities serving). According to the results, it turned out that 91% of Warsaw residents have access to recreation areas, 81% Warsaw residents live within 1200 metres of walking access to parks and green areas, 60% of Varsovians lives within a 500-metre walking distance of parks and green spaces, and 25% people live within 1200 m walking distance access to forests and areas Vistula riverfront (See: Figure 15).

Although the greenery surrounds the city, Warsaw is facing a problem of urban sprawl and city development which is reaching green spaces reducing resilience of the environment and the city as a whole against climatic and ecological threats. As a result, the environment loses its biodiversity and becomes, among other things, more vulnerable to droughts, attacks by pests and diseases or the encroachment of invasive species that displace native plants and animals. In this context, taking actions towards greater access to nature and food self-production seems to indicate a direction of development.

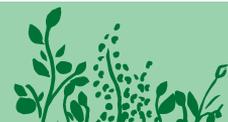

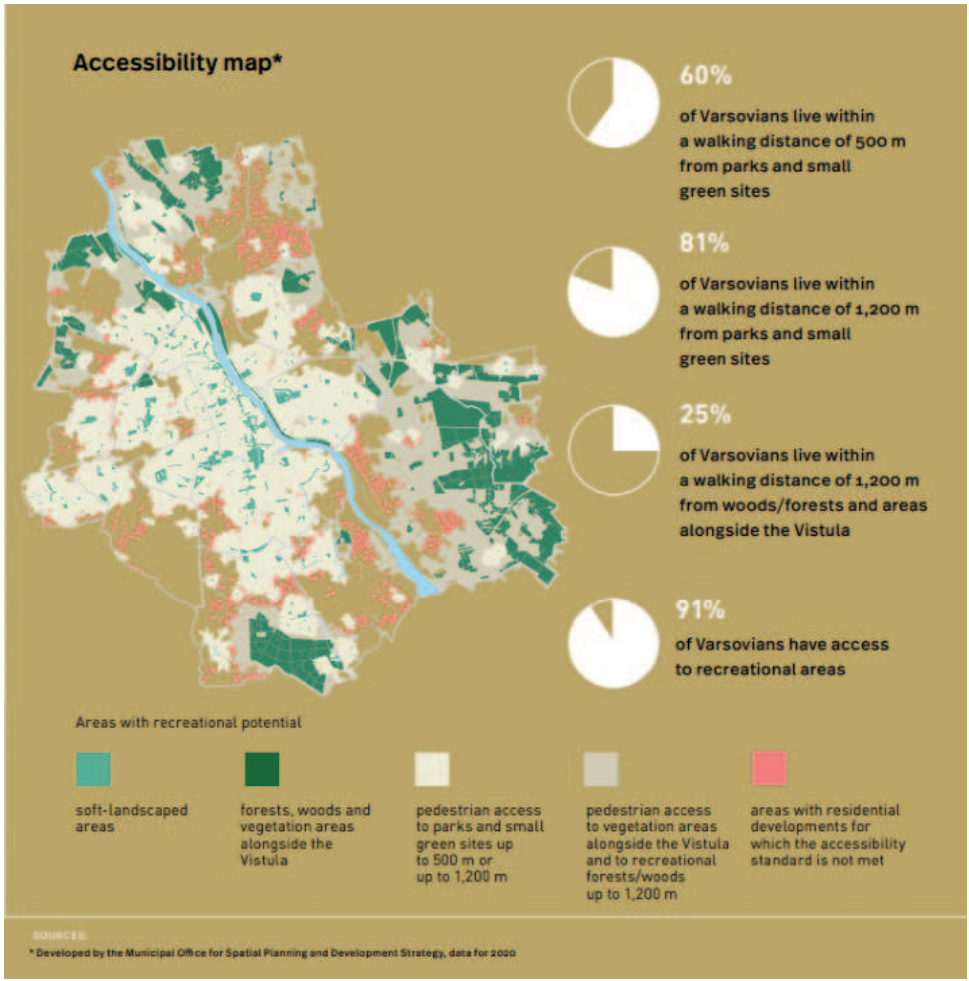

**FIGURE 15.** Map of accessibility areas

Source: Municipal Office for Spatial Planning and Development Strategy (n.d.). *Report Warsaw planning reports II. The environment we live in,* (p. 14). https://architektura.um.warszawa.pl/-/raporty-z-planowania-studium, 16.03.2024

## 4.4. Participants

Presented research is a part of a unique project which joins urban food self-production, environmental issues, technical advanced solutions, and social relations. Its uniqueness is based on hydroponic cultivation in a single block of flats where hydroponic cabinets using solar energy and rainwater are to be situated within the corridors of the building. Each individual participating in the project was to be provided with their own dedicated cabinet for the cultivation of plants. Given the substantial financial outlay involved, the pilot project was limited to the installation of twenty hydroponic cabinets in a selected block of flats. Consequently, the research sample is purposive and includes residents who have expressed a willingness to look after individual food-growing cabinets. Given our interest in deeper understanding of the individuals who form the core of one of the first links of innovation diffusion and given that both groups volunteered to take part in the experiment, we analyse the results of the interviews from both waves. This is because our focus is on understanding the rationale behind these individuals' decision to take part in the experiment. The subsequent research will be primarily focused on the analysis of the experiment itself.

In the first phase, the project encountered difficulties in contacting individual residents due to unforeseen legal barriers. During the course of the interviews, it became clear that it would not be possible to implement it in Łódź. This is most likely why three participants out of the 20 selected did not respond to phone calls and emails and were eventually not interviewed. Consequently, fifteen individual interviews and two parallel interviews with two residents (partners planning to grow food together in one cabin) were conducted in the first round, resulting in a total of seventeen interviews (coded TG2 to TG20) with nineteen participants, ($M_{age}$ = 39.4, $SD$ = 9.9, Range 27–72) including 11 women ($M_{age}$ = 40.5, $SD$ = 12.2, Range 27–72) and 8 men ($M_{age}$ = 37.9, $SD$ = 6.0, Range 31–50). The interviews were carried out between October and November 2022 and lasted between 1.5 and 2 hours.

The second set of interviews was conducted from September 2023 to January 2024. We conducted seventeen individual interviews and three parallel interviews with two residents (usually partners, spouses or family members planning to grow food together in one cabin), resulting in a total of twenty interviews (coded TG21 to TG40) with twenty three participants, ($M_{age}$ = 47.6, $SD$ = 17.2, Range 28–78) including 15 women ($M_{age}$ = 50.8, $SD$ = 17.8, Range 29–78) and 8 men ($M_{age}$ = 42.2, $SD$ = 15.8, Range 28–77). The summary is presented in Table 1. These were conducted according to the same interview scenario and also lasted between 1.5 and 2 hours. The survey questionnaire contained questions centred around four main topic areas: (1) background information, (2) social relations with neighbours, (3) motivation and personal per-ception of growing food in hydroponic cabinets, (4) food waste. Participants were compensated with a monthly amount of approximately twenty-two euros for participating in project activities.

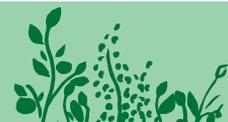

**TABLE 1.** Characteristics of study sample

| | First round study sample, planned location of the experiment – Łódź | Second round study sample, actual location of the experiment – Warsaw | Total study sample |
|---|---|---|---|
| Number of individual interviews | 15 | 17 | 32 |
| Number of interviews with two residents | 2 | 3 | 5 |
| Total number of interviews | 17 | 20 | 37 |
| Characteristics of participants interviewed | 19 participants $M_{age}$ = 39.4, $SD$ = 9.9, Range 27–72 | 23 participants $M_{age}$ = 47.6, $SD$ = 17.2, Range 28–78 | 42 participants $M_{age}$ = 43.7, $SD$ = 14.6, Range 27–78 |
| Characteristics of the women interviewed | 11 participants $M_{age}$ = 40.5, $SD$ = 12.2, Range 27–72 | 15 participants $M_{age}$ = 50.8, $SD$ = 17.8, Range 29–78 | 26 participants $M_{age}$ = 46.1, $SD$ = 16.1, Range 27–78 |
| Characteristics of the men interviewed | 8 participants $M_{age}$ = 37.9, $SD$ = 6.0, Range 31–50 | 8 participants $M_{age}$ = 42.2, $SD$ = 15.8, Range 28–77 | 16 participants $M_{age}$ = 40.1, $SD$ = 11.8, Range 28–77 |

## 4.5. Data analysis

Interviews were conducted and transcribed within the Microsoft Teams application. The analysis was performed using MAXQDA Analytics Pro 2022 software, based on the original Polish text, and then the quotes were translated into English. A general inductive approach strategy (Thomas, 2006) was implemented as the method of data analysis, with raw data being subjected to coding. The inductive approach employed has the objective of elucidating the process of data reduction by developing aggregate themes or categories from the raw data. This approach expected results to emerge from analysis of the raw data, not from assumptions or theoretical models drawn from existing literature. In accordance with the views of David R. Thomas, a general inductive approach is deemed appropriate for research when the analytical strategies and questions employed are applied in a context where the underlying meanings evident in the text, which are pertinent to the evaluation or research objectives, emerge. It is anticipated that the outcome of the analysis will be the identification of the most pertinent themes or categories in relation to the research objectives. The presentation of findings should be accompanied by a description of the most important themes (Thomas, 2006).

In accordance with the instructions provided by Thomas, the initial stage of the process is the preparation of the raw data files. In line with this directive, all interview transcriptions were imported into a unified project within the MAXQDA software used. In the second step, a meticulous examination of the text was conducted until it was fully read, and a comprehensive understanding of the themes and issues raised in the text was achieved. In the third step, we identified and de-fined the emerging themes. The more general categories were derived from the study objectives. The specific categories were derived from a comprehensive analysis of the raw data. This method is referred to as in vivo coding. The inductive coding employed involved the creation of categories based on actual phrases within specific text segments. The MAXQDA text quality analysis software was employed to facilitate the coding process.

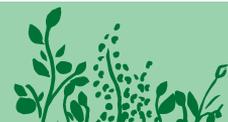

In the subsequent phase, categories that were thematically overlapping or redundant were eliminated. From these, the most relevant categories were then identified, which were used to construct a profile of the potential participant in the social urban gardening experiment in which they would take part. The categories were selected on the basis of a quantitative criterion, namely the counting of the most frequent indications of interview participants in a given category. Due to the recruitment of two independent groups of participants, one from Łódź and one from Warsaw, we conducted coding separately for each group. Subsequently, an analysis was conducted to ascertain whether there were any significant differences in each category as a condition of group membership.

The technical advanced solution adopted in the project is a response to nowadays trends directed to sustainability, re-connectedness to nature and self-sufficiency. The authors of this study intended that this individual hydroponic gardening would be connected with individual learning, however, corridors as a common room will provide space for social interactions between residents which oriented the whole process to social learning. This research is therefore embedded in Bandura's *social learning theory* and Rogers' *diffusion of innovation theory* which fulfil themselves creating widespread framework for comprehension of how people learn about food self-production and how new ideas and technologies spread within societies through shared experiences, interactions and knowledge exchange contributing to sustainable and transformative process of social change.

The basic assumption is connected with the statement that individuals learn by observing others, engaging in shared activities and reflecting on their experiences, which can influence the adoption of innovations. The whole innovation-decision process takes place in five phases. These are (1) acquainting oneself with knowledge, (2) convincing, (3) decision-making, (4) putting into effect, and (5) verifying (Rogers, 1983). By examining the preliminary experiences of Urban Living Lab's participants, we can more accurately evaluate the procedure, ascertain the level of implementation of the innovation, and devise a more effective plan for the subsequent stages of the project.

# Motivations for hydroponic farming in participants' opinions

## 5.1. Curiosity

The majority of participants in the interviews provided a number of reasons for their decision to participate in the urban hydroponic food growing project. The most frequently cited reason was curiosity. This factor was mentioned by one in two interview participants, with it also being the leading factor in the second group to join the project (Figure 16). Ten interview participants indicated this reason as decisive, an example being the statement: "I'm very interested in this crop growing. This is my main motivation. Indeed, it is such a bit of a cosmic thing, so uncommon, original, ecological" (TG32, 261).

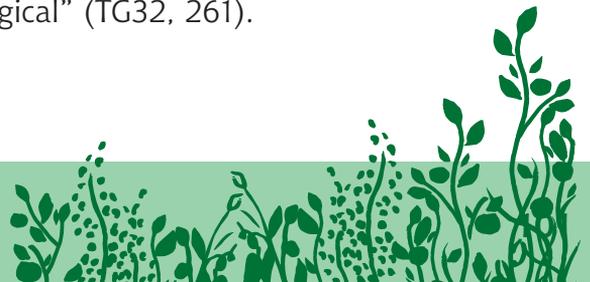

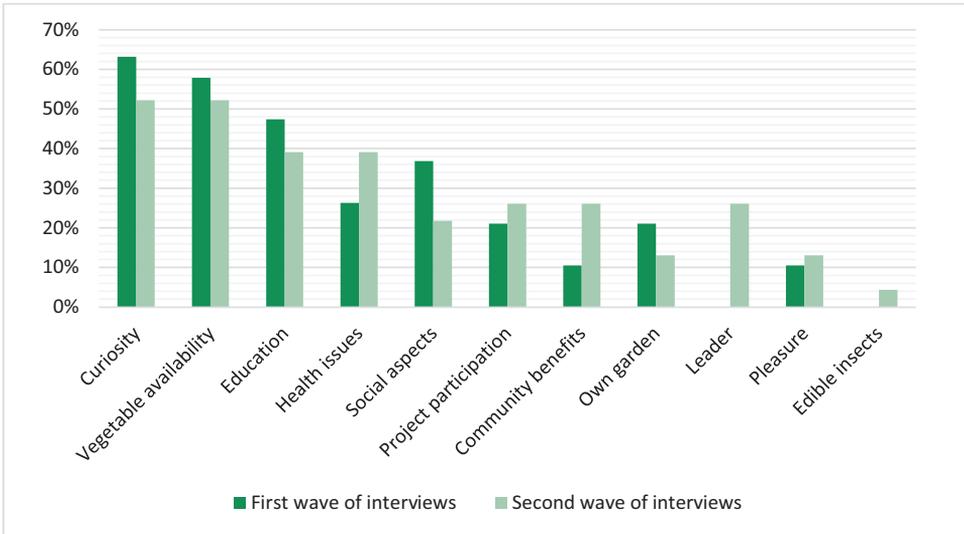

**FIGURE 16.** Motivations for hydroponic gardening. Percentage calculated in relation to each group ($N_1 = 19$, $N_2 = 23$)

Authors' own elaboration

Participants declared that their curiosity was due to the novelty of the issue. However, the arguments differed slightly from group to group. In the first group, three people expressed the opinion that there had not yet been such a project in Poland, that it was the first time they had encountered a similar initiative and that they were eager to experience the novelty offered by the project or even presented the forthcoming hydroponic cultivation as a vision for the future: "Certainly curiosity, because it's so innovative in general that I haven't encountered it, and whoever I've told about it hasn't encountered it either. It is even a bit like science fiction basically" (TG14, 145). Participants in the second group were slightly more curious about the technical solutions. They pointed to innovation in the form of the possibility of powering the crop with energy from photovoltaic panels, or hydroponic technology itself, as "another concept which, precisely because it grows on the eighth floor and not at ground level, is ideologically interesting, just to think about" (TG37, 154).

Seven people presented their motivation as curiosity about the possibility of testing innovations. The vast majority of this subgroup (Figure 17) belonged to the group of participants from Warsaw. They declared that they had heard about hydroponic technology or even have a deeper knowledge of it, but it still represents an innovation to them, and they would like to take part in testing this technology. Participants' statements indicate a positive attitude towards hydroponic technology: "I generally like new things a lot, or rather all kinds of gadgets, so I am very curious to see how this cabinet will work. Will this kind of hydroponics work for us? Generally, there are a lot of people who have hydroponic growing. I have heard about it, but I've never tried it, so I'm curious, and I've heard good things. I know those who have done it this way have been very happy, but I've never tried it myself, so I would love to test this solution" (TG18, 96).

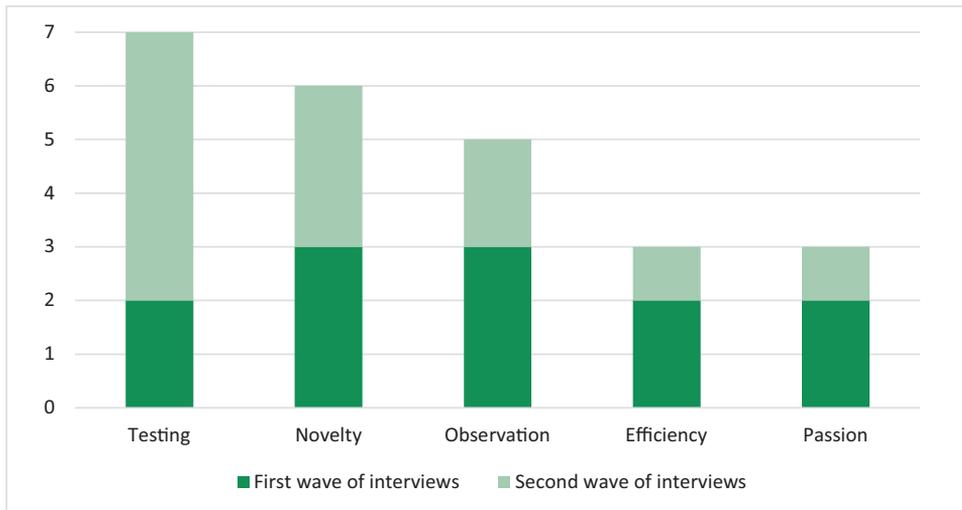

**FIGURE 17.** Curiosity as motivation for hydroponic gardening
Authors' own elaboration

The five participants wanted to find out empirically how hydroponic food cultivation works. They reacted with curiosity not only about being able to observe the process, but also to empirically verify which plants could grow, "when I decided to take part in the project, I was mainly curious about what such cultivation would look like, how it would roll; how the plants would develop under such conditions, because I had never dealt with this type of cultivation before and it was such a novelty for me" (TG33, 127). According to one participant, the idea of growing food together in the block seemed remarkably interesting. When she additionally learnt that the process was to be carried out in the form of hydroponic cultivation, her level of curiosity increased enough that she decided to apply in order to take part in the project.

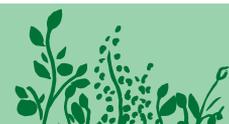

A further three participants were interested in gaining knowledge about the productivity of hydroponic farming. One participant expressed that he was familiar with the topic of hydroponics and felt that on a large scale it could contribute to more efficient production of healthy food worldwide. In the context of the project, participants interviewed wanted to know whether growing food from a single cabin could meet their current nutritional needs. An example of this was the following statement from a participant "I wonder if anything more can be planted in a cabinet, because you won't survive eating only lettuce for too long. Let us agree, eating only lettuce won't be very healthy. So, curiosity about what can be done with it, how to grow it and how much can we get from such a small space" (TG11, 217).

For a further three participants, curiosity for this particular project was a manifestation of a general interest in innovative activities taking place in the city. Respondents emphasised that they are interested in what is happening in the city, they follow the different projects that are taking place, especially those they find interesting. One participant shares this passion with his wife, so together they decided to get involved in a project that they find remarkably interesting.

## 5.2. Vegetable availability

The next most frequently cited reason for signing up to participate in an Urban Living Lab experiment was the desire for an easy and quick access to vegetables. Twenty-three participants identified this factor as a significant influence on their decision-making, including eleven from the initial review cohort and twelve from the subsequent review cohort. As with the previous theme, the availability of vegetables was viewed from a variety of perspectives. Two issues were most frequently mentioned: firstly, the proximity of the crop, and secondly, the opportunity for participants in the experiment to decide what vegetables would be planted. Interviewers identified a key benefit of having vegetables and fruit readily available in close proximity to their residences. It would be enough just to simply step outside and pick them. There is no necessity to visit a shop or market to purchase vegetables, as residents have immediate access to them. This aspect was more frequently highlighted by participants in the first group. Four out of five participants identified the benefits of proximity to vegetables (Figure 18). Among the benefits of the proximity to the crop, the lack of need to drive to the shop and the associated time and fuel savings were mentioned, as well as other issues related to the process of shopping in the shop, related to environmental sustainability: "you go to the shop, they print that receipt there, then you don't know what to do with it then you don't know where to put it, to the mixed or paper waste" (TG19, 208).

The undoubted advantage of urban cultivation was self-determination. Here, too, the statements were not evenly distributed. Seven participants declared that they were excited about being able to plant the kind of plants they wanted, which they would then consume. For three of them it was indirectly related to finance, i.e. they would not have to be guided by the price of the product when choosing a seedling, as they would when shopping in a shop. One participant was guided by her own and her family's preferences, i.e. she was looking forward to growing her favourite strawberries, which her children and she would later use to pre-pare desserts. Another participant was interested in growing radishes. Another participant was happy to have a wider choice of vegetables, as she had previously undertaken vegetable growing activities on her balcony, which limited the cultivation to a few basic plants. Two participants were interested in growing a particular species of vegetable, not easily available in shops – coriander and watercress.

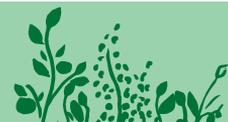

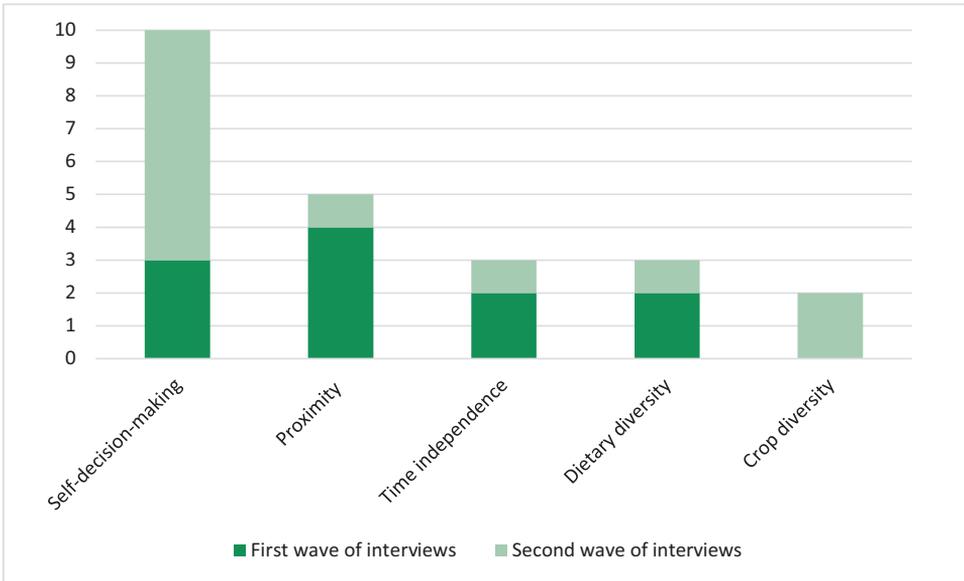

**FIGURE 18.** Vegetable availability as motivation for hydroponic gardening
Authors' own elaboration

Another aspect taken up by the three participants is access to vegetables as time independent. When growing in the corridor of a block of flats, vegetables are available regardless of the time of day or night, "nobody locks it like a vegetable shop" (TG18, 120). There are also no restrictions due to the time of year, as is the case with some seasonal vegetables or fruit. It is also valuable not to have to store vegetables artificially, i.e. in the case of daily cooking, participants can decide immediately before preparing a meal what to prepare as part of it and use the vegetables that are in the cupboard.

For three participants, the possibility to have variety in the meals they eat proved to be important. Participants declared that they regularly eat vegetables and fruit, but not in the quantities they find satisfactory. This fact may, on the one hand, be due to the increase in prices, as participants communicated that they eat significantly less vegetables than they did just two or three years ago. On the other hand, they choose solutions that require less effort "Many people forget about vegetables because they prefer to pour ketchup on everything, and vegetables are relegated to the background" (TG5, 133). The availability of vegetables will eliminate the identified barriers affecting the inadequate richness of the diet. The higher consumption will be not only due to the availability of vegetables, but also to the accompanying emotional considerations. It will be satisfying for residents to be able to eat vegetables and herbs that they have planted themselves, nurtured each day and then harvested, right behind their door.

A final aspect related to the availability of vegetables is the ability to grow multiple types of plants in one cabin. The benefit is not just access itself, but access to a variety of vegetables. An example is the following statement: "I think I would like to grow just the most commonly used plants, herbs; something that we practically use and use on a daily basis – all sorts of greens, lettuces, radishes. It would be nice if there was a whole spectrum of them" (TG33, 127).

## 5.3. Education

The results of the interviews show that educational issues were an important driver for residents to decide to start urban food growing. Eighteen respondents indicated this factor. However, educational needs differed between respondents (Figure 19). Nine participants saw a benefit in acquiring new and additional knowledge. The process of acquiring new information related to innovative food cultivation would take place in parallel and, in the opinion of some participants, "by the way". Participation in the project could provide more motivation to acquire new knowledge or deepen existing knowledge in a practical way, through project activities and information obtained from the project team, "I hope I can ask for some advice on the occasion on how to grow it at home, not only in this cabin that will take care of it all beautifully, but also about growing it on the window or balcony" (TG25, 186).

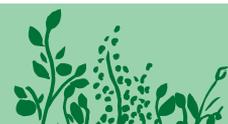

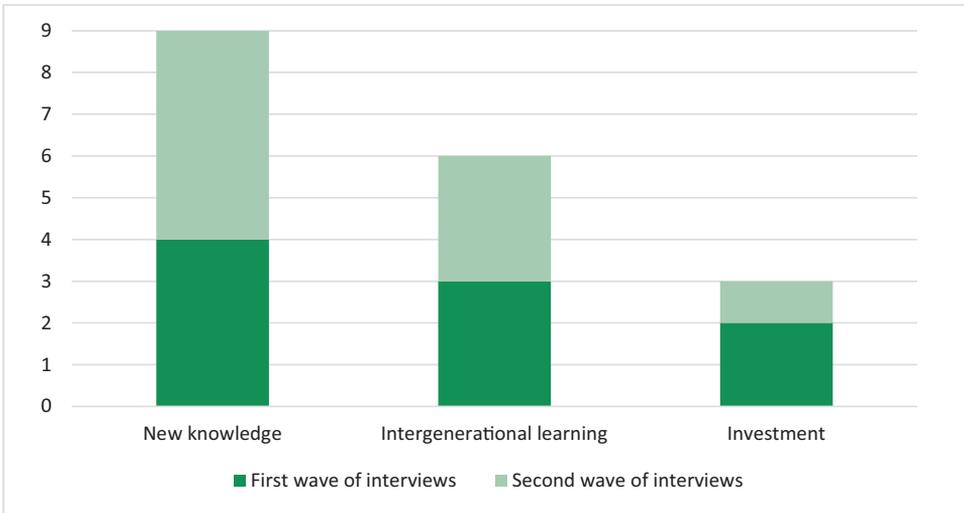

**FIGURE 19.** Education as motivation for hydroponic gardening
Authors' own elaboration

The men saw participation in the project as an opportunity to broaden their knowledge of technical topics related to hydroponic food growing, they wanted to learn the mechanical principles of the cabinets in practice as something new to them. They were curious about the construction of the cabin, the principles of operation, the technology used, which could pay off by trying to build such a system after the project: "To find out how to do it and maybe technically do something similar in the future. For me to build such an aquarium myself, I think it will not be that difficult, if I'm sentenced to the city, of course. Maybe then something will grow for me eventually" (TG8, 133).

For the six participants, learning was important because of the opportunity to pass on knowledge to their children. They would like to make their children aware of where vegetables come from in order to teach them respect for food, but also to introduce them to environmental knowledge in a practical way. The transfer of knowledge would be accompanied by the children's enjoyment of watching the plants sprout, watching them grow, watching the plants being watered, learning what nutrients they need or how they are harvested, "I mainly do this just for the sake of my kids, to encourage them and make them a little bit more aware of where vegetables come from. So that they know how it all develops, grows, so that they know that it is not like that, that it's taken off the shelf, that it's dedicated time, work, energy" (TG5, 130). In the eyes of the interview participants, the project is seen as an activity that shows children that they as parents should change their approach to food production, to buying and to consumption, to teach this approach to their children. Participants emphasised that they would feel very unreliable telling their child about their beliefs, at the same time knowing that out of lack of interest or passivity they did not take part in such an innovative activity taking place in their block. They would thus lose the opportunity to involve children in the learning process.

In contrast, for three participants, the opportunity to learn from the project was seen as an investment in the future. Some of them linked the need to learn and acquire new skills to their plans to move out of the city when they can afford to buy a plot of land and build a house there, "I don't want to live in the city all my life, I'm sure I will escape one day, the question is when. And when I escape, I would at least like to know how to plant a radish" (TG17, 169). Others saw this as the start of other activities. The perceived opportunity to expand knowledge could pay off with new interests, a new hobby, for example.

## 5.4. Health issues

For participants in the interviews, health issues were an important is-
sue. These can be divided into two main streams (Figure 20). The first
was related to the quality of the vegetables grown. Being able to grow
one's own vegetables was linked to the quality of nutrition, which
in turn allows one to take care of one's health. It was important to
the participants that the vegetables were protected from chemical
sprays. This is because shop-bought food is seen as a source of addi-
tional chemical ingredients of unknown proportion and effect on their
health. Being able to grow your own vegetables allows you to control
the growth process of the vegetables and therefore their quality, "If
I can have my own vegetables and herbs and know that it's healthy,
unsprayed, then I think that's a very good initiative. I think it is more
health-oriented that way. By joining this programme, I just had this
idea of healthy eating, that something of your own is always better
than bought" (TG15, 129). Although one participant also expressed doubt on this point. Namely, he hoped that the vegetables obtained from the crop would be organic, despite being grown in the artificial conditions of an urban block of flats.

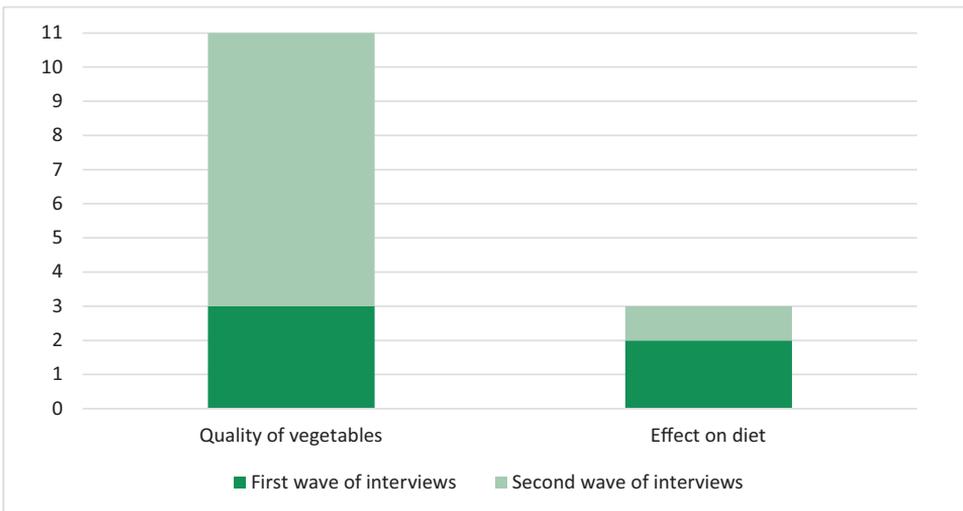

**FIGURE 20.** Health issues as motivation for hydroponic gardening
Authors' own elaboration

Another group of people declared that they hoped that growing vegetables themselves would motivate them to pay more attention to health issues by changing their eating habits. The sight of vegetables on the staircase might convince their family members to eat more herbs and vegetables, "because it's also very different when you take care of it yourself than when you just go to the shop and buy it" (TG23, 101).

## 5.5. Social aspects

Another factor that motivated participants in the interviews to decide to participate in the Urban Living Lab educational experiment was the social aspects (Figure 21). On the one hand, the jointly implemented activities could have been a pretext to establish closer relations with neighbours. The joint initiative would have been an opportunity to gather and collaborate and could have resulted in deeper relationships. Participants in the interviews, especially from the first wave of interviews, emphasised that they currently have a lot of work responsibilities and if they do not have a common topic such as children or animals, for example, it is difficult for them to find the motivation to take the time to go out with the initiative to establish closer relations with their neighbours. An example of this is what a participant said: "This is also one of the reasons why we applied for this programme, because our relationships with our neighbours simply do not exist. They don't exist for various reasons. We don't have any contact, maybe with two people we will talk for a while, but they are not the kind of relations I know from my childhood what relations between neighbours used to be" (TG13, 102). There were also statements from interview participants that participating in the project would be a motivation and an opportunity to spend more time with their children, "I was hoping first of all that it would be something I could do together with my daughter as a family. That's the most important thing" (TG22, 346).

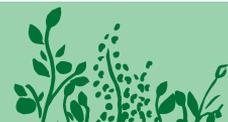

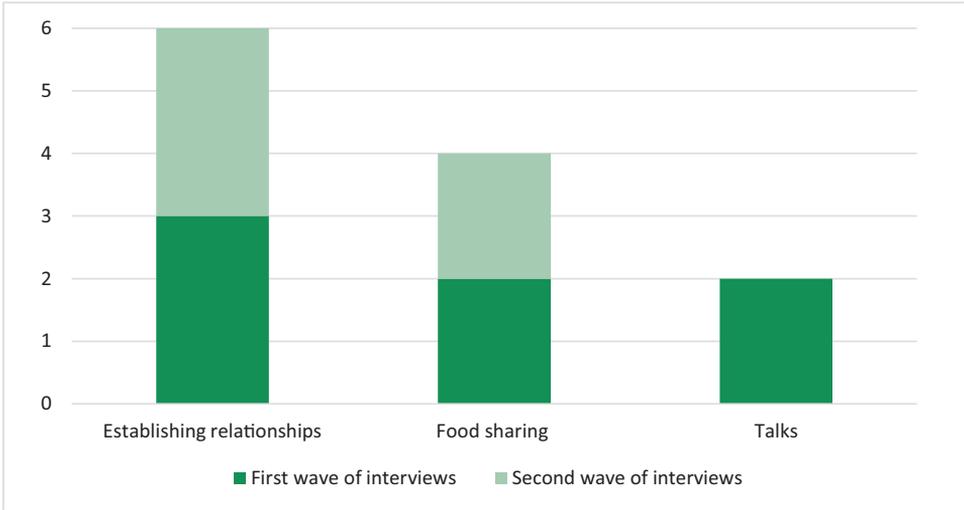

**FIGURE 21.** Social aspects as motivation for hydroponic gardening
Authors' own elaboration

The form of the project as a group project was attractive to some participants. They suggested that the project could, in a sense, work as a kind of co-operative. Neighbours could agree on what they grow and, in the event of an excess harvest, exchange with others. This would be a form of linking the creation of social relationships based on the needs of the individual. Moreover, the exchange described would not only involve sharing surplus crops with each other, but also the food prepared from them, "I thought I could make an agreement with my neighbour that we would make all sorts of salads here and all this stuff from what we grow" (TG35, 111). Through the mutual exchange of products, the community of residents would have access to a more diverse range of vegetables, fruits and herbs and the dishes prepared from them.

The opportunity to meet at the hydroponics stalls, to talk about growing topics, the emergence of common interests would become an opportunity to make new friends that would enrich the quality of life and well-being of the residents of the block. They could provide a pretext not only for establishing deeper relationships, but also for simple inter-neighbourly conversations, especially important for people who have not lived in the block for long, "When we were moving it was Mr […] who told us about this project. The desire to get to know the people in the block, to establish some kind of relationship, which was such a fundamental factor why we agreed to participate in this project" (TG38, pos. 194).

## 5.6. Other factors

One in four participants in the interviews felt that participation in the project itself was a factor that attracted their attention (Figure 22). The opportunity to engage in project activities is an interesting and innovative challenge. Instead of watching TV, they would like to actively influence their lives. The predominant statements here were about wanting to take part in something that is within one's reach and not standing on the side-lines, "I prefer to do something rather than watching" (TG5, Item 141); wanting to influence the surrounding reality, "for me it's an opportunity to just take part in a project and maybe affect it in some positive way or negatively, it's hard to say" (TG9, 124), gaining a variety of life experiences that diversify the daily routine, "it's also cool to be part of cool experiments. I treat it as an experiment to take part in; an experience to be gained. I feel it's on a par with going to the theatre, the cinema or go-karting. For me, it's just an experience I want to have" (TG24, Item 93). Still, the uniqueness of the venture was something that particularly attracted the participants, on the one hand something that is a novelty for themselves, but also atypical on a national scale, "This is something that nobody else in Łódź will have for the time being, so I have the opportunity to test something technologically new myself. I still see it as a technological innovation bordering on agriculture and science fiction" (TG17, 169).

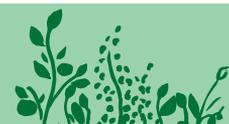

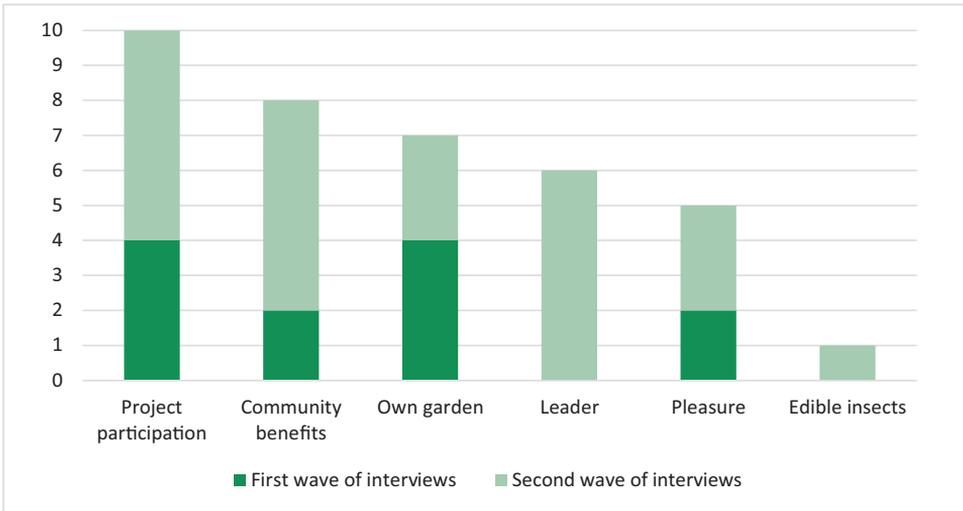

**FIGURE 22.** Other factors motivating hydroponic gardening
Authors' own elaboration

Interestingly, financial issues were not important to the participants in the interviews. In their view, the savings from access to vegetables would not be significant for the household budget, as they would still need to buy some basic vegetables such as potatoes, carrots, parsley, cauliflower, and broccoli. On the other hand, issues of benefit, important from the point of view of the whole residential community, proved to be significant. Eight participants stressed that it was important to them that, as a result of their participation in the project, a rainwater harvesting system and solar panels would be installed in their block, which would be at their disposal once the project was completed.

A proportion of the participants in the interviews indicated that it was not feasible for them to purchase a house with an allotment, either due to financial constraints or their partner's preference. In such cases, the aspiration of owning a house with a garden was superseded by the desire to either purchase or rent a flat with a balcony or terrace. Participating in the project affords four residents the opportunity to create a substitute for their own garden. "When I heard about this project, I just thought it could be such a great solution, as if we could have a house with a garden without having a house with a garden after all" (TG18, 28).

The role of the local leader proved to be an important aspect of the decision-making process. He was a concrete person in the case of the second group. On the one hand, participants in the interviews emphasised that he was able to encourage people to take part in the project. Participants declared that the leader infected them with passion for the project, persuaded them to cooperate as a community, "he cared a lot, because he lacked people, so that there would be 20 people, because they wanted to set it up. And we finally say yes, let it be. Because yes, the community wouldn't get it" (TG26, 732). Participants mentioned that the leader went from flat to flat and talked to all the residents, answered questions that arose and cleared up doubts.

Concerns were also raised at the community meetings, mainly about technical issues, about the fire risk caused by the additional electrical installation on the one hand and making fire-fighting difficult on the other. Residents feared that a possible fire-fighting operation would be hampered by the working photovoltaic installation and, on the other hand, the cabins in the corridors would take up space for evacuation. In this situation, the role of the leader not only caused doubts to be dispelled with talks, but also with concrete actions, "he just had to bang his fist in front of these board members and get on the board, because otherwise it would have never happened" (TG30, 75).

Another motive of importance to one in eight interview participants was the pleasure of contact with some form of nature. Participants looked forward to the mere opportunity to grow vegetables, to watch them grow, they saw the possibility of growing something with their own hands and using it to prepare meals, eating fresh herbs and vegetables, showing it off to family, friends as a process that would bring them satisfaction. The vision of having a crop without too much effort also seemed pleasant, "that's why we also applied in part because from what we were told, all this automation is supposed to take care of it for us on a daily basis, so it takes some of the responsibility off of us and we would have to spend less time on this crop, so it would be a good option for us. That's why we are very much looking forward to it" (TG22, 200).

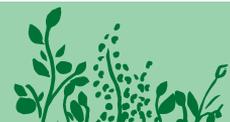

One participant identified the opportunity to grow edible insects, a voluntary part of the project, as key. The project would serve as an opportunity for him to receive technical and content support, which he could continue after the project ended.

# Educational contexts of respondents' experiences – towards gardening innovation participation

## 6.1. Informal learning in the decision making phase

### 6.1.1. Spontaneous decision without learning

A common thread running through many of the statements made by participants of interviews was that the decision to join the Urban Living Lab project was made on an ad-hoc basis, without further consultation or seeking information on what such a process might entail (Figure 23). Nineteen participants declared that they did not take any steps to acquire knowledge about the urban hydroponic food growing process, either before or after their decision. This was due to two main reasons.

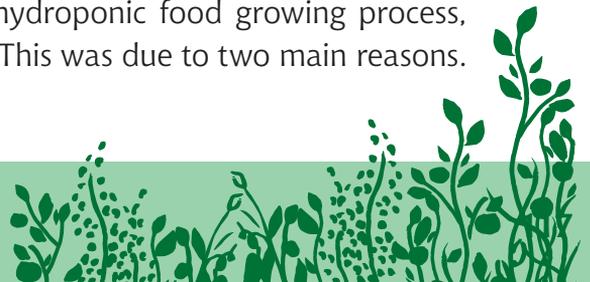

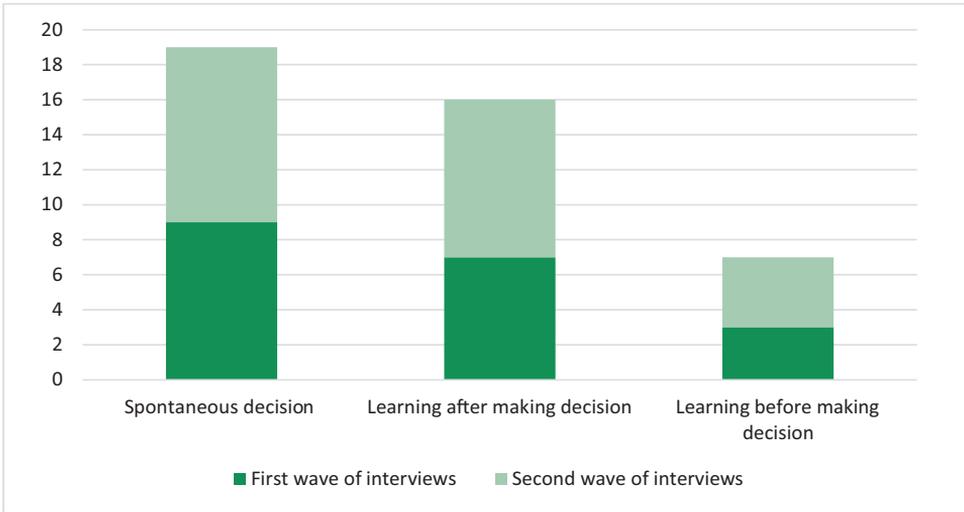

**FIGURE 23.** Informal learning in the decision-making phase
Authors' own elaboration

The participants did not perceive the task as challenging or difficult; rather, they regarded it as innovative and interesting. They treated it as a light-hearted activity or as an opportunity to diversify their daily lives, "A neighbour posted the information on this community page that such a project is going to be run and a question who would like to take part in it and without thinking I applied because I just wanted to do something interesting, different that I hadn't done before" (TG7, 208). Furthermore, the ease of recruitment, as evidenced by the simple online application questionnaire, contributed to the formation of this attitude. The questionnaire encouraged participants to make a quick, impulsive decision, which in turn reinforced the aforementioned attitude.

Secondly, participants contended that it is not necessary to learn during the initial phase of the project, given that the entire process will be monitored by the project consortium. Consequently, residents of the blocks assume that they will receive adequate support, both in terms of content and technical assistance. Besides, if they acquire wrong knowledge, it can hinder them in their gardening activities,

"I also think there's no point in preparing if I don't know what the conditions are. So, what if I read something, then I think of something and then I'm disappointed? I'd rather wait for someone to tell me something, what I'm going to grow there, how much space I have for it, how much nutrients to use there, how big plants I can grow, how I can divide these cabinets into some smaller shelves. There's no point in thinking about it. We'll wait until it's there" (TG1, 183).

Some of the participants doubted whether it was worthwhile to undertake educational activities, as the mere decision of the participants to participate in the experiment did not mean that the project would be implemented. This decision and the signing of the final binding agreements lay with the community management. Participants therefore preferred to wait with all activities until the project started, as the following statement indicates: "Since it's not there, I can't say. I will look into it once there is a chance that anything will actually happen there. That's when I'm very keen and I'll be reading different things about it. But until there is it…" (TG35, 118).

### 6.1.2. Learning after making decision

One-third of the participants in the interviews indicated that they began seeking information on hydroponic plant cultivation after submitting the project application form. In the first cohort of interviews, three of the respondents attempted to obtain information about the project via the project website or by attending online meetings organised by the project team. Four additional individuals attempted to procure information via the Internet. In contrast, participants in the second cohort were eager to receive guidance from a local leader who could provide a clear explanation of the fundamental aspects of the planned initiative or could facilitate attendance at a meeting organised by members of the project team. Nevertheless, the acquisition of knowledge for these participants was superficial, limited in time to approximately three hours. It was intended to provide them with a general overview of what was to come, "The consultation was when the contract came up. It was obvious that we didn't want to step into something, to put it ugly" (TG12, 159), with the assumption that they would receive full educational support from the project team at a later stage of the project.

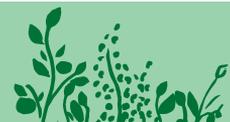

The respondents indicated that a more comprehensive understanding of the project at this stage was not necessary, given the information they had received. They did not anticipate any difficulties in joining the project activities. This is indicated by the following statement from one of the participants: "Well, at the beginning I didn't know anything, but then there were meetings with members of the project team, where the ladies explained to us that generally the handling of this will be minimal. They will supervise everything remotely, we would only have to replace or change something there, so it would not be very difficult to handle it" (TG21, 176). Furthermore, participants highlighted that the unconventional nature of the subject matter made it challenging for them to locate the information they were seeking, thus limiting their ability to gain the knowledge that would have adequately prepared them to participate in the project.

### 6.1.3. Learning before making decision

One in six interviewees indicated that they had sought information on hydroponic cultivation prior to joining the project. Concurrently, some of them had amassed this knowledge prior to becoming aware of the project. The collective body of knowledge was derived from perusing a multitude of articles pertaining to urban food cultivation, rather than the project itself. Two, however, were more deliberate in their pursuit of knowledge, reading material provided by a neighbour and other available material on the Internet, "I also searched somewhere, read on some forums, there was something on Facebook, I watched a video, I just gathered such basic information, I thought a year is not a lot, you can play with it" (TG8, 124) or "I try not to look for information on Facebook, but rather some articles that are written by some-one who has any idea about the subject" (TG38, 305). Nevertheless, as with the learning process that commenced upon the decision to join the project, the knowledge acquired was of a rudimentary nature. The second group comprised individuals with relevant knowledge derived from their shared professional backgrounds.

## 6.2. Farming backgrounds reported by participants

### 6.2.1. Early experiences

Our focus was also on the interview participants' experiences of their agricultural activities. Based on the collected statements, we identified four subgroups of participants in the experiment, namely those who had no early farming experiences at all; participants who had such experiences occasionally; participants who had them very often; and those who came from and grew up among farming families (Figure 24). The percentage distribution of the different subgroups was remarkably similar in both waves of interviews conducted.

In particular, the first subgroup, the least numerous, were participants who did not have access to any form of agricultural land, thus declaring that they did not have the opportunity to grow food during their childhood and from this period of their lives they have no agricultural experience. Most were of urban origin, with no close or extended family in the countryside. Nor did their parents or grandparents have a plot of land to grow food for their own use. These people lived in a block of flats where, due to the lack of a balcony enabling at least minor sowing or their parents' lack of interest in growing, there was no custom or family tradition of undertaking agricultural activities, as the following statements by participants indicate, "Nothing. Not in the sense that I don't recall them, they just don't exist. I was born in the city; I grew up in the city. My parents and grandparents are from the city, we didn't have a plot of land, well only my mum's brother has two apple orchards. Apart from going to my uncle's orchard of apples, my experience of growing fruit trees ends there" (TG17, 56); "No, I didn't have a grandmother in the countryside, nor a grandfather or anything. I used to go to the countryside, it was a summer village, but there were no crops there, the animals were there" (TG30, 51); "I don't have any experience. If we planted something there, it was maybe in the time when I was in the kindergarten or primary school, well maybe that's it, but I don't remember such situations" (TG38, 33).

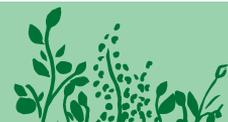

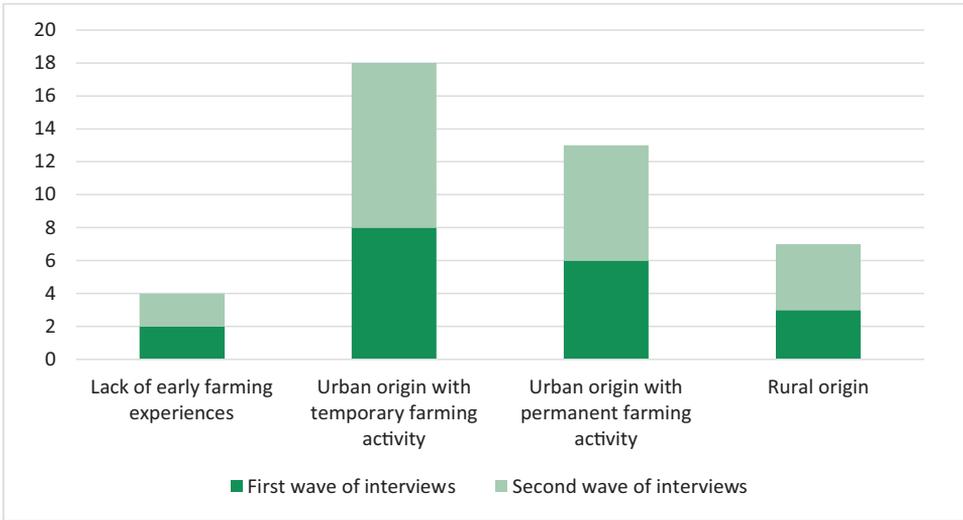

**FIGURE 24.** Early farming experiences of participants ($N_1 = 19$, $N_2 = 23$)
Authors' own elaboration

Another subgroup was made up of participants who had lived in the city in their early childhood but had occasional experience of growing food. They made up 43% of the interviewees. The farming activities of these interview participants most often involved working on an allotment owned by their parents, which they visited at weekends. In turn, some participants had family, most often grandparents, in the countryside. Farm work was usually part of their holiday activities. Among the representatives of this group, one can distinguish those who, as their statements indicate, positively perceived such a form of activity, "I had such a hobby as a child. When I was 4 years old, my parents bought such a recreational plot of land near Łódź, where we spent weekends. When I was little, I really wanted to have my own garden bed. I grew chives there, one bush of strawberries and potatoes, which I grew a lot of and was very proud of. I don't know where this curiosity came from, but I always had the desire to grow my vegetables. That bed was very small, by the standards of a small child, but I grew it for myself over the years, and we had the allotment for a long time. It was kind of my bed; I did everything to it myself. It was age appropriate, but I remember that I really enjoyed planting, weeding and I always had such fun doing it" (TG18, 32). There were also some participants for whom farm work was not perceived positively, "I was never passionate about the garden, I was more just going there to be in the green, not to grow vegetables or fruit" (TG28, 52).

Another subgroup consisted of those who had lived in the city while growing up and who regularly spent time gardening. This happened through access to a home garden or a garden or access to an allotment located where they lived, for example in the form of the Family Allotment Gardens (in Polish ROD, Rodzinne Ogrody Działkowe) described in Chapter 3. Sometimes participants spoke of their experiences of growing food there with a distanced attitude, "To tell you the truth, I don't remember if I participated in the work in the garden, although I think I was somehow dragged to it by my mother, yes, but to tell you the truth, I don't remember such activities taking up a big part of my time. It wasn't that" (TG39, 32).

Some participants saw both negatives and benefits in having an allotment, "I grew up in the city, but my parents had a garden where my mum tried to grow all sorts of things. My grandfather still planted fruit trees, so in the summer it was like 'now we're picking cherries', well at some point everyone hates it, but then there are jams and you can tell what this apricot jam can be delicious, then we have full cupboards of dried apples, you kind of appreciate the effect.

On the other hand, it was known that there was always some pretty big work involved" (TG4, 18).

In turn, for some people, the opportunity to do gardening work was seen as a method of shaping themselves, "I felt mostly useful. I had the joy of doing something physically. Anyway, the quick results of this work are there too. I felt independent as a child, that I could be relied on. It was a pleasure for me. […] We had a flat in such a small tenement and on the other side of the entrance, because there were two entrances to the tenement, there was a little garden – I mean a lawn, let's say – and part of this lawn was fenced off. It was, let's say, 3 metres long and maybe 10 metres wide, and my other grandmother brought soil there, agricultural soil, and mainly some decorative plants were planted there, but when mum came to visit, she planted strawberries and lovage. I picked the strawberries when they were just fruiting, and mum always sent me for lovage. Lovage for the broth and it was my job to bring lovage to the broth" (TG32, 73–85).

Another interesting statement from an interview participant was the form of motivation presented for children to participate in agricultural work. It was customary in his family to receive small financial gratification, "As far as weeding was concerned, sometimes I used to catch up with my grandfather in the greenhouse in the form of weeding for a small wage. The children had something to occupy themselves with, my grandfather was happy because he had cheap labour, as we worked on this basis with my cousins. But it was a pittance" (TG28, 39).

At times, participants explicitly mentioned the possibility of giving up family trips to the allotment, linked to the excuse of growing up. When older, going to the allotment was seen as a form of support for the immediate family, rather than a keen interest in growing greens or food. As an example, the following statement from one of the interview participants can be quoted: "I remember since I was a little boy the allotment was there, because my grandparents have always had it. I don't even know when I started going there more often. When I was 10 years old, you know, when I grew a bit older and so on, I used to go there 2–3 times a week. When I was little, we used to go there, because mainly my grandparents took care of me, after school and so on. And then, the older you got, the less often you went because you had other things to do than go to the allotment with your grandparents. In that later period, my role mainly just came down to pumping, watering, I don't know how you had to harvest. Blowing up and other things, it was no longer my job" (TG21, 80).

The most distinct emotional attitude towards food cultivation issues was expressed by individuals of rural origin. On the one hand, they appreciated the quality of food from their own crops. Among the childhood memories recalled by the respondents was the abundance of fruits and vegetables. Participants remember the possibility of pulling carrots straight from the ground and eating them, sometimes even without washing. The smell of fresh fruits such as plums, mirabelles, currants, chokeberries, or vegetables, especially the scent of freshly picked tomatoes or cucumber salad, remained vivid in their memory.

On the other hand, individuals raised in the countryside saw gardening work not so much as pleasure or satisfaction, but rather as a natural duty or even an unpleasant necessity: "I don't know if I liked this work. You just had to do it and didn't think about it. I think my generation is somewhat differently raised than, for example, modern ones. They now, for example, ask why I should do this, and I don't want to do it, and so on. And back then, you had to help your parents, and for example, there were no such questions: why should I do this, do I want to do it? Because I knew I had to do it, because I would eat it, right?" (TG40, 149); "Of course, I helped with field work, but I didn't like it when my parents went to pick strawberries or tomatoes" (TG6,32). The issue of helping parents with food cultivation work was perceived by the respondents as the hardships of living in the countryside, leading to later abandonment of it and a kind of escape to the city, perceived during that time as easier and more enjoyable to live in.

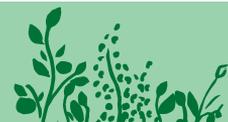

## 6.2.2. Later experiences

As suggested by the statements of the interview participants, their attitudes and attitudes towards growing food changed in adulthood, but these were largely influenced by their living conditions and access to resources to undertake farming activities. Overwhelmingly, residents were trying, to a greater or lesser extent, to undertake such activities (Figure 25).

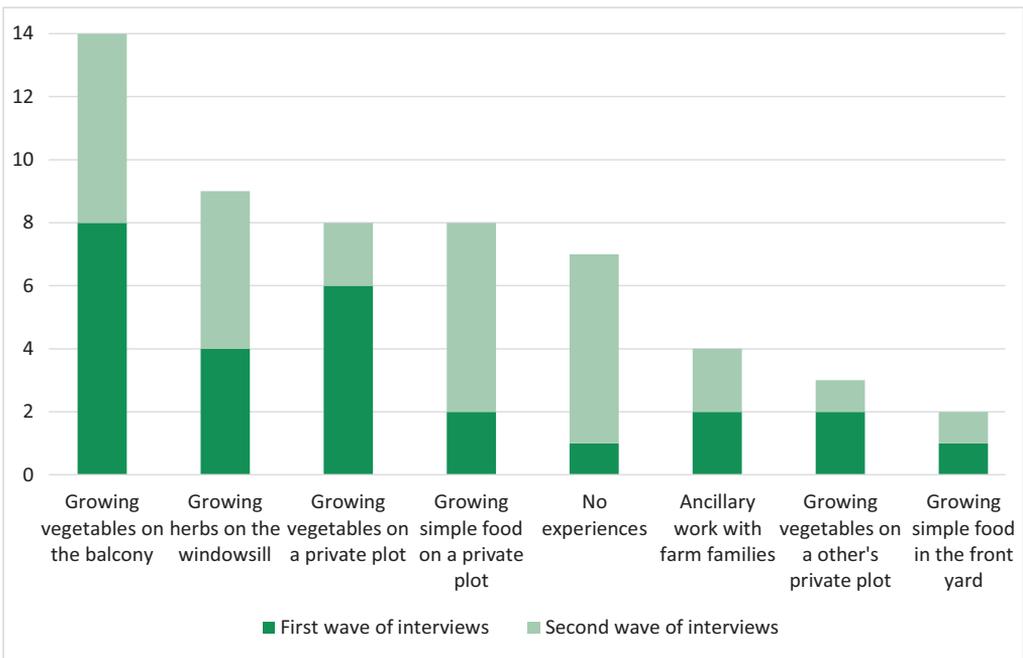

**FIGURE 25.** Later farming experiences of participants ($N_1 = 19$, $N_2 = 23$)
Authors' own elaboration

One-third of the study participants declared that they cultivate vegetables, fruits, and herbs on their balcony or terrace, with the percentage of such individuals being higher for the first wave of interviews (42%) than for the second wave (26%). Herbs are most commonly grown on balconies, along with small fruits such as strawberries or wild strawberries, or small vegetable bushes such as cocktail tomatoes and peppers. Some of the surveyed individuals limited themselves to herbs such as basil, chives, mint, rosemary, and lettuce, which they considered the easiest to grow because balcony conditions are challenging due to very limited space and lack of choice in planting location, as usually the entire balcony receives the same amount of sunlight. Some balconies were excessively sunny, while others were insufficiently sunny. Nevertheless, for many participants who chose it as the place to try planting food, the balcony was the only place where they could undertake such activity. The difficult conditions on the balcony also led some residents to give up after several unsuccessful attempts, as one participant expressed: "I once tried to plant something on the balcony, but it just doesn't grow at all. So why should I try when it doesn't grow? I'll just go to the health food store and buy a bunch of basil. I don't know, maybe it contradicts the fact that I love healthy things and so on, but I'm completely uninterested in it" (TG30, 45).

Due to the possibility of planting throughout the entire calendar year or due to the difficulties of balcony cultivation, every fifth participant engaged in activities related to herb cultivation on the windowsill. Even seemingly unlikely herbs were cultivated, as illustrated by the following example: "Three years ago, we moved to a new apartment and I have a balcony facing south, and we just have sunlight that operates quite strongly all the time. In the first year, I tried to grow tomatoes here on the balcony, but unfortunately, it didn't work out well because they started to burn, but we managed to gather some, nonetheless. But those plants weren't happy on my balcony. So, I just gave up, for now, unfortunately. I don't know about this apartment. Maybe eventually, the bedroom, because it's on the northwest side, so there's a little less sun there. Well, but that's it. However, I tried to grow strawberries here. It was going quite well, only they were attacked by pests, which I couldn't deal with, although the first harvests were very nice. I also had lettuce in the beginning in the kitchen, not on the balcony, and the lettuce grew great, and I just took some leaves, then again. I also had spring onions, scallions, in the kitchen on the windowsill, and garlic" (TG23, 40–42).

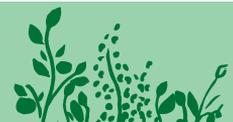

Another group of participants engaging in gardening activities consists of individuals with access to some form of land that can be used for this purpose. Four out of ten participants declared having such an opportunity. In particular, nearly every fifth participant stated that they own their own plot of land, where they regularly cultivate fruits and vegetables, with the percentage of such individuals being higher for the first wave of interviews (32%) than for the second wave (9%). Many of these individuals inherited plots from their grandparents or parents, which implies that they are not located very close to their place of residence, but they are already well-maintained and require continuing work, as illustrated by the following example: "This is not a new garden; it's 20 years old and it just grows lushly. And it sometimes annoys me that I can't keep up with maintaining order there. It's not that I like it, I just own it and I respect it and try to keep it in some reasonable condition. That's how I would call it. So, with this cultivation, I'm more concerned about efficiency" (TG3_2, 83).

A similar group owns their own plot of land but is less interested in cultivation matters, so they limit themselves to choosing plants that do not require much work, such as raspberry bushes, blueberries, or fruit trees. In this group, the proportions were reversed, with the percentage of such individuals being lower for the first wave of interviews (11%) than for the second wave (26%). Some interviewees use the plot mainly for recreational purposes due to its soil attributes: "Later, when I became the owner of my own plot, the plot is basically a forest plot, so attempts to grow anything there failed because the soil didn't suit it. There are many different trees there, pines, birches, and so on. I wanted to create some corner for carrots, radishes, and nothing unfortunately came out of it. However, I managed to grow chives, parsley, and dill in pots, that's what grows there" (TG39, 25).

Seven interview participants, six of them from the second wave, declared that they have not engaged in food cultivation activities in adult-hood. Firstly, these were individuals from rural backgrounds who rely on fruits and vegetables cultivated by their parents, as the amount of harvested food is sufficient to meet the needs of the entire family. One participant expressed, "There are a bit more of these vegetables that mom cultivates. I don't bother with things like zucchinis because there is always such a bumper crop there that we are practically sent as couriers where we live, because my mom doesn't have a way to process it all at the moment. Two people, like my parents, are not able to consume it all. The scale there is definitely wider" (TG2, 86). The abundance of food may also lead to a decrease in motivation and interest in independent cultivation work, as another participant mentioned, "It has accompanied us, both me and my husband, forever. It was natural for us. And it's natural that basically every weekend, and even more often due to having a grandchild, we go to our grandparents who gave us lots of these things, so it's quite natural for us. However, I myself have never dealt with it. And it will be a big challenge because neither I nor my husband have acquired such plant cultivation skills, and we can even dry out a cactus, so despite the patterns inherited from home, we don't fully find ourselves in it, so yes. It's close to me, but not in my own execution when it comes to plant and food cultivation" (TG14, 30).

The second group comprises individuals who declare that they do not have time to engage in additional activities due to numerous professional obligations, which prevent them from regularly tending to plants. One participant mentioned, "So now I'm just starting to work towards returning to my roots. Because I've been completely cut off from it for so many years. A bit out of desire, and a bit because I had to work a lot because my workload is over three hundred hours a month. That's two full-time jobs, plain and simple. So, day and night. So, I just wouldn't have time for that, plain and simple. Or I'd fall asleep in the car driving there. And now something is just starting to change, and something new is emerging" (TG37, 84). Sometimes, this choice is preferred as a kind of escape from the challenging work remembered from childhood or as a necessity, which is harder to accept, as expressed by another partici-pant, "I don't hide that I regret it. I think that a necessary condition to have such a garden is either actually owning a plot of land or actually being in one place, not constantly travelling between two places. So, I assume that if I were in this place for longer or had children, then I would probably set up such a garden on the balcony" (TG24, 21).

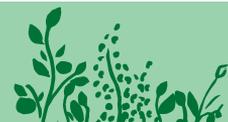

The next group consists of four individuals who do not have their own cultivation but still assist their parents or grandparents in agricultural work. These activities are carried out irregularly, mostly during periods of increased agricultural activity when parents need additional help. They arise from a sense of duty or willingness to help the family. Another three individuals used plots owned by their friends or family, most often in the form of Family Allotment Gardens, and jointly cultivated vegetables and fruits for their own needs. In this case, the activities undertaken were motivated by interests, hobbies, a desire to spend time together with family or friends.

The last group declaring experiences with vegetable cultivation in adulthood comprises three individuals who have access to communal garden plots, meaning they have apartments on the ground floor. The possibility of using such a garden usually leads to the care of grass or flowers, but it also allows for the cultivation of small, easy-to-care-for food items. One participant mentioned, "This year, I'm growing individual herbs. Those for everyday use. Also, in such a garden plot by the block, because I live on the ground floor, so I have a tiny piece allocated for mint and a few other small things" (TG28, 62).

## 6.3. Learning through farming activities

### 6.3.1. Early years learning

In many interviewees' statements, the conviction prevailed that despite spending a significant amount of time on agricultural work, they did not perceive an increase in knowledge about food cultivation as a result of these activities. The chores performed in childhood were a response to family needs, thus representing a form of obligation rather than genuine childhood interests. An example can be seen in the following statement from a participant, illustrating the process of knowledge sharing more between adults than between adults and children: "You also need to have some knowledge, and I know that with those neighbours from the plot, they exchanged experiences together, gave each other advice. Grandma talked about cuttings, they carried these cuttings, did something. But I was somewhat in the background, yes. Maybe I helped a bit, but rather as a child when someone told me what to do like 'water now', I don't know, I liked weeding, pulling out those perennials. When they told me to weed, I weeded" (TG29, 15).

The lack of interest was evident in many respondents' statements, indicating that it led to the mechanical performance of agricultural tasks, without encouraging accompanying educational activities. "They were eager to share [knowledge], but I was reluctant to accept it. It was an age when I wasn't really interested" (TG22, 212). Learning took place rather through observation or unintentionally, almost incidentally. "There was a lot of it. But also, observation issues, what they do sometimes, fertilisation, crop rotation, harvesting techniques" (TG28, 57); "In those times, the Internet was not so accessible, so I didn't search. It wasn't as easily accessible, and I wasn't super interested, so I didn't reach out to sources. I just listened to what my grandparents told me. […] Such matters simply emerged in practice" (TG28, 56–60); "It wasn't a topic that was very important at the time, so I helped out there. But no, it didn't arouse great curiosity. Mainly, I listened to what grandparents talked about among themselves. About how it was, about some crop rotations, that last year something grew in a field, so this year something else needs to be planted so that the soil doesn't become depleted, to alternate sowing fields, so I was involuntarily a witness to it, listened to these conversations, so I learned something, but it wasn't like I was actively seeking it" (TG22, 136).

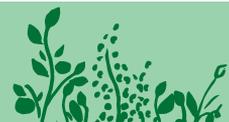

The lack of interest in learning issues was also evident on the part of adults, most likely busy with intensive and hard work on the farm, not inclined to make additional efforts to transmit knowledge to children or grandchildren. "Grandpa was quite stern, so it was like you have to gather it and that's it, there's no time for that. When we're already here, we have to gather it" (TG35, 53); "Well, you helped your parents a bit, but a young person isn't very interested in what they have to do. When parents call you to help with something, you know something, but there wasn't much interest. I didn't really know anything about plants from practice" (TG6, 55).

The situation was somewhat different in the case of multi-generational rural families, where parents assumed that their children would continue their work, and at the same time, the children shared the same interests. In these families, knowledge was actively passed on, not only through observation but also through active peer learning. "We learned from our parents. It was our parents who showed us all these activities, how to deal with them, which of the plants was the right one, the one we cultivated, and which was a weed and not needed in this garden. That was all the knowledge from them. So, it was mostly mom who showed us, because that was mom's area – dad dealt with heavier work in the fields and with the animals. But it was mom who directed us around the beds and explained, showed how to sow seeds, at what distance, and so on. Later, how to weed them. All activities, skills, all knowledge came from mom. We were there with siblings. I have five siblings, so firstly my older sister did something, s – and then probably I gained knowledge from them, but I don't remember exactly. But that's the way it more less worked back then" (TG33, 34).

## 6.3.2. Lifelong learning

Learning in adulthood, although present, was not a deep process for most of the study participants. An exception was made for individuals professionally involved in food cultivation or displaying deep interests in agricultural topics. Besides the knowledge, skills, and competencies acquired during formal vocational education, these individuals declared that they regularly read guides and industry magazines and participate in formal courses. "I try to stay up to date with various pieces of information. Agricultural technology is interesting to me. [...] So when I come across something, I immediately familiarise myself with it because it interests me; I try to look broadly at various things because it's interesting. They won't take this [knowledge] away from me" (TG9, 73).

The most common form of learning how to cultivate edible plants for adults was using Internet resources. This learning method was declared as predominant by six out of ten interview participants (Figure 26). According to the respondents, the advantage of the Internet as a source of knowledge is its availability and ease of use, "because it's always at hand" (TG5, 47). Interview participants did not use any specific websites; they relied on entering keywords into search engines and reading available information on two or three randomly chosen pages dedicated to the given topic. When the information started to repeat, they considered it credible.

The main reason respondents do not delve deeply into the topic of cultivation methods is the belief that this process is not complicated and will not pose excessive difficulties, as indicated by the following statements: "I didn't delve into it that much. Just bought a pot and without delving into agricultural details of how it should look. No, I didn't dig that deep. Besides, I thought it was simple" (TG8, 26); "When I tried with those tomatoes, we just plant and see what comes out of it, on the basis of planting and seeing what grows out of it" (TG13_1, 74). For some of them, the cultivation process does not seem difficult due to practical knowledge acquired from childhood and early youth, obtained from observing family members and actively participating in agricultural work, "But we don't use any professional literature to educate ourselves on this topic. [...] Sometimes we watch TV programs, but it's more about nice gardens, aesthetics, not fruits and vegetables, or food farming" (TG3_1, 50).

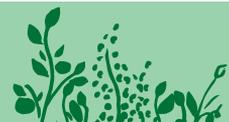

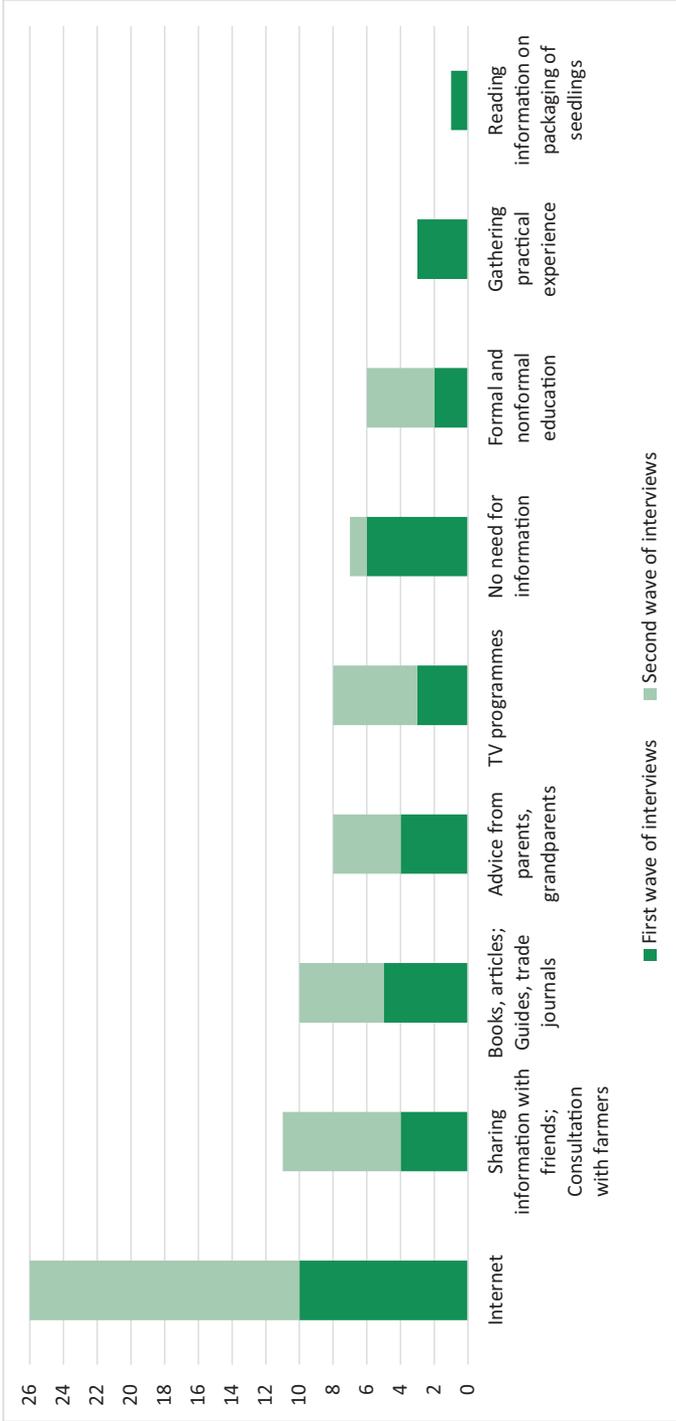

**FIGURE 26.** Source of information declared by participants ($N_1$ = 19, $N_2$ = 23)

Authors' own elaboration

Another issue indicating superficial learning is the lack of time, "But it's not like I actively seek this type of information – it's more of a re-action to a specific situation than a deliberate action. We can attribute it to a lack of time" (TG17, 302). Another statement by another re-spondent, "I'll put it this way, I never really had time for it. Since 2014, I haven't been working in Łódź, I work outside of Łódź. I said goodbye to the Łódź job market and commuted to work from Łódź, so I really don't have time for it. When do I have time? I have time on weekends. Just like today – I left home at 5 a.m. and came back at 7 p.m. And that's how the day goes. Unless I go straight to training after work, then I come back at 10 p.m. Then I make sandwiches for the morning, go to bed, and go to work in the morning" (TG15, 55).

Another way to acquire information about food cultivation is sharing knowledge with friends, often during gatherings or conver-sations at work, or asking friends for advice in case of specific doubts or the desire to get answers to specific questions. Some respondents also stated that they try to consult doubtful issues with farmers selling seedlings at the market, "I usually turn to the Internet, that's the first source of information. Possibly later, if I had any doubts, I would probably seek help among my colleagues from work who have hous-es and I often know that they also grow plants, but mainly the Inter-net" (TG34, 57).

In addition to sharing knowledge with friends, respondents stated that in order to acquire knowledge on topics related to food cultivation, they use books, guides, or articles, although, as with other sources of knowledge, they use them selectively or in a basic scope, as indicated by the following statement, "And I even bought a book about growing in an apartment. Urban gardening – something like that, a green one with a big carrot. Well, I didn't read the whole book, just what interest-ed me the most. However, this year I didn't sow those seeds, just more seedlings. So, I didn't use it very intensively" (TG16, 69).

Often participants try to acquire knowledge from various sources to make sure if a particular solution can be effective, "nowadays I often talk to friends, and if a topic interests me more, I just check it on the Internet on Google or on YouTube. It's not like it's my greatest passion, but if something really interests me, I check it out. If I had to make pots for winter (for example), so that some herbs or something wouldn't freeze, I would definitely check it out on the Internet and do thorough research. Probably 2 or 3 hours is not super thorough research. Then I would go to a gardening store and talk to some professionals there, how to prepare the substrate and so on, and only then would I make a decision on how to shape it, so there is always checking first, and then there is a decision" (TG24, 35).

People from rural areas or those raised in families with gardening traditions clearly value the skills of their parents or grandparents. They stated that in case of any doubts, they first call their parents or grandparents to ask for clarification on a particular issue, or even rely solely on family opinion, drawing knowledge and skills from them, "Here I admit that I take the easy way out, I simply call my mom or grandma because they have been doing this for x years. Grandma probably for 50 years, mom a bit less, but equally long. So mainly this way. No other sources. The experience and what I see in the work of my mom and grandma, the scale they are in, how the plants develop, and so on is enough for me to be sure that the work I put in and the money I invest will pay off" (TG2, 69).

After four respondents from each wave of interviews pointed to television programs dedicated to gardening issues as a source of knowledge or inspiration. Among the less popular sources were participation in courses, training sessions, or workshops; one person mentioned using the knowledge gathered during school education. Three more individuals relied on the experiences they gained during practical work in the garden under the guidance of parents or grandparents or later worked in their own garden. These individuals stated that the most valuable knowledge and skills for them are acquired through trial and error. One person, on the other hand, only used the instructions provided on the packaging of purchased seeds or seedlings.

Interestingly, seven interview participants, including six from the first wave of interviews, declared that they do not seek any information on food cultivation. One specified that they do not do it due to lack of time, while others due to lack of necessity. This partially applies to individuals who indicated that they do not have experience in food cultivation, "I didn't look. I didn't look because we've only been living here for 2 years, in this building where we have a balcony. Previously, we lived in an apartment where there was no balcony at all, so I didn't really assume that there was the possibility of growing anything there" (TG22, 204) or to those who did not need such knowledge because they did not have difficulties with the minor cultivations they had, "to be honest, we didn't really [look for it – ED] since it was growing. We saw that it was growing nicely. Maybe if it didn't grow so nicely, we would read something, try. But why bother if it works?" (TG12, 62).

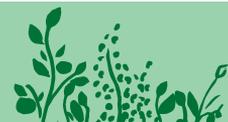

# Education for self-sufficiency in urban gardening – research summary

The presented monograph focuses on the role of education and social awareness in promoting urban food cultivation and increasing its acceptance and popularity among city residents. Our attention is centred on the initial stage of the project, during which a group of scientists – enthusiasts – along with the community of a selected residential block, participated in co-creating independent urban food production supported by modern technology, which we associate with the process of innovation diffusion. Our intention was to describe the learning process accompanying the residents who expressed willingness to participate in a unique project in Poland involving hydroponic cultivation of herbs and vegetables in the corridors of their residential block. We sought to obtain answers to our research questions regarding the motivations of participants to engage in the educational social experiment, the role of education in the decision-making process to participate in this project and learning during the acquisition of various gardening-related experiences, both within family and neighbourhood relationships. Below, we present a synthetic summary of the research results, described in detail in the previous chapters of the monograph.

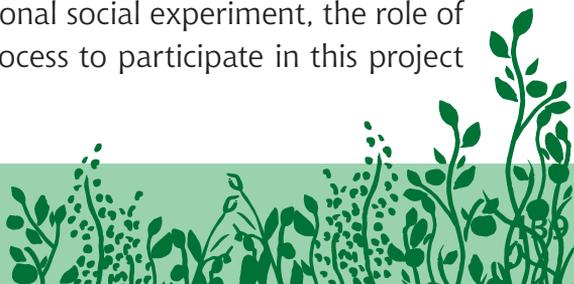

In response to the first research question, our study identified eleven most frequently mentioned motivations for joining the hydroponic food cultivation initiative. The first one is curiosity about the project, which addresses a topic that is still new to residents. For the first group of respondents, curiosity stemmed primarily from unfamiliarity with the issue, which even appeared as science fiction, leading to a desire to participate in the venture. Respondents were also interested in how this previously unknown innovation would be implemented as a process. On the other hand, for the second group, interest in the topic arose from technical matters that could be tested and verified. Curiosity also extended to the empirical exploration of the effectiveness and efficiency of the solution in the form of a relatively small hydroponic cultivation in the limited space of the residential block's corridor. Curiosity is the motivating factor that contributed to the decision to participate in the project.

Both the novelty of the issue, the opportunity to test innovative solutions, the acquisition of knowledge in the field of hydroponic food cultivation, or general curiosity about innovative activities fit the characteristics of innovators, who in Rogers' (1983) theory of diffusion of innovations are the first group of people ready to implement new ideas and solutions. This group of people may be perceived as a starting point of change which leads to transformation towards sustainability. It is important to emphasise, however, that many factors influence the process of change, such as financial, motivational, or time factors, but from the perspective of the conducted research, it seems that the most important element is the search for solutions that would address individual problems observed in the surrounding reality. This also means that there is a real need to reorient our education, regardless of the educational level, towards supporting strengths, developing creativity, and problem-solving.

The results of our study show that the second most common factor influencing the adoption of the proposed technical solution was the *availability of fresh vegetables*, defined in terms of time – twenty-four hours a day – and space – just behind the door. The issue of autonomy was also important for the participants. The ability to decide which specific plants to grow gave them a sense of belonging to the future project and a certain degree of independence. Additional obligations that participants committed to by applying to participate in the experiment, such as sending scans of grocery receipts, the necessity of weighing and recording future harvests, or participating in conducted research – filling out surveys, participating in interviews – seemed acceptable to them, considering the vision of their own space and influence on the project's results through independently made decisions, consistent with the project's needs and expectations.

The broadly defined availability of vegetables presented above refers to a sense of independence "from", which indicates the need for self-sufficiency and its fulfilment. This fits into the narrative of current trends, in which self-sufficiency appears as a key challenge of reality. The importance of self-sufficiency is particularly significant in the urban context. Land scarcity and needed space are major obstacles to self-sufficiency in urban gardening. Urban food cultivation often faces limited space, requiring local authorities to develop solutions that support local gardeners in their pursuit of self-sufficiency. Furthermore, the complete set of solutions also requires agreements regarding water, securing the area against theft, and responsibility for actions taken, creating the need to develop sustainable resource management practices in urban gardening.

Presenting the results of the conducted research in previous chapters of the monograph also reveals the residents' interest in *educational issues*, although various priorities emerged within this framework. Participation in the offered social experiment was seen as an opportunity to acquire new, additional gardening knowledge and skills in a practical way, happening, as interview participants say, "on the side." Since many of them lead busy professional lives, it is difficult for them to find time among their daily duties for additional learning. Participation in the project would therefore be, in their opinion, an opportunity for learning within educational activities conducted by members of the project team or through peer learning.

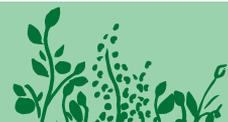

The aspect of learning was important for interview participants both in relation to themselves, as an expansion of current knowledge or the development of skills that they could use in their daily lives, for example, in the context of professional or personal self-development. For some participants, learning was important in relation to family relationships, i.e., the possibility of providing educational support to their children. For other participants, learning in the project was perceived as an investment in the future. According to the results, it may be claimed that education may be perceived as a self-development motivational factor which includes other people to the learning process either by sharing knowledge and gaining it from others. It seems that it reflects the social learning process where other people may be perceived as a source of information. Reaching self-sufficiency in urban gardening necessitates tackling social aspects like community involvement, the exchange of knowledge, and the sharing of resources. Establishing robust community connections and encouraging cooperative practices can enhance the self-sustainability of urban gardening projects by utilising the collective efforts and skills present within the community. However, it may be stated that such educational practices should be a part of the educational process from a very early age.

When considering the decision to join the innovative project, participants also paid attention to *health issues* in two aspects. On the one hand, for them, participating in the project meant the chance to have access to decent quality food products in the form of fresh herbs and vegetables that would be pesticide-free. Being able to control the cultivation, even in the case of unusual growing conditions that were not yet known in detail, of vegetables inspired confidence in future potential gardeners. Secondly, participants expected that easy access to, and frequent sight of, vegetables would result in them consuming a more nutrient-rich diet, regularly ensuring that herbs and vegetables contributed to the meals they prepared, providing them with greater variety and enjoyment. According to the results, urban food self-production seems to influence the perception of eating habits and the quality of food as a source of health. This concept focuses predominantly on a physical aspect of health which plays a vital role especially in times of various crises, such as a Covid-19 pandemic when it was difficult to get fresh food and the food supply chains were disrupted. However, it is worth emphasising that the practice of urban gardening is also a part of social health while communicating with other people, as well as mental health, which can be seen as an antidote to separation from nature, oneself, and other human beings. This component is much connected to the possibility of creating *social networks* among urban dwellers.

For some of the participants, taking part in the experiment would be a chance to establish inter-neighbourly relationships, which they either do not have time for due to the intensity of their working lives or because of difficulties in establishing relationships. The common goal that brings them together during the project activities, the common topic of conversation, the opportunity to exchange food harvests from the cabins are seen by the participants as a chance not only to establish relationships, but also to strengthen existing ones. The fact that participants indicate the need to build social relationships shows that there is clearly a paucity of them. Participants from the second wave of interviews who are representatives of the intergenerational block indicated that current relationships are no longer as strong as they used to be, even though some of them have lived together for quite a long time or have moved back into the block in adulthood.

The findings provide an insight into a great need of being a part of a learning community. The respondents indicated all strategies of social learning processes which indicate how people learn about food self-production, which included learning by observing other people, collaborative learning, immersion in a social support network, and self-experience, reflection, and feedback. The above-mentioned elements are an essential part of a holistic learning towards sustainability.

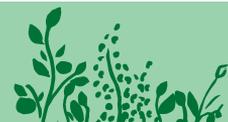

The *opportunity to participate in the project* is another motivating factor that encouraged interview participants to apply for participation in the social experiment. This project was seen by residents as a unique undertaking in which they could become pioneers at the city and country level. The project was associated with diversifying daily routines, the possibility of influencing their own lives and their surroundings. For some participants, it was intriguing that the project was funded by Norwegian sources, which, in their opinion, elevated the status of the event. The willingness to participate in a project that responds to real threats in today's reality shows, on the one hand, a huge demand for seeking solutions in the field of self-sufficient food cultivation, and on the other hand, indicates the need to develop nutritional awareness in this area, which should become an integral component of nutritional education.

Another factor is *financial benefits*, but in relation to the community of the block, not individual household budgets, although this gain was also indirectly taken into account. Specifically, issues related to savings associated with a reduced need to purchase food products were not significant, as these will still be incurred according to participants' opinions. The biggest benefit, as noted by block residents, is the installation of photovoltaic panels, which will remain after the project ends. Thanks to them, in the future, the energy demand for servicing shared areas in the block, such as lighting in staircases and corridors, electricity consumption for maintenance work in common areas, will be covered. Thus, the vision of reducing future electricity bills was visible to the residents. The above-mentioned attitude reveals the environmental awareness of the research participants in terms of the energy transition.

Another factor influencing the decision to participate in the Urban Living Lab is the conviction that the planned cultivation will be a substitute for their own garden. Access to greenery in an urban setting is an important aspect, especially for those residents who, for financial reasons, cannot afford to buy a plot of land. As the experiment itself showed later on, the participants sought contact with greenery and had the habit of frequently observing the plants growing in the cabins. Having one's own garden indicates a great need of re-connectedness to nature, especially in the urban area where the greenery is limited. It may reflect various aspects of providing physical, mental, and social health while communing with nature.

Despite the identified recurring factors influencing the decision to participate in the Urban Living Lab in both examined groups, there were also differences. In the second group, a *local leader* emerged clearly, capable of uniting the local community around a common initiative. Encouraging the undecided residents to participate in the project was a key element of the innovation diffusion process. His passion, energy, personality traits, and substantive preparation in the field of construction and architecture allowed him to dispel doubts among other residents about the project's assumptions and construction changes resulting from its implementation. This highlighted the significant role played by individuals who possess not only enthusiasm but also substantive knowledge in the implementation of innovation. According to Rogers' theory (1983), the presence of a leader highlighted his important function in the innovation diffusion process.

Another motif emerging from the conducted interviews is the *pleasure of contact with nature*. Research results indicate that activities related to urban food cultivation, especially in unusual conditions, appear to residents of the block as a potential source of pleasure. Emotional experiences proved to be extremely important, including: reactions to the possibility of daily observing growing vegetables and herbs, satisfaction derived from the possibility of independently growing one's own vegetables, joy during meals prepared with self-grown herbs and vegetables, especially those that can look aesthetically pleasing due to species and colour diversity; satisfaction and pride that can be shared with family and friends, which is particularly important in the age of prevailing social media. The need to be in nature is ingrained in humans, which is difficult in urban conditions. Such solutions, although they will never replace nature, create space for re-connectedness to nature.

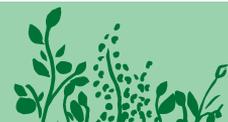

Interestingly, among this relatively small research group, there were three individuals interested in the possibility of *breeding edible insects* as a voluntary and additional project activity. Among them, one indicated this aspect as the main factor in deciding to join the Urban Living Lab. However, it is worth emphasising that Poland is not a country where consuming insects is popular. These results may be surprising, creating space for further in-depth research and analysis.

Responding to the second research question, an interesting conclusion from the study was that the decision-making process to join the project was accompanied by a very limited learning process. Despite the fact that participation in the experiment involved taking responsibility for a valuable cabin located outside their apartment and engaging in activities that represented a certain unknown, residents either did not seek information about the hydroponic food cultivation process at all or were satisfied with seeking basic information, often only after submitting their application. This means that the desire to participate was so strong that it pushed the natural need to gather information about the vegetable cultivation process in these technologically innovative conditions into the background.

The research results suggest that residents did not feel the need to acquire information about hydroponic food cultivation or sought it primarily after submitting their application mainly because they did not have enough free time due to their intense professional work and also because they trusted the project team, assuming that they would provide the necessary knowledge during the implementation stage of the project. The knowledge sought by the residents at this stage was rather superficial; they explained this by the fear that seeking information about hydroponic food cultivation would result in acquiring incorrect knowledge that would hinder the implementation of planned activities during the experiment.

Responding to the third research question regarding the experiences of interview participants related to agricultural activities, we distinguished two main periods of their lives. The first one covered experience gained during childhood and early adulthood, i.e., the period of dependence on parents and caregivers. The obtained results indicate that although these cases are the least numerous, individuals who did not accumulate any experiences related to food cultivation during childhood applied for the project. These were urban-origin individuals without close or distant family in the countryside. Their parents or grandparents also did not have a plot of land suitable for growing food for their own use. These individuals lived in an apartment building where, due to the lack of a balcony suitable for even small sowings or the lack of interest from parents in cultivation matters, there was no custom or family tradition of engaging in agricultural activities.

Another distinguished subgroup of interview participants were individuals who, in early childhood, lived in the city but had occasional experiences related to food cultivation. The obtained results suggest that the farming activities of these participants were most often associated with weekend work on a plot owned by their parents or as part of holiday activities in the form of farm work during leisure time spent with family in the countryside. These activities were largely perceived positively because, as children, they were most often involved in simple agricultural work, which they saw more as a form of leisure than actual work. These activities often ended with additional attractions, such as an evening bonfire. Individual participants who did not speak positively about agricultural experiences from childhood and early adulthood declared that during that time they were not interested in food cultivation and saw the time spent on the plot or in the countryside at their grandparents' as an opportunity to immerse themselves in greenery, contrasting with the urban hustle and bustle and dense development.

The results obtained indicate that the group of participants interested in joining the project consisted of individuals who, in childhood,

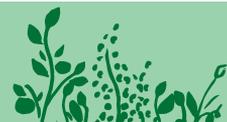

lived in urban areas allowing for regular time spent in the garden, either in the form of a backyard garden or a plot located close to their place of residence. In the case of this group of interview participants, increasingly less enthusiastic statements about experiences related to food cultivation began to resonate. These activities were rather initiated by their parents or grandparents interested in encouraging children to help with occasional gardening tasks. Nonetheless, participants appreciated the benefits of access to fresh and tasty fruits and vegetables grown in the plot or garden as adults. The benefit for them was the opportunity to shape their own independence and responsibility, and for some, the economic benefits, as their work sometimes led to minor financial gratification.

In the case of both of the above-mentioned groups of participants, the research revealed the occurrence of a phenomenon related to a gradual decline in interest in gardening activities with age. The changing interests during adolescence led participants to be increasingly less inclined to visit the plot or help in the backyard garden, limiting their food cultivation-related activities.

The conducted research indicates that individuals with a rural background had the most pronounced emotional attachment to food cultivation. These individuals valued the quality and availability of food from local crops, nostalgically recalling the opportunity to pick and eat vegetables and fruits "straight from the field", enjoying their scent and intense flavour. At the same time, these individuals viewed gardening tasks not as pleasure or satisfaction, but rather as a natural duty or even a necessity associated with physical effort. As children, they clearly participated in diverse types of agricultural work, including those requiring significant physical exertion.

The second period identified by us covered the experiences of interview participants acquired in adulthood, i.e., the period of independent living. The research results indicate that during this period, attitudes and approaches to food cultivation underwent changes, but these changes were dependent on living conditions and access to re-sources enabling agricultural activities. With varying degrees of intensity, residents tried to engage in such activities.

Seven interview participants declared that they did not engage in food cultivation activities in their adult lives. Firstly, this was due to the lack of need for such cultivation associated with access to fresh fruits and vegetables. These were individuals from rural backgrounds who relied on the harvests of their parents and grandparents because the amount of food collected was sufficient to meet the needs of the entire family. Secondly, the lack of gardening activity stemmed from a lack of time to engage in additional activities due to numerous and time-consuming professional obligations.

Our results suggest that one-third of the study participants declared cultivating small fruits such as strawberries or raspberries, vegetables such as cherry tomatoes, bell peppers, and herbs on their balcony or terrace. Balcony conditions such as surface area and sunlight availability determined the type of plants grown, sometimes limiting the choice to those easily cultivated, such as lettuce, basil, mint, chives, or rosemary. Nonetheless, the balcony was the only place where many participants could engage in gardening activities, sometimes ending in failure and abandonment of continuation in the following season. An alternative to the balcony or terrace, for every fifth interview participant, was growing herbs on the windowsill inside the apartment, not only in the kitchen but also in the living rooms.

The obtained results indicate that four out of ten interview participants cultivated fruits and vegetables on their own plot, inherited from their grandparents or parents or purchased independently. In the first case, the plots or gardens had been cultivated for many years, thus requiring the continuation of gardening practices started by the family. In the second case, participants mostly limited themselves to the care of plants that did not require extensive work, such as raspberry bushes, blueberries, fruit trees, or mainly used the plot for recreational purposes.

The results of our research indicate that individuals who volunteered for the project are those who, although they do not have their own crops, help with agricultural work to their parents or grandparents, often during periods of intensive agricultural work when parents need additional assistance. The motivation for undertaking this activity is a sense of duty or a desire to help the family. Among those interested in hydroponic food cultivation in the block were also three individuals who use plots belonging to friends or family, where they grow vegetables and fruits for their own needs together. In this case, gardening activities stem from interests, hobbies, and a desire to spend time with loved ones. The last group declaring experiences in vegetable cultivation in adulthood consists of three individuals who have access to garden plots adjacent to their ground-floor apartments. The opportunity to use such a plot usually encourages the maintenance of grass or flowers, but also allows for the cultivation of small, easy-to-care-for food plants, such as eggplants.

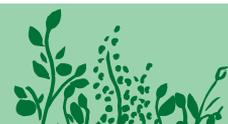

Responding to the fourth research question regarding educational experiences during activities related to food cultivation, as in the previous question, we referred to the childhood and adulthood periods. The results of the conducted research show that in the case of childhood, despite spending a significant amount of time on agricultural work, adults did not consider this time as effective in terms of acquiring knowledge about food cultivation. The work performed in childhood was a response to the needs of the family; it was a form of obligation rather than genuine childhood interests, and thus, in the respondents' opinion, it was almost mechanically performed, without accompanying educational curiosity. Learning took place rather through observation than instruction and in an unintentional manner, almost incidentally. Children did not strive to acquire knowledge from adults. Adults also did not show increased interest in imparting knowledge, as they were most likely absorbed in intense and strenuous physical work, which did not incline them to make additional efforts for intergenerational education.

This process differed in multi-generational rural families, where parents aimed for their children to continue working in agriculture, while children shared the same interests. In these families, knowledge was actively transmitted, not only through observation but also through active peer learning. Parents showed how to care for plants to ensure bountiful yields. The knowledge and skills acquired in this way paid off in adulthood in the form of continuing these activities as part of a chosen profession or hobby. In adulthood, these individuals clearly valued the competencies of their parents or grandparents. The obtained results indicate that in case of any doubts, these individuals first call their parents or grandparents for clarification on a given issue or rely solely on the opinion of their family, drawing knowledge and skills from them. Additionally, besides formal vocational education, these individuals declared that they regularly read guides and professional journals and participate in formal courses to enhance their agricultural skills.

In the conducted research, voices appeared stating that some respondents did not seek any information about food cultivation. Among the reasons for the lack of such activity were issues of lack of time due to intense professional work or lack of need because these individuals did not have conditions for home food cultivation and did not have a plot. Some respondents declared that they do not undertake educational activities because they do not have difficulty with small-scale cultivation. Only failures in the form of poor plant growth or other unsatisfactory results would prompt them to seek the necessary information.

A similar dependency can also be observed regarding individuals who, while engaging in educational activities, find this process rather superficial and serve more to dispel doubts or solve immediate problems than to a planned lifelong learning process. The research results show that for this purpose, participants most often use internet resources. Moreover, the source of knowledge is not dedicated expert portals presenting the issue comprehensively, but rather search engine results obtained after entering the appearing problem or query. According to the respondents, the advantage of the Internet is its accessibility and ease of use.

According to the respondents, the food cultivation process is not overly complicated and should not pose difficulties for them, so it does not require acquiring advanced knowledge. Additionally, participants feel that they have the necessary knowledge from childhood and ear-ly youth, which they acquired by observing the work of close family members and actively participating in agricultural work, so they do not need additional support. The motif of lack of time also repeated itself, preventing them from delving into the topic of food cultivation.

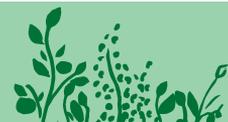

Among the other sources used by study participants to acquire information about food cultivation are books, guides, television programs, or articles in magazines, with participants also using these sources sporadically or to a basic extent, as in the case of online sources. An example of basic learning may be using only the instructions provided on the packaging of purchased seeds or seedlings. Less popular were also participation in courses, training sessions, or workshops, with one person indicating the use of knowledge accumulated during school education.

Our research shows that among the respondents, social learning is also valued in the form of seeking advice from friends, sharing knowledge with acquaintances, often during social gatherings or conversations at work. Some respondents mentioned consulting certain issues with farmers selling seedlings at the market. Other individuals drew from experiences gained during practical work in the garden under the guidance of parents or grandparents or later work in their own garden. These individuals stated that the most valuable knowledge and skills for them are acquired through trial and error.

The presented research results regarding the educational context fit into the concept of lifelong learning, which assumes learning throughout one's life. They reflect changes in interests and attitudes towards food cultivation at various stages of life. They perfectly illustrate that project participants showed different levels of engagement in the learning process, with a clear dependence on previous experiences, professional interests, and time availability. However, it is encouraging that regardless of the experience, both positive and negative, the respondents were open to testing new solutions and a willingness to make changes in their lives, suggesting the need to support participants in a systematic and continuous learning process through education methods tailored to their needs and lifestyle.

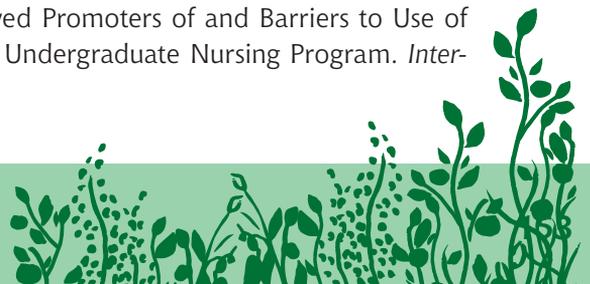

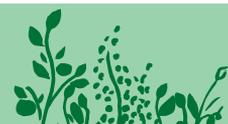

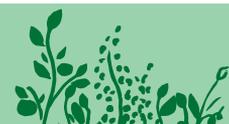

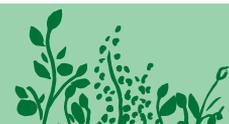

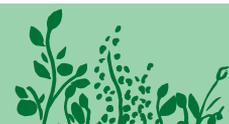

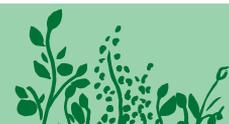

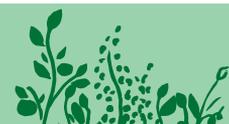

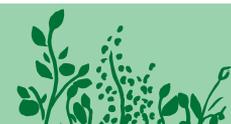

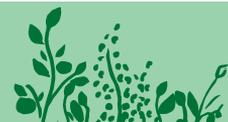

# Tables

# Figures

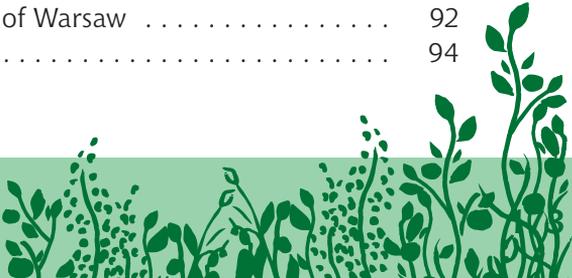

The monograph "Urban Food Self-Production in the Perspective of Social Learning Theory: Empowering Self-Sustainability" by Ewa Duda and Adamina Korwin-Szymanowska offers an innovative exploration into urban food self-production through a blend of social learning theory and the diffusion of innovation. This work is particularly noteworthy for its novelty in addressing urban sustainability and social change through self-sufficient food production. Some people may say that it is just returning to the roots. To the wisdom of our grandparents who had to be frugal due to financial hardship and lack of resources. They already grew food in the cities. Yes, however the rationale, the efficiency and the scale were completely different. The embedment in the Social Learning Theory and the showcasing of a unique study makes this monograph stand out as the source of new knowledge.

From the review by Professor Anna Odrowąż-Coates

The monograph explores a socially important issue related to the functioning of modern cities and their inhabitants. It provides insight into contemporary considerations and actions concerning innovative socio-technological solutions undertaken for sustainable food production and consumption. Due to its interdisciplinarity, it can be a valuable source of both scientific inspirations and practical actions.

From the review by Professor Aleksandra Tłuściak-Deliowska

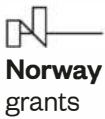 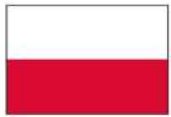 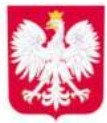 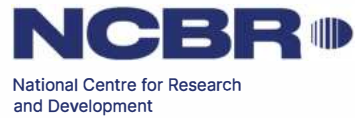

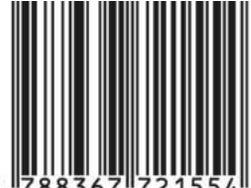

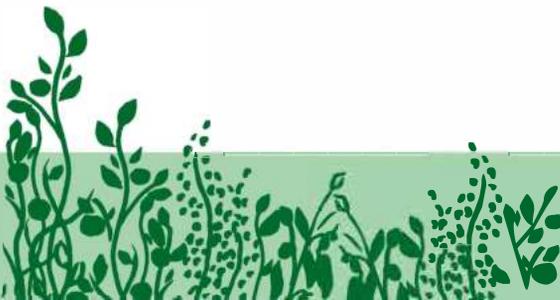